\documentclass[journal=jacsat,manuscript=article]{achemso}

\usepackage[version=3]{mhchem} 
\usepackage{chemformula} 
\usepackage[T1]{fontenc} 
\usepackage{longtable}
\usepackage{adjustbox}
\usepackage{lscape}
\usepackage{hyperref}
\usepackage{nameref}

\newcommand*\arcsec{\ensuremath{^{\prime\prime}}}

{\small
\author{Niels F.W. Ligterink}
\email{niels.ligterink@unibe.ch}
\affiliation[]{\small Physics Institute, University of Bern, Sidlerstrasse 5, 3012 Bern, Switzerland}
\author{Aida Ahmadi}
\affiliation[]{Leiden Observatory, Leiden University, PO Box 9513, 2300 RA Leiden, The Netherlands}
\author{Bijaya. Luitel}
\affiliation[]{Yale-NUS College, \#01-220, 16 College Avenue West, 138527, Singapore}
\author{Audrey Coutens}
\affiliation[]{Institut de Recherche en Astrophysique et Planétologie, Université de Toulouse, UPS-OMP, CNRS, CNES, 9 av. du Colonel Roche, 31028 Toulouse Cedex 4, France}
\author{Hannah Calcutt}
\affiliation[]{Institute of Astronomy, Faculty of Physics, Astronomy and Informatics, Nicolaus Copernicus University, Grudziadzka 5, 87-100 Torun, Poland}
\author{\L{}ukasz Tychoniec} 
\affiliation[]{ESO, Karl-Schwarzschild-Strasse 2, 85748 Garching bei München, Germany}
\author{Harold Linnartz}
\affiliation[]{Laboratory for Astrophysics, Leiden Observatory, Leiden University, PO Box 9513, 2300 RA Leiden, The Netherlands}
\author{Jes K. J{\o}rgensen}
\affiliation[]{Centre for Star and Planet Formation, Niels Bohr Institute \& Natural History Museum of Denmark, University of Copenhagen, {\O}ster Voldgade 5--7, 1350 Copenhagen K., Denmark}
\author{Robin T. Garrod}
\affiliation[]{Departments of Chemistry and Astronomy, University of Virginia, Charlottesville, VA 22904, USA}
\author{Jordy Bouwman}
\affiliation[]{Leiden Observatory, Leiden University, PO Box 9513, 2300 RA Leiden, The Netherlands}
\alsoaffiliation[]{Laboratory for Atmospheric and Space Physics, Department of Chemistry, Institute for Modeling Plasma, Atmospheres, and Cosmic Dust (IMPACT), University of Colorado, Boulder, CO 80303, USA}
}

\title[The prebiotic molecular inventory of Serpens SMM1]
  {\Large The prebiotic molecular inventory of Serpens SMM1: \\ II. The building blocks of peptide chains}


\begin{document}
\setcounter{secnumdepth}{2}

\newpage

\begin{abstract}
This work aims to constrain the abundances of interstellar amides, by searching for this group of prebiotic molecules in the intermediate-mass protostar Serpens SMM1-a. ALMA observations are conducted toward Serpens SMM1. A spectrum is extracted toward the SMM1-a position and analyzed with the CASSIS line analysis software for the presence of characteristic rotational lines of a number of amides and other molecules. NH$_{2}$CHO, NH$_{2}$CHO $\nu_{12}$=1, NH$_{2}^{13}$CHO, CH$_{3}$C(O)NH$_{2}$ $\nu$=0,1, CH$_{2}$DOH, CH$_{3}$CHO, and CH$_{3}$C(O)CH$_{3}$ are securely detected, while trans-NHDCHO, NH$_{2}$CDO, CH$_{3}$NHCHO $\nu$=0,1, CH$_{3}$COOH, and HOCH$_{2}$CHO are tentatively identified. The results of this work are compared with detections presented in the literature. A uniform CH$_{3}$C(O)NH$_{2}$/NH$_{2}$CHO ratio is found for a group of interstellar sources with vast physical differences. A similar ratio is seen for CH$_{3}$NHCHO, based on a smaller data sample. The D/H ratio of NH$_{2}$CHO is about 1--3\% and is close to values found in the low-mass source IRAS~16293--2422B. The formation of CH$_{3}$C(O)NH$_{2}$ and NH$_{2}$CHO is likely linked. Formation of these molecules on grain surfaces during the dark cloud stage is a likely scenario. The high D/H ratio of NH$_{2}$CHO is also seen as an indication that these molecules are formed on icy dust grains. As a direct consequence, amides are expected to be present in the most pristine material from which planetary systems form, thus providing a reservoir of prebiotic material.
\end{abstract}

\section{Introduction}

Peptide chains, amino acids connected by a R$_{1}$-C(=O)N-R$_{2}$R$_{3}$ group, are an essential component of life as we know it, since these molecules can fold into proteins, which, in turn, are the engines that life runs on. During protein biosynthesis, ribosomes catalyze condensation reactions between amino acids, resulting in the formation of a peptide bond, also called an amide, and a water molecule\citep{lucas-lenard1971}. However, to understand how the peptide chains formed before life was present, abiotic processes need to be studied.   

Several mechanisms have been proposed for abiotic peptide synthesis, involving catalysis by minerals\citep{mckee2018}, in hydrothermal vents\citep{imai1999,lemke2009}, with UV light\citep{simakov1996}, or by impact shocks\citep{sugahara2014}. The last mechanism is particularly interesting, because it hints at an extraterrestrial origin of peptide chains. In fact, in the laboratory it has been demonstrated that dipeptides such as glycine-glycine and alanine-leucine can form in interstellar ice analogues\citep{kaiser2013}. Incorporation of these interstellar peptides into planetary building material and subsequent delivery by impactors like comets, could have helped start life on the Early Earth. 

To understand interstellar peptide chemistry, observations of star-forming regions can help elucidate how these molecules form. While observations of dipeptides and larger molecules are beyond current capabilities, several peptide-like molecules have been detected with radio and mm-wave observations. Formamide (NH$_{2}$CHO) has the same chemical structure as a peptide bond and is the simplest peptide-like molecule. It is the first peptide-like molecule to be detected in the interstellar medium (ISM) and is routinely found in extraterrestrial environments, ranging from star-forming regions to comets\citep{rubin1971,coutens2016,biver2021}. Since the detection of formamide, several other peptide-like molecules have been detected, such as acetamide (CH$_{3}$C(O)NH$_{2}$)\citep{hollis2006}, N-methylformamide (CH$_{3}$NHCHO)\citep{belloche2017,belloche2019,ligterink2020a,colzi2021}, and urea (also known as carbamide, NH$_{2}$C(O)NH$_{2}$)\citep{belloche2019,jimenez-serra2020}. Searches for and tentative detections of several other peptide-like molecules have been reported, such as glycolamide (HOCH$_{2}$C(O)NH$_{2}$)\citep{sanz-novo2020,colzi2021}, cyanoformamide (N$_{2}$C(O)CN)\citep{colzi2021}, and propionamide (CH$_{3}$CH$_{2}$C(O)NH$_{2}$)\citep{li2021}.

Observations of these molecules provide insight into the chemical processes that result in their formation, which, in turn, indicate whether larger amides or even peptides are expected to form in the ISM. For example, correlations found between interstellar abundances of NH$_{2}$CHO and CH$_{3}$C(O)NH$_{2}$ hint that there is a physico-chemical link between these two molecules \citep{ligterink2020b,colzi2021}. The similarity in the ratios of these molecules in a sample of sources with very different physical characteristics, is explained by an early formation during the star formation cycle, presumably on ice-coated interstellar dust grains. Furthermore, this observation hints that these molecules form in related chemical reactions or even directly act as a precursor. 

While observations of peptide-like molecules are valuable, detections of these molecules are still relatively sparse. Furthermore, detections are biased toward luminous high-mass protostars and star-forming regions. Peptide-like molecules detected toward low-mass protostars almost exclusively consists of NH$_{2}$CHO, with as only exception a tentative detection of CH$_{3}$C(O)NH$_{2}$ toward the low-mass protostar IRAS~16293--2422B\citep{ligterink2018a}. Identifications toward low-mass and Sun-like protostars will extend our knowledge of interstellar amide formation.

In this publication, observations of Serpens SMM1-a (hereafter SMM1-a), an intermediate-mass Class 0 protostar in the Serpens cloud (D $\approx$ 436.0$\pm$10 pc\citep{ortiz-leon2017}), are used. In particular, the outflows of the SMM1 region are well studied\citep{dionatos2013,hull2016}, but also its molecular inventory\citep{oberg2011,goicoechea2012,ligterink2020c,tychoniec2021}. The comparatively high luminosity of $\sim$100 $L_{\odot}$ makes this an interesting source in the low-end protostellar mass range to search for weak spectral lines of prebiotic and peptide-like molecules. 

This work presents the first detection of NH$_{2}$CHO, CH$_{3}$C(O)NH$_{2}$, and various isotopologues toward SMM1-a, combined with a tentative detection of CH$_{3}$NHCHO and upper limits of various other peptide-like molecules. The observations and analysis are presented in Sect. \nameref{sec:obs&analysis}, followed by the results in Sect. \nameref{sec:results}. These results are discussed in Sect. \nameref{sec:discussion} and conclusions are given in Sect. \nameref{sec:conclusion}.

\section{Observations \& Analysis}\label{sec:obs&analysis}

The observations of Serpens SMM1 and the analysis methods are introduced in \citet{ligterink2020c} and only a brief overview will be given here. 

Observations of SMM1 were conducted on 27-03-2019 in ALMA cycle 6 for project \#2018.1.00836.S (PI: N.F.W. Ligterink). The source was observed toward the phase centre $\alpha_{\rm J2000}$ = 18:29:49.80 $\delta_{\rm J2000}$ = +01:15:20.6 in select frequency windows between 217.59 and 235.93 GHz. The spectral resolution was 488.21 kHz ($\sim$0.33~km~s$^{-1}$, several bands between 217.59 and 233.65 GHz) and 1952.84 kHz ($\sim$1.25~km~s$^{-1}$, between 234.06 and 235.93 GHz) for the continuum band, while the spatial resolution was $1.32\arcsec\times1.04\arcsec$. The spectrum of the SMM1-a hot core was extracted toward position $\alpha_{\rm J2000}$ = 18:29:49.793, $\delta_{\rm J2000}$ = +1.15.20.200 in a beam of $1.32\arcsec\times1.04\arcsec$. The background temperature was determined to be $\sim$5~K. 

The CASSIS\footnote{Based on analysis carried out with the CASSIS software (http://cassis.irap.omp.eu; \citet{vastel2015}. CASSIS has been developed by IRAP-UPS/CNRS.} line analysis software was used to analyze the spectra. Line lists taken from the Cologne Database for Molecular Spectroscopy (CDMS)\citep{muller2001,muller2005,endres2016}, the JPL database for molecular spectroscopy\citep{pickett1998}, and from the literature were used to identify and fit spectral lines. These databases are also used to identify the lines that are largely free of blending transitions originating from other molecules. Largely clean lines are defined as transitions that have an almost negligible contribution of a blending transition (approximately 10\% of the observed line intensity) or only blend with another species in the wing of the transition. Only largely clean lines were used for subsequent line fitting. A synthetic spectrum consisting of the emission of all 26 securely and tentatively detected species detected toward SMM1-a in this work and in \citet{ligterink2020c} was generated to check against and verify that detected lines were unblended. An overview of the spectroscopic data of the species discussed in this paper are given in the spectroscopy supporting information section, Table \ref{tab:SMM1_lines}. 

For a given column density ($N_{\rm T}$), excitation temperature ($T_{\rm ex}$), source velocity ($V_{\rm LSR}$), line width ($\Delta V$), background temperature ($T_{\rm BG}$ = 5.2~K, derived from the average continuum flux density of 0.41 Jy beam$^{-1,}$\citep{ligterink2020c}), and source size, which is assumed to be equal to the beam size ($\theta_{\rm source}$ = 1.2$\arcsec$), a synthetic spectrum was generated, assuming Local Thermodynamic Equilibrium (LTE) conditions. This synthetic spectrum was initially fit by-eye to the rotational lines identified in the observed spectrum. Next, a $\chi^{2}$ minimization routine in combination with the Monte-Carlo Markov Chain (MCMC) algorithm was applied to find the best fit to the observed spectrum. The column density was given as a free parameter over two orders of magnitude around the by-eye fit column density, while $T_{\rm ex}$ = 50 -- 350~K, $\Delta V$ = 1.0 -- 4.0 km s$^{-1}$, and $V_{\rm LSR}$ = 6.0 -- 9.0 km s$^{-1}$. In certain cases, $T_{\rm ex}$, $\Delta V$, or $V_{\rm LSR}$ could not properly be fit and instead estimated values were given. In most cases, these estimates were based on an average of values found for other molecules where a proper fit was possible or based on related species, for example isotopologues. 

A secure identification is based on the detection of at least 3 largely unblended lines for species that have prominent spectral lines and are routinely observed in interstellar observations (e.g., NH$_{2}$CHO, CH$_{3}$OH), while 5 or more lines are required to claim a detection of species that are less routinely observed or have comparatively weak spectral lines (e.g., CH$_{3}$C(O)NH$_{2}$ or glycolaldehyde, HOCH$_{2}$CHO). If no lines are identified of a species, an upper limit is determined that is based on the 3$\sigma$ noise level of a line of the species that is found in a line-free segment of the spectrum.

From the average continuum flux density (0.41 Jy beam$^{-1}$\citep{ligterink2020c}), the H$_{2}$ column density can be estimated following equation\citep{schuller2009,ahmadi2018}:

\begin{equation} 
    N(H_{\rm 2}) = \frac{F_{\rm \nu}R}{B_{\rm \nu}(T_{\rm D}) \Omega \kappa_{\rm \nu} \mu m_{\rm H}},
    \label{eq:h2column}
\end{equation}

where $F_{\rm \nu}$ is the flux density, $R$ the gas-to-dust mass ratio, $B_{\rm \nu}(T_{\rm D})$ the planck function for dust temperature $T_{\rm D}$, $\Omega$ the beam solid angle, $\kappa_{\rm \nu}$ the dust absorption coefficient, $\mu$ the mean molecular weight, and $m_{\rm H}$ the atomic hydrogen mass. $R$ is assumed to be 100, $\kappa_{\rm 220 GHz}$ = 0.9 cm$^{2}$ g$^{-1,}$\citep{ossenkopf1994}, and $\mu$ = 2.8. The dust temperature is assumed to be coupled to the gas temperature. The excitation temperature of the dense gas tracer CH$_{3}$CN\citep{ligterink2020c} is adopted, resulting in $T_{\rm D}$ = 190~K. From this, a hydrogen column density of $N(H_{2})$ = (1$\pm$0.2)$\times$10$^{24}$ cm$^{-2}$ is estimated.

\section{Results}\label{sec:results}

\begin{figure*}
\includegraphics[width=\hsize]{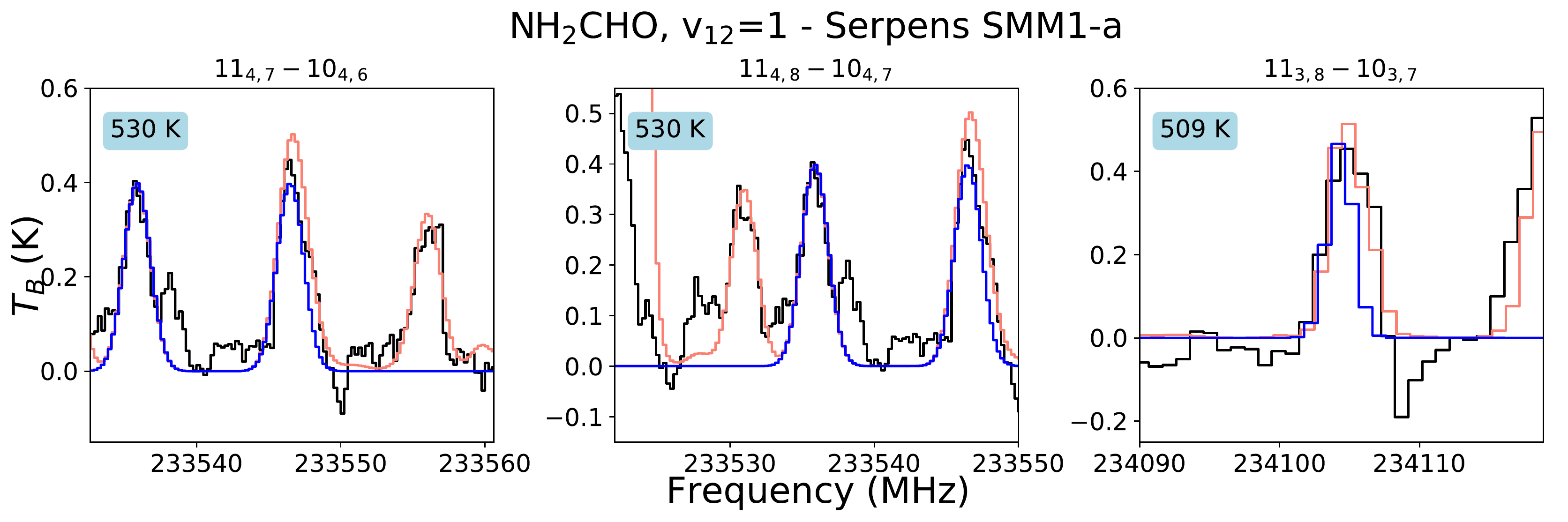}
\caption{Identified lines of NH$_{2}$CHO, $\nu_{12}$=1 toward SMM1-a. The observed spectrum is plotted in black, with the synthetic spectrum of the species overplotted in blue, and the synthetic spectrum of all fitted species combined in red. The transition is indicated at the top of each panel and the upper state energy is given in the top left of each panel. Detected transitions of NH$_{2}$CHO $\nu$=0, NH$_{2}^{13}$CHO, NH$_{2}$CDO, and NHDCHO are presented in the supporting information section.}
\label{fig:lines_NH2CHOv12}
\end{figure*}

\begin{figure*}
\includegraphics[width=\hsize]{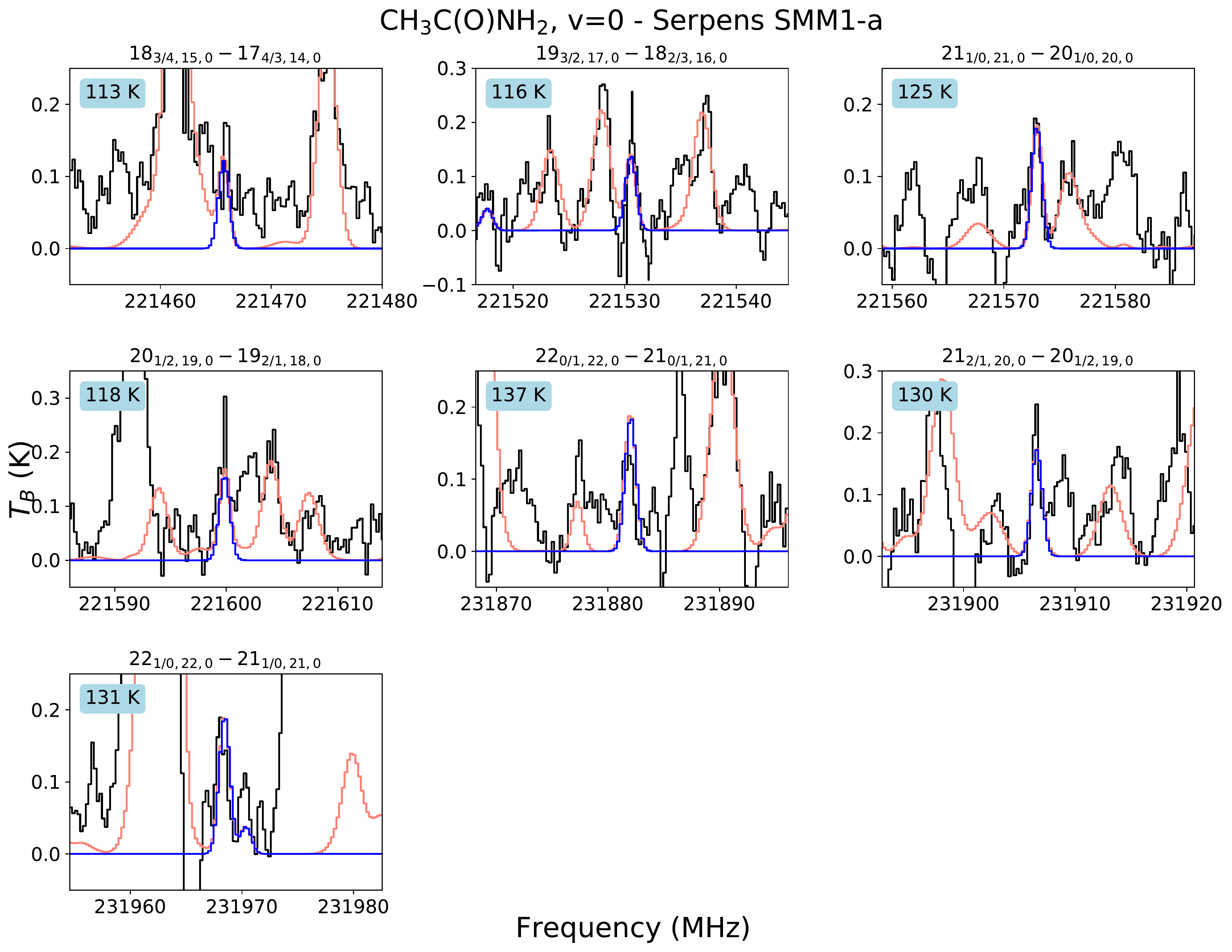}
\caption{Identified lines of CH$_{3}$C(O)NH$_{2}$, $\nu$=0 toward SMM1-a. The observed spectrum is plotted in black, with the synthetic spectrum of the species overplotted in blue, and the synthetic spectrum of all fitted species combined in red. The transition is indicated at the top of each panel and the upper state energy is given in the top left of each panel. Detected transitions of CH$_{3}$C(O)NH$_{2}$, $\nu$=1 are presented in the supporting information.}
\label{fig:lines_CH3CONH2v0}
\end{figure*}

\begin{figure*}
\includegraphics[width=\hsize]{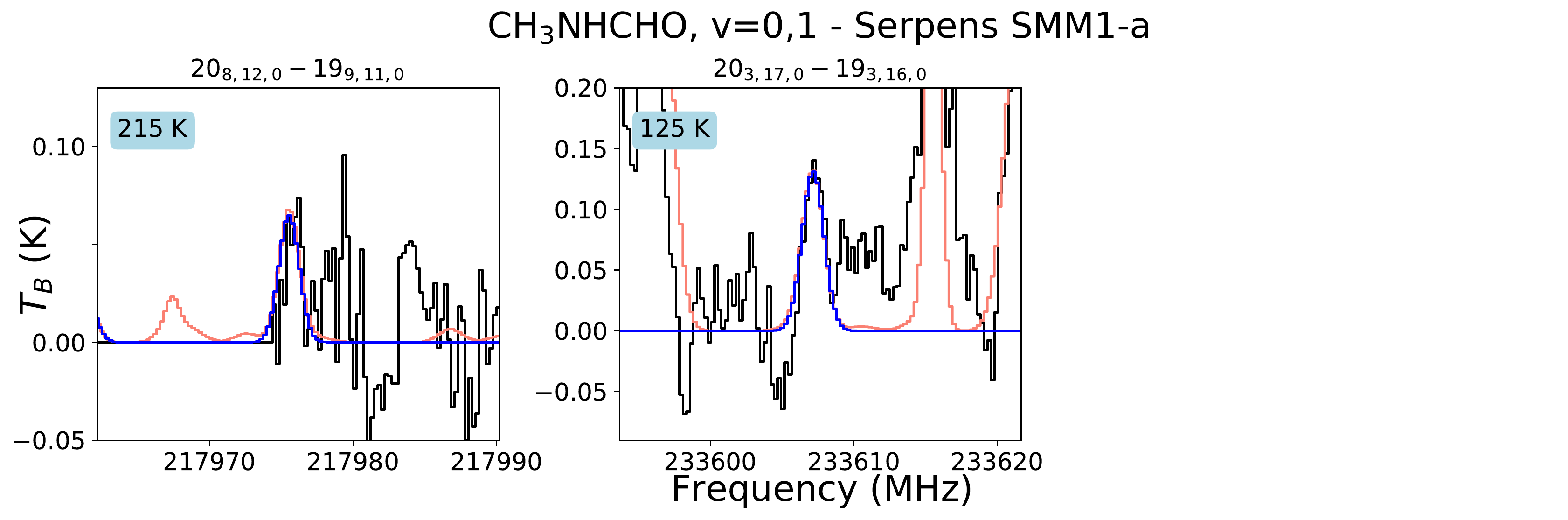}
\caption{Identified lines of CH$_{3}$NHCHO, $\nu$=0,1 toward SMM1-a. The observed spectrum is plotted in black, with the synthetic spectrum of the species overplotted in blue, and the synthetic spectrum of all fitted species combined in red. The transition is indicated at the top of each panel and the upper state energy is given in the top left of each panel.}
\label{fig:lines_CH3NHCHOv01}
\end{figure*}

\begin{figure*}
\includegraphics[width=\hsize]{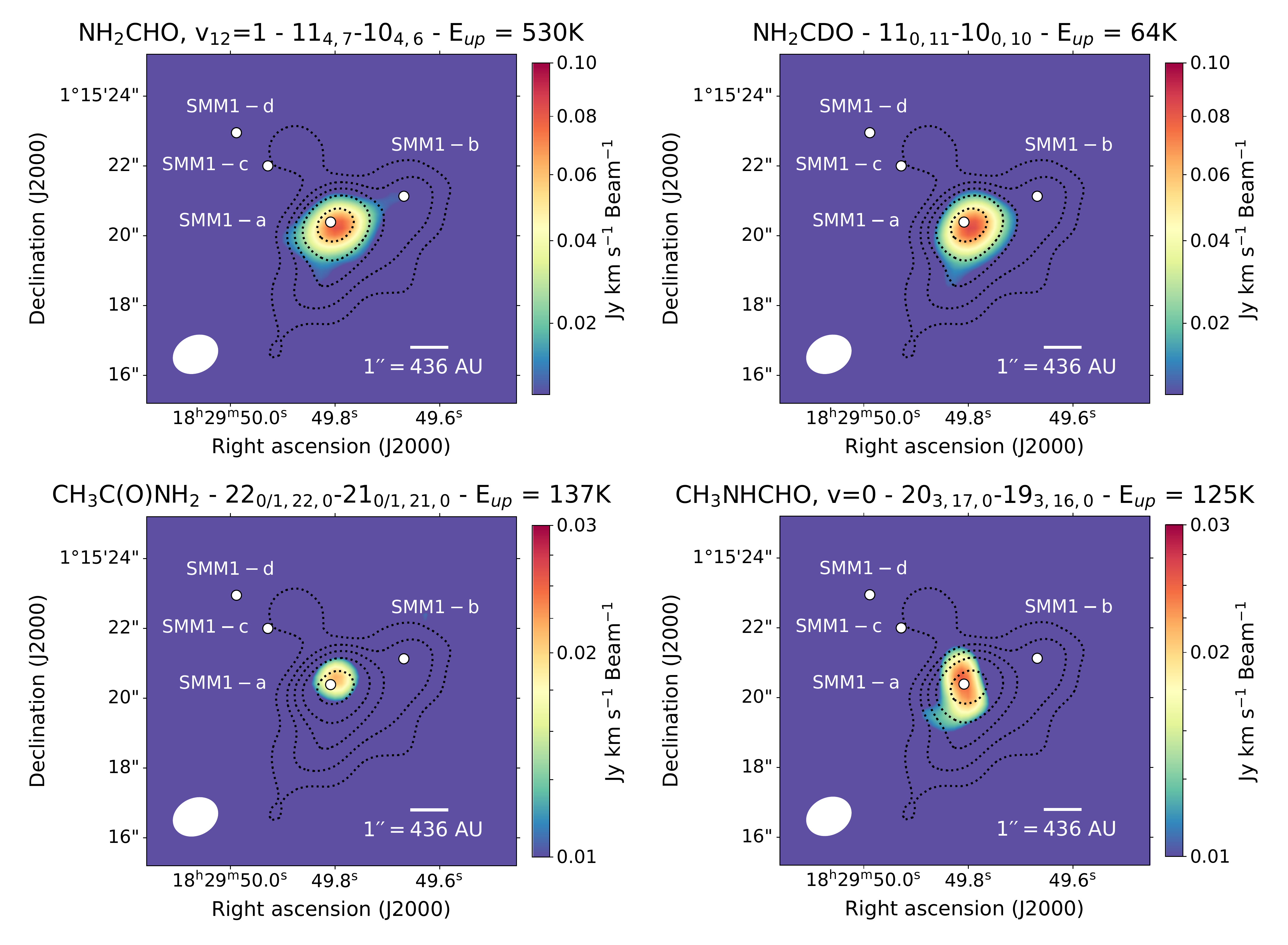}
\caption{Moment 0 maps of lines of NH$_{2}$CHO, $\nu_{12}$=1, NH$_{2}$CDO, CH$_{3}$C(O)NH$_{2}$, and CH$_{3}$NHCHO toward SMM1. All lines are integrated over eight velocity bins, centred on the peak frequency of each line as determined toward SMM1-a. Positions of protostars in the SMM1 region are indicated and the beam size (1.32$\arcsec\times$1.04$\arcsec$) is visualised in the bottom left corner. Dust continuum contours are given by the black dotted line at the levels of 0.02, 0.05, 0.1, 0.2, and 0.5 Jy Beam$^{-1}$.}
\label{fig:mom0}
\end{figure*}

In the SMM1-a spectrum several lines originating from amides are identified, leading to the secure detection of NH$_{2}$CHO $\nu_{12}$=1, CH$_{3}$C(O)NH$_{2}$ $\nu$=0,1, NH$_{2}$CHO $\nu$=0, and NH$_{2}^{13}$CHO (see Figs. \ref{fig:lines_NH2CHOv12} and \ref{fig:lines_CH3CONH2v0} and supplementary information Figs. \ref{fig:lines_CH3CONH2v1}, \ref{fig:lines_NH2CHO}, and \ref{fig:lines_13-NH2CHO}, while CH$_{3}$NHCHO $\nu$=0,1, NH$_{2}$CDO, and trans-NHDCHO are tentatively identified (see Fig. \ref{fig:lines_CH3NHCHOv01} and supplementary information Figs. \ref{fig:lines_NH2CDO} and \ref{fig:lines_trans-NHDCHO}). Moment 0 maps of the emission of selected NH$_{2}$CHO $\nu_{12}$=1, NH$_{2}$CDO, CH$_{3}$C(O)NH$_{2}$ $\nu$=0, and CH$_{3}$NHCHO $\nu$=0 lines are shown in Fig. \ref{fig:mom0}. These maps show compact emission of these amides toward the peak continuum of SMM1-a. The difference in the extent of the emission between the NH$_{2}$CHO isotopologues and the larger CH$_{3}$C(O)NH$_{2}$ and CH$_{3}$NHCHO molecules, probably results from the lower line brightness of the latter two species, but may also be due to different chemical responses to the environment, for example the gas density\citep{calcutt2014}. The spatial resolution of these observations is insufficient to resolve finer features in the distribution of these molecules throughout the source and to give a definitive answer to this question.

The excitation temperature could only be determined for CH$_{3}$C(O)NH$_{2}$, $\nu$=0, albeit with a large error bar, at $T_{\rm ex}$ = 235$\pm$80~K. For the CH$_{3}$C(O)NH$_{2}$, $\nu$=1 fit the excitation temperature of its ground vibrational state is adopted. The remaining species were fit with $T_{\rm ex}$ = 200~K, the average temperature that could be securely determined from the detected molecular species toward SMM1-a\citep{ligterink2020c}. Other excitation temperatures are possible, which can affect the derived column density. For $T_{\rm ex}$ of 100 to 300~K, temperatures that are generally realistic for hot core and corinos, the column density can vary by about a factor of two. This temperature variation may also affect CH$_{3}$C(O)NH$_{2}$, for which an excitation temperature is determined. The column densities of its ground and first excited vibrational states differ substantially at (2.2$\pm$1.1)$\times$10$^{14}$ and (6$\pm$3)$\times$10$^{14}$ cm$^{-2}$, respectively. Due to the large error bar on the determined excitation temperature it is possible that this species is excited at a substantially different temperature. While no other $T_{\rm ex}$ was found for which the column densities of the two states converged, the column densities are consistent within the uncertainties.

For amides for which accurate spectroscopic laboratory data are available, but no convincing lines are found in our observed spectra, upper limits on the column density have been derived. This applies to the following amides (between brackets the line frequency is listed from which the upper limit is calculated): $^{15}$NH$_{2}$CHO (233~523.8 MHz), cis-NHDCHO (234~776.2 MHz), NH$_{2}$CH$^{18}$O (231~843.5 and 235~678.6 MHz), NH$_{2}$C(O)NH$_{2}$ (221~615.8 MHz), NH$_{2}$C(O)CN (217~621.9 and 218~497.0 MHz), and HOCH$_{2}$C(O)NH$_{2}$ (234~277.2 MHz). 

Finally, lines of several oxygen-COMs are also detected. While not the aim of this work, these species are analysed to account for line blending with the amides. Furthermore, some species are of considerable prebiotic interest, such as glycolaldehyde (HOCH$_{2}$CHO), the simplest sugar-like molecule. In the SMM1-a spectrum CH$_{2}$DOH, acetaldehyde (CH$_{3}$CHO), and acetone (CH$_{3}$C(O)CH$_{3}$) are securely identified (see supplementary information Figs. \ref{fig:lines_CH2DOH}, \ref{fig:lines_CH3CHO}, and \ref{fig:lines_CH3COCH3}), while acetic acid (CH$_{3}$COOH) and HOCH$_{2}$CHO are tentatively detected (see supplementary information Figs. \ref{fig:lines_CH3COOH} and \ref{fig:lines_HOCH2CHO}). 

The number of identified lines and fit parameters of all species are presented in Table \ref{tab:abundances}, while the line parameters of the identified transitions are listed in Table \ref{tab:SMM1_lines} in the supporting information.

\begin{landscape}
\begin{table}[h]
\caption{Best fit parameters for the molecules investigated in this work.} 
\label{tab:abundances}      
\centering          
\resizebox{\columnwidth}{!}{\begin{tabular}{l c r c c c r r r}    
\hline\hline  
Molecule & Lines & $N_{\rm T}$ & $T_{\rm ex}$ & $V_{\rm LSR}$ & $\Delta V$ & [X] / [NH$_{2}$CHO]$^{a}$ & [X] / [CH$_{3}$OH]$^{b}$ & [X] / [H$_{2}$] \\ 
 & \# & (cm$^{-2}$) & (K) & (km s$^{-1}$) & (km s$^{-1}$) & & & \\ 
\hline                         
   NH$_{2}^{13}$CHO   &   2   &   (7.5$\pm$1.2)$\times$10$^{13\phantom{,c]}}$   &   [200]   &   6.8$\pm$0.2   &   2.6$\pm$0.4   &   --   &   (7$\pm$3)$\times$10$^{-5}$   &   (8$\pm$4)$\times$10$^{-11}$ \\ 
   NH$_{2}^{12}$CHO$^{c}$   &   7   &   [(3.9$\pm$1.3)$\times$10$^{15,c}$]   &   --   &   --   &   --   &   1.0$\times$10$^{0}$   &   (3.5$\pm$1.7)$\times$10$^{-3}$ & (3$\pm$2)$\times$10$^{-9\phantom{0}}$\\
   NH$_{2}$CHO, $\nu_{12}$=1   &   3   &   (1.0$\pm$0.2)$\times$10$^{15\phantom{,c]}}$   &   [200]   &   6.9$\pm$0.1   &   2.8$\pm$0.2   &   (2.6$\pm$1.0)$\times$10$^{-1}$   &   (9$\pm$4)$\times$10$^{-4}$ & (1.0$\pm$0.5)$\times$10$^{-9\phantom{0}}$\\
   NH$_{2}$CDO   &   2   &   $\sim$(1.4$\pm$0.3)$\times$10$^{14\phantom{,c]}}$   &   [200]   &   6.4$\pm$0.2   &   3.0$\pm$0.5   &   $\sim$(3.6$\pm$1.4)$\times$10$^{-2}$   &   $\sim$(1.3$\pm$0.5)$\times$10$^{-4}$ & $\sim$(1.4$\pm$0.7)$\times$10$^{-10}$ \\ 
   trans-NHDCHO   &   3   &   $\sim$(4.3$\pm$1.6)$\times$10$^{13\phantom{,c]}}$   &   [200]   &   7.4$\pm$0.6   &   3.3$\pm$0.4   &   $\sim$(1.1$\pm$0.6)$\times$10$^{-2}$   &   $\sim$(4$\pm$2)$\times$10$^{-5}$ & $\sim$(4$\pm$3)$\times$10$^{-11}$ \\ 
   CH$_{3}$C(O)NH$_{2}$, $\nu$=0   &   7   &   (2.2$\pm$1.1)$\times$10$^{14\phantom{,c]}}$   &   235$\pm$80   &   6.4$\pm$0.2   &   1.5$\pm$0.4   &   (6$\pm$3)$\times$10$^{-2}$   &   (2.0$\pm$1.2)$\times$10$^{-4}$ & (2.2$\pm$1.5)$\times$10$^{-10}$ \\
   CH$_{3}$C(O)NH$_{2}$, $\nu$=1   &   5   &   (6$\pm$3)$\times$10$^{14\phantom{,c]}}$   &   [235]   &   6.9$\pm$0.5   &   [1.5]   &   (1.4$\pm$1.0)$\times$10$^{-1}$   &   (5$\pm$3)$\times$10$^{-4}$ & (6$\pm$4)$\times$10$^{-10}$\\
   CH$_{3}$NHCHO, $\nu$=0   &   1   &   $\sim$(6$\pm$3)$\times$10$^{14\phantom{,c]}}$   &   [200]   &   7.3$\pm$0.6   &   2.3$\pm$0.5   &   $\sim$(1.5$\pm$0.9)$\times$10$^{-1}$   &   $\sim$(6$\pm$3)$\times$10$^{-4}$ & $\sim$(6$\pm$4)$\times$10$^{-10}$ \\ 
   CH$_{3}$NHCHO, $\nu$=1   &   1   &   $\sim$(6$\pm$4)$\times$10$^{14\phantom{,c]}}$   &   [200]   &   7.5$\pm$0.8   &   2.3$\pm$0.5   &   $\sim$(1.5$\pm$1.0)$\times$10$^{-1}$   &   $\sim$(5$\pm$4)$\times$10$^{-4}$ & $\sim$(6$\pm$4)$\times$10$^{-10}$ \\
   $^{15}$NH$_{2}$CHO   &   0   &   $\leq$1.5$\times$10$^{14\phantom{,c]}}$   &   [200]   &   [7.0]   &   [3.0]   &   $\leq$3.8$\times$10$^{-2}$   &   $\leq$1.4$\times$10$^{-4}$ & $\leq$1.5$\times$10$^{-10}$ \\ 
   cis-NHDCHO   &   0   &   $\leq$1.0$\times$10$^{14\phantom{,c]}}$   &   [200]   &   [7.0]   &   [3.0]   &   $\leq$2.6$\times$10$^{-2}$   &   $\leq$9.1$\times$10$^{-5}$ & $\leq$1.0$\times$10$^{-10}$ \\ 
   NH$_{2}$CH$^{18}$O   &   0   &   \phantom{$\sim\sim$*}$\leq$1.5$\times$10$^{14\phantom{,c]}}$   &   [200]   &   [7.0]   &   [3.0]   &   $\leq$3.8$\times$10$^{-2}$   &   $\leq$1.4$\times$10$^{-4}$ & $\leq$1.5$\times$10$^{-10}$\\
   NH$_{2}$C(O)NH$_{2}$   &   0   &   \phantom{$\sim\sim$*}$\leq$5.0$\times$10$^{13\phantom{,c]}}$   &   [200]   &   [7.0]   &   [3.0]   &   $\leq$1.3$\times$10$^{-2}$   &   $\leq$4.5$\times$10$^{-5}$ & $\leq$5$\times$10$^{-11}$\\
   NH$_{2}$C(O)CN   &   0   &   \phantom{$\sim\sim$*}$\leq$2.0$\times$10$^{14\phantom{,c]}}$   &   [200]   &   [7.0]   &   [3.0]   &   $\leq$5.1$\times$10$^{-2}$   &   $\leq$1.8$\times$10$^{-4}$ & $\leq$2.0$\times$10$^{-10}$ \\
   HOCH$_{2}$C(O)NH$_{2}$   &   0   &   \phantom{$\sim\sim$*}$\leq$5.0$\times$10$^{13\phantom{,c]}}$   &   [200]   &   [7.0]   &   [3.0]   &   $\leq$1.3$\times$10$^{-2}$   &   $\leq$4.5$\times$10$^{-5}$ & $\leq$5$\times$10$^{-11}$ \\
\hline
   CH$_{3}$OH$^{b}$   &   5   &   [1.1$\times$10$^{18}$]$^{b}\phantom{,}$   &   --   &   --   &   --   &   --   &   1.0$\times$10$^{0\phantom{-}}$ & (1.1$\pm$0.3)$\times$10$^{-6\phantom{0}}$ \\
   CH$_{2}$DOH   &   4   &   (1.2$\pm$0.2)$\times$10$^{16\phantom{,c]}}$   &   [250]$^{d}$   &   7.5$\pm$0.1   &   2.3$\pm$0.2   &   (3.1$\pm$1.1)$\times$10$^{0\phantom{-}}$   &   (1.1$\pm$0.4)$\times$10$^{-2}$ & (1.2$\pm$0.6)$\times$10$^{-8\phantom{0}}$ \\ 
   CH$_{3}$CHO   &   6   &   (1.5$\pm$0.2)$\times$10$^{15\phantom{,c]}}$   &   170$\pm$10   &   7.8$\pm$0.1   &   3.4$\pm$0.2   &   (3.8$\pm$1.4)$\times$10$^{-1}$   &   (1.4$\pm$0.5)$\times$10$^{-3}$ & (1.5$\pm$0.7)$\times$10$^{-9\phantom{0}}$\\  
   CH$_{3}$C(O)CH$_{3}$   &   4   &   (8$\pm$3)$\times$10$^{15\phantom{,c]}}$   &   250$\pm$40   &   7.4$\pm$0.2   &   3.4$\pm$0.3   &   (2.1$\pm$1.0)$\times$10$^{0\phantom{-}}$   &   (7$\pm$4)$\times$10$^{-3}$ & (8$\pm$5)$\times$10$^{-9\phantom{0}}$ \\
   CH$_{3}$COOH   &   2   &   $\sim$(2.8$\pm$0.7)$\times$10$^{15\phantom{,c]}}$   &   180$\pm$40   &   7.4$\pm$0.2   &   3.6$\pm$0.6   &   $\sim$(7$\pm$3)$\times$10$^{-1}$   &   $\sim$(2.5$\pm$1.1)$\times$10$^{-3}$ & $\sim$(2.8$\pm$1.4)$\times$10$^{-9\phantom{0}}$\\
   HOCH$_{2}$CHO   &   4   &   $\sim$(1.6$\pm$0.2)$\times$10$^{15\phantom{,c]}}$   &   220$\pm$20   &   6.8$\pm$0.3   &   4.6$\pm$0.3   &   $\sim$(4.1$\pm$1.5)$\times$10$^{-1}$   &   $\sim$(1.5$\pm$0.6)$\times$10$^{-3}$ & $\sim$(1.6$\pm$0.7)$\times$10$^{-9\phantom{0}}$ \\  
   H$_{2}$CCO    &   0   & $\leq$1.5$\times$10$^{16\phantom{,c]}}$ & [200]   &   [7.0]   &   [3.0]   &   $\leq$3.8$\times$10$^{0\phantom{-}}$   &   $\leq$1.4$\times$10$^{-2}$ & $\leq$1.5$\times$10$^{-8\phantom{0}}$  \\
\hline
\hline                  
\end{tabular}}
\textsuperscript{Values of $T_{\rm ex}$, $V_{\rm LSR}$, and $\Delta V$ in square brackets are estimated. The $\sim$ symbol indicates a tentative identification and $\leq$ an upper limit.}
\textsuperscript{$^{a}$The [X]/[NH$_{2}$CHO] values are determined based on the NH$_{2}$CHO column density, which is derived from the column density of the optically thin NH$_{2}^{13}$CHO isotopologue.}
\textsuperscript{$^{b}$The CH$_{3}$OH main isotopologue column density is taken from \citet{ligterink2020c}, which is based on the CH$_{3}^{18}$OH column density of (2.0$\pm$1.1)$\times$10$^{15}$ cm$^{-2}$ multiplied by a $^{16}$O/$^{18}$O ratio of 560 \citep{wilson1999}.}
\textsuperscript{$^{c}$The NH$_{2}^{12}$CHO column density is determined from the NH$_{2}^{13}$CHO column density, multiplied by a $^{12}$C/$^{13}$C ratio of 52.5$\pm$15.4 \citep{yan2019}. The number of lines listed are those identified of NH$_{2}$CHO.}
\textsuperscript{$^{d}$The fit of CH$_{2}$DOH makes use of the $T_{\rm ex}$ determined for CH$_{3}^{18}$OH by \citet{ligterink2020c}.}
\end{table}
\end{landscape}

\section{Discussion}\label{sec:discussion}

The detected molecules presented in Sect. \nameref{sec:results} substantially expand the number of identified species in Serpens SMM1\citep{oberg2011,ligterink2020c,tychoniec2021}. Particularly noteworthy is the detection of CH$_{3}$C(O)NH$_{2}$, making SMM1-a the lowest luminosity and mass object where acetamide is securely identified, with only a tentative detection of the molecule in the low-mass protostar IRAS~16293--2422B\citep{ligterink2018a}. Also, the tentative identification of deuterated formamide and CH$_{3}$NHCHO in SMM1-a add constraints to two molecules for which not many detections are found in the literature\citep{coutens2016,belloche2017,belloche2019,ligterink2020b,colzi2021}. 

In the following three sections, comparisons are made with the molecular inventories of other extraterrestrial sources. An overview of these sources, including relevant molecular abundances and column densities, is found in the supplementary information Table \ref{tab:literature}. The list covers a diverse set of objects, such as the comets 67P/Churyumov-Gerasimenko (hereafter 67P/C-G) and 46P/Wirtanen or the giant molecular cloud G+0.693, where the molecular complexity is driven by shocks resulting from a cloud-cloud collision\citep{zeng2020}. Data from the low-mass protostars Perseus B1-c (hereafter B1-c), Serpens S68N (hereafter S68N), Per-emb 44, Per-emb 12 B, Per-emb 13, Per-emb 27, Per-emb 29, and IRAS~16293--2422B are used. High-mass protostars and star-forming regions are represented in G328.2551–0.5321 (hereafter G328), W3(H2O), NGC~6334I, G31.41+0.31, GAL 034.3+00.2, Orion KL, NGC 7538 IRS1, NGC~6334-29, AFGL~4176, GAL 10.47+00.03, G10.6-0.4, and Sagittarius B2 (hereafter Sgr B2). Most of these objects have been observed with interferometric observations, but data for 46P/Wirtanen, G+0.693, W3(H2O), GAL 034.3+00.2, NGC~7538 IRS1, NGC~6334-29, GAL 31.41+0.31, GAL 10.47+00.03, Sgr B2(M), and Sgr B2(N) has been obtained with a variety of single dish telescopes. Data for 67P/C-G was obtained by the Rosetta space exploration mission.

\begin{figure*}[h]
\centering
   \includegraphics[width=1.0\hsize]{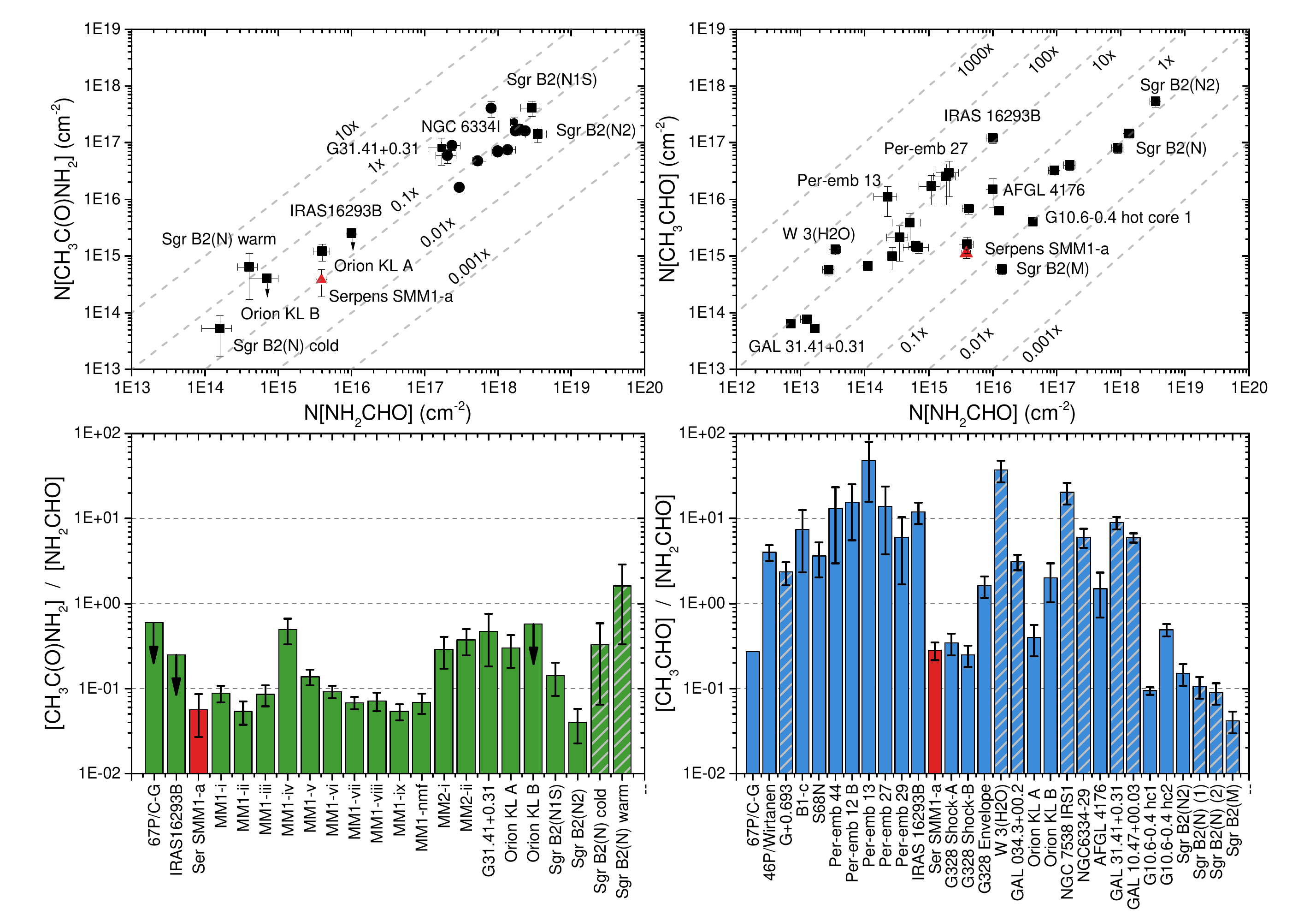}
\caption{Scatter plots and bar graphs of the CH$_{3}$C(O)NH$_{2}$ / NH$_{2}$CHO ratio (left) and CH$_{3}$CHO / NH$_{2}$CHO ratio (right) toward various interstellar sources. Serpens SMM1-a is indicated with a red triangle or red bar. For readability the NGC~6334I sources are grouped together and indicated with a circle and abbreviated as MM1-i to MM2-ii. Also for readability, only a select number of sources are labeled in the CH$_{3}$CHO / NH$_{2}$CHO scatter plot. The sources in the bar plot are roughly ordered by luminosity. Single dish observations are indicated with striped bars. Column densities taken from the literature\cite{goesmann2015,altwegg2017,biver2021,requena-torres2008,zeng2018,jimenez-serra2020,vangelder2020,nazari2021,yang2021,coutens2016,lykke2017,ligterink2018a,csengeri2019,bisschop2007,ligterink2020b,colzi2021,widicus-weaver2017,cernicharo2016,bogelund2019b,law2021,belloche2013,belloche2017,belloche2019,halfen2011} are presented in Table S3 in the supporting information.}
\label{fig:ratios_amides}
\end{figure*}

\subsection{The NH$_{2}$CHO -- CH$_{3}$C(O)NH$_{2}$ link}

Since the first detection of acetamide it has been suggested that CH$_{3}$C(O)NH$_{2}$ forms in reactions starting from NH$_{2}$CHO\citep{hollis2006}. If this is the case, a column density correlation between the two species may be observed. In the left side panels of Fig. \ref{fig:ratios_amides} the results of this study and literature data are collected and the ratios between NH$_{2}$CHO and CH$_{3}$C(O)NH$_{2}$ are presented in a scatter plot and bar graph. Literature data are taken from a variety of sources, ranging from comets to protostars and star-forming regions, with vastly varying luminosities and masses.  

From Fig. \ref{fig:ratios_amides} it becomes apparent that NH$_{2}$CHO and CH$_{3}$C(O)NH$_{2}$ column densities strongly correlate. A CH$_{3}$C(O)NH$_{2}$ / NH$_{2}$CHO ratio of approximately 0.1 with a scatter of less than an order of magnitude is found, in particular when single dish data are omitted. The correlation is in especially prominent when contrasted to CH$_{3}$CHO / NH$_{2}$CHO ratios, which easily scatter over 2 to 3 orders of magnitude, see right side panels of Fig. \ref{fig:ratios_amides}. Furthermore, it is interesting to point out that this ratio is similar in a diverse set of objects with very different physical characteristics, such as luminosity and protostellar mass, as these sources range from the low end of the mass range and low-luminosity, such as IRAS~16293--2422B ($\sim$3$L_{\odot}$\citep{jacobsen2018}) and Ser SMM1-a ($\sim$100 $L_{\odot}$) to complex, multi-protostar, high-mass star-forming regions with luminosities of 1000s or 10000s $L_{\odot}$ (e.g., NGC~6334I, Sgr B2(N)). In line with the conclusions of \citet{colzi2021}, derived from a similar set of data, this points to a chemical link between the two species and/or a similar response of reaction pathways to physical conditions\citep{quenard2018}. Formation of CH$_{3}$C(O)NH$_{2}$ and NH$_{2}$CHO in the ice mantles of grains in dark clouds or during warm-up toward the protostellar stage followed by release to the gas-phase due to thermal desorption is a likely scenario. This is in line with modeling work of \citet{quenard2018}, who find that NH$_{2}$CHO production at high temperatures (i.e., in or near a hot core/corino) is primarily the result of desorbing ice mantles. In contrast, the large differences seen in the CH$_{3}$CHO / NH$_{2}$CHO ratio, hint at a more complex interplay of reaction pathways and response to physical conditions for one or both of these species. Prominent gas-phase formation pathways are known for CH$_{3}$CHO, such as CH$_{3}$CH$_{2}$ + O $\rightarrow$ CH$_{3}$CHO + H\citep{charnley2004,garrod2021}, and its gas-phase formation has been hinted at in other interstellar sources, such as the L1157-B1 shock\citep{codella2017}.

Formation of acetamide by gas-phase reactions and the possible chemical link with formamide have been relatively well investigated, mainly by computational studies \citep{hollis2006,quan2007,halfen2011,redondo2014,yang2015,foo2018,kothari2020}. While solid-state formation of acetamide has been seen in various experiments\citep{berger1961,bernstein1995,henderson2015,ligterink2018a}, formation pathways and the chemical link with NH$_{2}$CHO are less well established.

Reactions involving the NH$_{2}$CO (carbamoyl) radical have been suggested as solid-state pathways to form amides in the literature\citep{agarwal1985,ligterink2018a}. One solid-state link between formamide and acetamide is found in the reaction between the methyl and carbamoyl radical: 
\begin{equation} 
    \ce{CH3(s) + NH2CO(s) -> CH3C(O)NH2(s)}
    \label{eq:acetamide_1}
\end{equation}
The carbamoyl radical links acetamide and formamide, since this radical can be formed by abstraction, radiolysis, and photolysis of NH$_{2}$CHO, via the NH$_{2}$ + CO radical addition, via the radical reaction CN + H$_{2}$O\citep{rimola2018}, or by hydrogenation of HNCO\citep{haupa2019}. Variations on above reaction may be possible, for example by exchanging CH$_{3}$ with the methylene (CH$_{2}$) radical and an additional hydrogenation step. To further elucidate reactions involving carbamoyl, matrix isolation experiments should be conducted to determine to which extent this and related reactions contribute to the formation of acetamide with formamide.

Another chemical link between formamide and acetamide is found in reactions starting from cyanides. With quantum chemical calculations, \citet{rimola2018} found a formamide formation pathway starting from the CN radical and water:
\begin{equation} 
    \ce{CN(s) + H2O(s) -> HN=COH(s)}
    \label{eq:formamide_1}
\end{equation}
\begin{equation} 
    \ce{HN=COH(s) -> NH2CO(s)}
    \label{eq:formamide_2}
\end{equation}
\begin{equation} 
    \ce{NH2CO(s) + H2O(s) -> NH2CHO(s) + OH(s)}
    \label{eq:formamide_3}
\end{equation}

There exists an analogue of this reaction for the formation of acetamide, starting from acetonitrile (CH$_{3}$CN) and water, which also has been proposed as a gas-phase reaction pathway\citep{foo2018}:

\begin{equation} 
    \ce{CH3CN(s) + H2O(s) -> CH3C(OH)=NH(s)}
    \label{eq:acetamide_2}
\end{equation}
\begin{equation} 
    \ce{CH3C(OH)=NH(s) -> CH3C(O)NH2(s)}
    \label{eq:acetamide_3}
\end{equation}

Important to note is that above reactions start with a closed shell species (Eq. \ref{eq:acetamide_2}, while the pathway to form NH$_{2}$CHO begins with the CN radical (Eq. \ref{eq:formamide_1}. However, if the CH$_{3}$CN reaction proceeds, this could explain a chemical link between NH$_{2}$CHO and CH$_{3}$C(O)NH$_{2}$, if there is a relation between (H)CN and CH$_{3}$CN formed in interstellar ices and their abundances are high enough. In the literature, there is some experimental evidence that acetamide can be formed from CH$_{3}$CN. \citet{bulak2021} investigated the photoprocessing of CH$_{3}$CN:H$_{2}$O ice mixtures with laser desorption post-ionization time-of-flight mass spectrometry and identified mass signatures that can be assigned to acetamide. However, an unambiguous identification was not made. Furthermore, \citet{duvernay2010} investigated the solid-state formation and photoprocessing of alpha-aminoethanol (CH$_{3}$C(OH)NH$_{2}$) a molecule that shows some chemical similarity to the CH$_{3}$C(OH)=NH intermediate produced in reaction \ref{eq:acetamide_2}. Photoprocessing of this species resulted in the formation of acetamide, hinting that reaction \ref{eq:acetamide_3} may result in the same product. 

Several other reactions that form CH$_{3}$C(O)NH$_{2}$ are possible, albeit that none of these show a direct chemical link with NH$_{2}$CHO. Among them, the reaction between ketene (H$_{2}$CCO) and ammonia (NH$_{3}$) is an interesting one that warrants further investigations. The gas-phase reaction of these two molecules has been investigated\citep{kothari2020}, but limited experimental evidence for a solid-state equivalent exists. \citet{haupa2020} investigated the hydrogen abstraction of CH$_{3}$C(O)NH$_{2}$, resulting in the formation of the CH$_{2}$C(O)NH$_{2}$ radical. Photoprocessing of this radical resulted in ketene formation. This can hint that a reverse pathway, starting from H$_{2}$CCO may be possible. It is worth noting that only one H$_{2}$CCO line is covered in the SMM1-a spectrum, which is not detected, making it difficult to assess ketene as a precursor of acetamide from these observations.

To conclude, observations indicate that the column densities of NH$_{2}$CHO and CH$_{3}$C(O)NH$_{2}$ might be correlated. However, the exact nature of this correlation is as of yet unknown and could be due to linked chemical processes, either in the gas-phase or in the solid-state, or production of these species under similar physical conditions. 

\subsection{Deuterated formamide}

Lines of two deuterated forms of formamide are tentatively detected in the SMM1-a spectrum, namely trans-NHDCHO and NH$_{2}$CDO, see Figs. S4 and S5. While the number of detected lines of these species is insufficient to claim a secure detection, they can be used to give an indication of D/H fraction of NH$_{2}$CHO in SMM1-a and be compared with the values of IRAS~16293--2422B, the only source for which formamide D/H is thoroughly analysed\citep{coutens2016}. 

\begin{figure*}[h]
\centering
   \includegraphics[width=1.0\hsize]{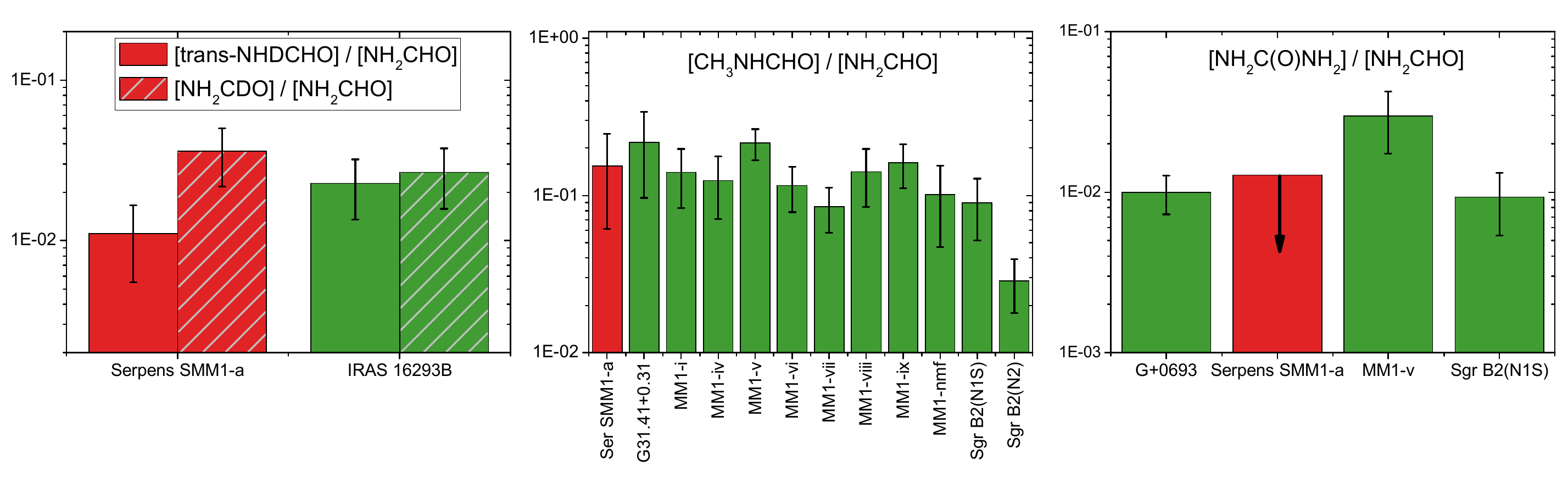}
\caption{Bar graph plots of deuterated formamide (left), N-methylformamide (middle), and carbamide (right) abundances with respect to NH$_{2}$CHO toward SMM1-a (red) and IRAS~16293--2422B\citep{coutens2016}, G31.41+0.31\citep{colzi2021}, NGC~6334I~MM1\citep{ligterink2020b}, Sgr B2(N)\citep{belloche2017,belloche2019}, and G+0.693\citep{jimenez-serra2020} (all in green). This work and \citet{coutens2016} use different values for the $^{12}$C/$^{13}$C ratio\citep{milam2005,yan2019}. For consistency, the NH$_{2}$CHO column density has been determined in both sources with the $^{12}$C/$^{13}$C = 52.5$\pm$15.4\citep{yan2019} ratio in order compare the formamide D/H ratios. For readability the NGC~6334I sources MM1-i to MM1-nmf are listed by their abbreviated names. The SMM1-a detection of deuterated NH$_{2}$CHO and CH$_{3}$NHCHO are tentative, while NH$_{2}$C(O)NH$_{2}$ is not detected and an upper limit is derived. The literature data used in this figure are presented in the supporting information Table S3.}
\label{fig:ratios_other_amides}
\end{figure*}

Deuterium fractions of $\sim$(1.1$\pm$0.5)$\times$10$^{-2}$ and $\sim$(3.6$\pm$1.0)$\times$10$^{-2}$ are found for trans-NHDCHO and NH$_{2}$CDO, respectively, toward SMM1-a, which agree reasonably well with the CH$_{2}$DOH / CH$_{3}$OH ratio of (1.1$\pm$0.4)$\times$10$^{-2}$. The difference in abundance between trans-NHDCHO and NH$_{2}$CDO of a factor of three is peculiar. This can indicate something about the underlying chemistry, but additional observations of deuterated formamide spectral lines are needed to further constrain the synthetic fit and in turn their abundances. Within the error bars, the SMM1 abundances with respect to NH$_{2}$CHO match those found toward IRAS~16293--2422, see left panel of Fig. \ref{fig:ratios_other_amides}. This hints that similar physico-chemical processes are taking place in both sources that result in the formation of formamide and its deuterium isotopomers. 

The formation of formamide is still strongly debated, with proponents for gas-phase and solid-state chemistry\citep{lopez-sepulcre2019}. Traditionally, deuteration levels of several percent, as found for SMM1-a, are seen as an indication that ice chemistry is producing these species\citep{tielens1983}. For formamide, \citet{skouteris2017} showed that the gas-phase reaction NH$_{2}$ + H$_{2}$CO $\rightarrow$ NH$_{2}$CHO + H can also be efficient in forming deuterated formamide. However, in this scenario, the high deuteration levels of H$_{2}$CO and NH$_{2}$ (or its precursor) have to be set on the grains. This raises the question whether one set of species can form on grains (H$_{2}$CO, NH$_{2}$) and the other will not (NH$_{2}$CHO), while laboratory evidence shows that both can simultaneously be formed in simulated interstellar ice mantles\citep{fedoseev2016}. Exclusive gas-phase formation of formamide in hot cores seems unlikely. 

\subsection{Other amides}

The last couple of years have seen a boost in searches for and detections of amides, especially N-methylformamide and carbamide. In the middle and right panel of Fig. \ref{fig:ratios_other_amides} the tentative detection of CH$_{3}$NHCHO and upper limit of NH$_{2}$C(O)NH$_{2}$ toward SMM1-a are compared with interstellar detections of these molecules available from the literature. 

The CH$_{3}$NHCHO / NH$_{2}$CHO ratio in SMM1-a matches with those found in the high-mass sources G31.41+0.31\citep{colzi2021}, NGC~6334I\citep{ligterink2020b}, and Sgr B2(N1S)\citep{belloche2019}. This supports the idea that the underlying physico-chemical processes that result in CH$_{3}$NHCHO formation in SMM1-a are similar to those in the high-mass objects. 

In ice mantles, several CH$_{3}$NHCO formation pathways are available\citep{belloche2017,frigge2018}, such as the radical-radical addition reactions
\begin{equation} 
    \ce{CH3NH(s) + CHO(s) -> CH3NHCHO(s),}
    \label{eq:nmf_1}
\end{equation}
and
\begin{equation} 
    \ce{HNCHO(s) + CH3(s) -> CH3NHCHO(s),}
    \label{eq:nmf_2}
\end{equation}
and the hydrogenation reaction
\begin{equation} 
    \ce{CH3NCO(s) + 2H -> CH3NHCHO(s).}
    \label{eq:nmf_3}
\end{equation}

The tentative correlation between CH$_{3}$NHCHO and NH$_{2}$CHO can hint that the formation of these two species is linked and that reaction \ref{eq:nmf_2} is involved in its formation. It would be of interest to determine the CH$_{3}$NHCO/NH$_{2}$CHO toward the giant molecular cloud G+0.693 to see if this correlation also holds at an early stage of star formation. At the same time, parallel detections of CH$_{3}$NCO and CH$_{3}$NH$_{2}$, of which currently a limited number exist in the literature, may provide insight into the other two formation pathways. 

Of the amides that are not detected in SMM1-a, the upper limit abundance of NH$_{2}$C(O)NH$_{2}$ can be compared with detections of this molecule in G+0.693 and Sgr B2(N1S) and a tentative identification in NGC~6334I~MM1-v \citep{belloche2019,jimenez-serra2020,ligterink2020b}, see right panel of Fig. \ref{fig:ratios_other_amides}. The upper limit abundances are found to be similar to those of the detections. This suggests that a slightly deeper search that targets prominent spectral lines of NH$_{2}$C(O)NH$_{2}$ may result in a detection of this molecule toward SMM1-a. 

Two other complex amides that are searched for in SMM1-a, NH$_{2}$C(O)CN and HOCH$_{2}$C(O)NH$_{2}$, are not detected and at present they have not been detected toward other sources either\citep{sanz-novo2020,colzi2021}. The NH$_{2}$C(O)CN / NH$_{2}$CHO upper limit ratios are similar in SMM1-a and G31.41+0.31 at $\leq$0.05 and $\leq$0.02\citep{colzi2021}, respectively. For HOCH$_{2}$C(O)NH$_{2}$ the upper limits toward SMM1-a, G31.41+0.31, and Sgr B2(N2) are $\leq$0.01, $\leq$0.004\citep{colzi2021}, and $\leq$0.007\citep{sanz-novo2020}, respectively, with respect to NH$_{2}$CHO. Compared to the abundances of CH$_{3}$C(O)NH$_{2}$ and CH$_{3}$NHCHO with respect to NH$_{2}$CHO, which are approximately 0.1, this shows that NH$_{2}$C(O)CN and HOCH$_{2}$C(O)NH$_{2}$ are substantially less abundant by at least a factor 5--25. 

It is interesting to note that NH$_{2}$C(O)CN and HOCH$_{2}$C(O)NH$_{2}$ are much less abundant, while their chemical complexity seems to be similar or marginally greater than that of CH$_{3}$C(O)NH$_{2}$ and CH$_{3}$NHCHO. This suggests that reactions forming these molecules are substantially less efficient than the pathway(s) leading to CH$_{3}$C(O)NH$_{2}$ and CH$_{3}$NHCHO or that these molecules are much more readily destroyed. 

\subsection{Prebiotic chemistry \& planetary exploration}

The results of this study suggest that SMM1-a has a rich inventory of amides, which have been primarily formed in the ice mantles of interstellar grains and are sublimated in the hot corino. Recent studies indicate that ice formed in the dark cloud is inherited into the planet-forming disk \citep{booth2021}. Chemical inventories, including NH$_{2}$CHO, of several hot corinos have indeed been shown to be similar to the composition of comet 67P/Churyumov--Gerasimenko \citep{drozdovskaya2019,belloche2020}. If amides are formed in the dark cloud or during the onset of the protostellar stage, when ice mantles warm up, they can survive into the planet-forming disk and be available as planet building material. Remnants of planet building, such as comets and asteroids, can bombard newly formed planets and seed it with a cocktail of prebiotic molecules, including amides\citep{chyba1992}. 

More studies of comets and asteroids are needed to determine the amount of amides that are present in these objects. Radio telescope observations may help with this by analyzing the coma gas\citep{biver2021}, but observational tools are generally not sensitive enough to detect larger amides, since they are usually less abundant and not very volatile. Space exploration missions are required to explore these chemical inventories in situ, for example with the ORganics Information Gathering Instrument (ORIGIN)\citep{ligterink2020a} space instrument, which is intended to detect large non-volative prebiotic molecules and biomolecules.

\section{Conclusions}\label{sec:conclusion}

In this publication the amide inventory of the hot corino around the intermediate-mass protostar Serpens SMM1-a is studied. The first detections of NH$_{2}$CHO $\nu$=0, NH$_{2}$CHO $\nu_{12}$=1, NH$_{2}^{13}$CHO, and CH$_{3}$C(O)NH$_{2}$ $\nu$=0,1 toward this source are presented, as are the tentative identifications of trans-NHDCHO, NH$_{2}$CDO, and CH$_{3}$NHCHO $\nu$=0,1. Serendipitous detections of the oxygen COMs CH$_{3}$CHO and CH$_{3}$C(O)CH$_{3}$ are made, while CH$_{3}$COOH and HOCH$_{2}$CHO are tentatively identified. Ratios of the detected molecules are derived with respect to NH$_{2}$CHO and compared with molecular inventories of a varied sample of other sources. 

For CH$_{3}$C(O)NH$_{2}$ and CH$_{3}$NHCHO a ratio of approximately 0.1 and a scatter of less than an order of magnitude is found. For CH$_{3}$C(O)NH$_{2}$, for which a larger sample of detections is available, this uniform ratio suggests that there is a link in the formation of this molecule and NH$_{2}$CHO and that the ratio must already be set at an early stage of star formation, presumably in ice mantles on dust grains that are present in dark clouds. The D/H ratio of NH$_{2}$CHO is found to be about 1--3\% and is similar to the ratio derived for IRAS~16293-2422B. This is a strong clue that ice mantle reactions are involved in the formation of NH$_{2}$CHO, but further studies are required to assess the extent to which gas-phase reactions are involved. A comparison of the upper limit abundance derived for NH$_{2}$C(O)NH$_{2}$ toward Serpens SMM1-a with those in other sources, suggests that this is a suitable source for follow-up searches for this relevant prebiotic molecule.

The rich amide inventory of Serpens SMM1-a is further evidence for the widespread availability of these molecules in star-forming regions. This increases the likelihood that of these molecules are inherited into comets and other planetesimals and subsequently delivered to planetary surfaces, where they can aid the formation of biomolecules. 

\begin{acknowledgement}

The authors thank E.G. B{\o}gelund and E.F. van Dishoeck for valuable discussions of data presented in this manuscript. The authors acknowledge assistance from Allegro, the European ALMA Regional Center node in the Netherlands. We thank the three anonymous reviewers for their helpful comments and suggestions. This paper makes use of the following ALMA data: ADS/JAO.ALMA\#2018.1.00836.S. ALMA is a partnership of ESO (representing its member states), NSF (USA) and NINS (Japan), together with NRC (Canada), MOST and ASIAA (Taiwan), and KASI (Republic of Korea), in cooperation with the Republic of Chile. The Joint ALMA Observatory is operated by ESO, AUI/NRAO and NAOJ. NFWL is supported by the Swiss National Science Foundation (SNSF) Ambizione grant 193453. JKJ is supported by the European Research Council (ERC) under the European Union's Horizon 2020 research and innovation programme through ERC Consolidator Grant ``S4F'' (grant agreement No~646908). AC received financial support from the Agence Nationale de la Recherche (grant ANR-19-ERC7-0001-01) and from the European Research Council (ERC) under the European Union’s Horizon 2020 research and innovation programme through the ERC starting grant Chemtrip (grant agreement 949278).

\end{acknowledgement}

\begin{suppinfo}

\section{Spectroscopy}\label{ap:spec_data}

Spectroscopic linelists used in the work are primarily taken from the CDMS database, but several linelists from the JPL spectroscopic database and literature sources are used as well. Table \ref{tab:spec_lit} gives an overview of the catalogues and identifiers for each molecule analyzed in this work and the main works these entries are based on. 

\begin{table}[h]
\caption{References of molecular spectroscopy used in this work}             
\label{tab:spec_lit}      
\centering          
\begin{tabular}{c c c}     
\hline\hline                           
Molecule & ID \& Catalogue & Literature \\
\hline
CH$_{2}$DOH & 33004 (JPL) & \cite{pearson2012,jacq1993,quade1980,mukhopadhyay1997,su1989,elhilali2011} \\
H$_{2}$CCO & 42501 (CDMS) & \cite{johnson1952,fabricant1977,brown1990} \\
CH$_{3}$CHO & 44003 (JPL) & \cite{kleiner1996} \\
NH$_{2}$CHO, $\nu$=0 & 45512 (CDMS) & \cite{motiyenko2012,kryvda2009,blanco2006,vorobeva1994,moskienko1991,gardner1980,hirota1974,kukolich1971b}\\
NH$_{2}$CHO, $\nu_{12}$=1 & 45516 (CDMS) & \cite{hirota1974,moskienko1991,vorobeva1994,kryvda2009,motiyenko2012} \\
NH$_{2}^{13}$CHO & 46512 (CDMS) & \cite{motiyenko2012,gardner1980,blanco2006,kryvda2009}\\
NH$_{2}$CDO & 46520 (CDMS) & \cite{kutsenko2013} \\
cis-NHDCHO & 46521 (CDMS) & \cite{kutsenko2013} \\
trans-NHDCHO & 46522 (CDMS) & \cite{kutsenko2013} \\
CH$_{3}$C(O)CH$_{3}$ & 58003 (JPL) & \cite{peter1965,vacherand1986,oldag1992,groner2002} \\
CH$_{3}$C(O)NH$_{2}$ & Literature &  \cite{ilyushin2004} \\
CH$_{3}$NHCHO & Literature  & \cite{belloche2017} \\ 
CH$_{3}$COOH & 60523 (CDMS) & \cite{ilyushin2013} \\
HOCH$_{2}$CHO & 60501 (CDMS) & M\"{u}ller et al. in prep. \\
NH$_{2}$C(O)NH$_{2}$ & 60517 (CDMS) & \cite{remijan2014,brown1975,kasten1986,kretschmer1996}\\
NH$_{2}$C(O)CN & 70504 (CDMS) & \cite{christiansen2005,winnewisser2005} \\
HOCH$_{2}$C(O)NH$_{2}$ & 75517 (CDMS) & \cite{sanz-novo2020} \\
\hline                  
\end{tabular}
\end{table}

\section{Observed lines}\label{ap:obs_lines}

\begin{landscape}
\begin{longtable}{l c l c c l}
\caption{Spectral information of molecules detected toward SMM1-a}\\
\label{tab:SMM1_lines}
Molecule & Transition & Frequency & $E_{\rm up}$ & $A_{\rm ij}$ & Minor\\  
& $J,K_{\rm a},K_{\rm c},(F_{1},F)$ & (MHz) & (K) & (s$^{-1}$) & blending\\
\hline
\endfirsthead
\multicolumn{6}{c}%
{\tablename\ \thetable\ -- \textit{Continued from previous page}} \\
\hline
Molecule & Transition & Frequency & $E_{\rm up}$ & $A_{\rm ij}$ & Minor\\  
& $J,K_{\rm a},K_{\rm c},(F_{1},F)$ & (MHz) & (K) & (s$^{-1}$) & blending species\\
\hline
\endhead
\hline \multicolumn{6}{r}{\textit{Continued on next page}} \\
\endfoot
\hline
\endlastfoot
\hline
CH$_{2}$DOH	&	10 1 10 0 -- 9 0 9 1\phantom{11} 	&	221 391.766	(0.0101)	&	120	&	2.85	$\times$10$^{-5}$ & -- \\
 & 10 1 9 0 -- 9 2 8 0\phantom{1}  	&	231 840.297	(0.0097)	&	124	&	1.66	$\times$10$^{-5}$ & CH$_{3}^{18}$OH		\\
 & 9 2 7 0 -- 9 1 8 0 	&	231 969.226	(0.0101)	&	113	&	8.64	$\times$10$^{-5}$ & CH$_{3}$CHO	\\
 & 8 2 6 0 -- 8 1 7 0 	&	234 471.033	(0.0098)	&	94	&	8.43	$\times$10$^{-5}$ & --		\\
\hline
H$_{2}$CCO$^{a}$  & 11 8 3 -- 10 8 2 & 221 546.0454 (0.0225) &	895 & 5.77$\times$10$^{-5}$ & -- \\
            & 11 8 4 -- 10 8 3 & 221 546.0454 (0.0225) &	895	& 5.77$\times$10$^{-5}$ & --\\
\hline
CH$_{3}$CHO	&	12 3 10 1 -- 11 3 9 1	&	231 748.7186	(0.0036)	&	93	&	3.94	$\times$10$^{-4}$ & CH$_{3}$OCHO		\\
 & 12 3 9 2 -- 11 3 8 2	&	231 847.5795	(0.0037)	&	93	&	3.94	$\times$10$^{-4}$ & CH$_{3}$OCHO		\\
 & 12 2 10 2 -- 11 2 9 2	&	234 795.4555	(0.0039)	&	82	&	4.28	$\times$10$^{-4}$ & CH$_{3}$OCHO (w)		\\
 & 12 2 10 0 -- 11 2 9 0	&	234 825.8718	(0.0039)	&	82	&	4.28	$\times$10$^{-4}$ & --		\\
 & 12 2 10 3 -- 11 2 9 3	&	234 902.9702	(0.0096)	&	287	&	4.31	$\times$10$^{-4}$ & --		\\
 & 12 1 11 3 -- 11 1 10 3	&	235 217.8542	(0.0087)	&	282	&	4.43	$\times$10$^{-4}$ & --		\\
\hline													
NH$_{2}$CHO, $\nu$=0	&	10 1 9 -- 9 1 8 	&	218 459.213	(0.0100)	&	61	&	7.48	$\times$10$^{-4}$ & --		\\
 & 11 8 3 -- 10 8 2 	&	233 488.887	(0.0009)	&	258	&	4.36	$\times$10$^{-4}$ & CH$_{3}$OCHO		\\
 & 11 8 4 -- 10 8 3 	&	233 488.887	(0.0009)	&	258	&	4.36	$\times$10$^{-4}$ & CH$_{3}$OCHO		\\
 & 11 9 2 -- 10 9 1 	&	233 492.678	(0.0010)	&	308	&	3.06	$\times$10$^{-4}$ & --		\\
 & 11 9 3 -- 10 9 2 	&	233 492.678	(0.0010)	&	308	&	3.06	$\times$10$^{-4}$ & --		\\
 & 11 7 4 -- 10 7 3 	&	233 498.065	(0.0100)	&	213	&	5.51	$\times$10$^{-4}$ & --		\\
 & 11 7 5 -- 10 7 4 	&	233 498.065	(0.0100)	&	213	&	5.51	$\times$10$^{-4}$ & --		\\
 & 11 6 6 -- 10 6 5 	&	233 527.795	(0.0100)	&	174	&	6.51	$\times$10$^{-4}$ & CH$_{3}$CHO (w)		\\
 & 11 6 5 -- 10 6 4 	&	233 527.795	(0.0100)	&	174	&	6.51	$\times$10$^{-4}$ & CH$_{3}$CHO (w)		\\
 & 11 5 7 -- 10 5 6 	&	233 594.501	(0.0100)	&	142	&	7.36	$\times$10$^{-4}$ & --		\\
 & 11 5 6 -- 10 5 5 	&	233 594.501	(0.0100)	&	142	&	7.36	$\times$10$^{-4}$ & --		\\
 & 11 3 8 -- 10 3 7 	&	234 315.498	(0.0100)	&	94	&	8.67	$\times$10$^{-4}$ & --		\\
\hline															
NH$_{2}$CHO, $\nu_{12}$=1	&	11 4 7 -- 10 4 6	&	233 551.991	(0.01)\phantom{00}	&	530	&	8.05	$\times$10$^{-4}$ & CH$_{3}$OCHO (w)		\\
 & 11 4 8 -- 10 4 7	&	233 541.38\phantom{0}	(0.0100)	&	530	&	8.05	$\times$10$^{-4}$ & --		\\
 & 11 3 8 -- 10 3 7	&	234 110.228	(0.0100)	&	509	&	8.64	$\times$10$^{-4}$ & CH$_{3}$OCHO		\\
\hline															
NH$_{2}^{13}$CHO	&	11 2 10 -- 10 2 9	&	231 844.16\phantom{0}	(0.0022)	&	79	&	8.77	$\times$10$^{-4}$ & --		\\
 & 11 3 9 -- 10 3 8	&	233 568.296	(0.0022)	&	93	&	8.58	$\times$10$^{-4}$ & --		\\
\hline															
NH$_{2}$CDO	&	11 0 11 -- 10 0 10	&	218 022.107	(0.0013)	&	64	&	7.49	$\times$10$^{-4}$ & --		\\
 & 12 1 12 -- 11 1 11	&	234 414.679	(0.0013)	&	76	&	9.33	$\times$10$^{-4}$ & CH$_{3}$C(O)NH$_{2}$		\\
 & 11 1 10 -- 10 1 9	&	235 136.546	(0.0012)	&	70	&	9.35	$\times$10$^{-4}$ & CH$_{3}$OCHO		\\
\hline
cis-NHDCHO$^{a}$	&	12 0 12 -- 11 0 11	&	234 776.217	(0.0014)	&	74	&	9.40	$\times$10$^{-4}$ & --		\\
\hline															
trans-NHDCHO	&	12 2 11 -- 11 2 10  	&	234 250.932	(0.0015)	&	85	&	9.12	$\times$10$^{-4}$ & --		\\
 & 12 4 8 -- 11 4 7 	&	235 718.042	(0.0014)	&	121	&	8.50	$\times$10$^{-4}$ & --		\\
 & 12 3 10 -- 11 3 9	&	235 859.775	(0.0014)	&	100	&	8.98	$\times$10$^{-4}$ & CH$_{3}$C(O)NH$_{2}$		\\
\hline															
$^{15}$NH$_{2}$CHO$^{a}$	&	11 1 10 -- 10 1 9	&	233 523.780	(0.0032)	&	70	&	9.18	$\times$10$^{-4}$ & --		\\
\hline															
NH$_{2}$CH$^{18}$O$^{a}$	&	12 1 12 -- 11 1 11	&	231 843.462	(0.0034)	&	76	&	9.03	$\times$10$^{-4}$ & --		\\
 & 12 0 12 -- 11 0 11	&	235 678.570	(0.0035)	&	74	&	9.53	$\times$10$^{-4}$ & --		\\
\hline
CH$_{3}$C(O)CH$_{3}$	& 20 2 18 1 -- 19 3 17 1 & 218091.4115 (0.0138) & 119 & 4.31$\times$10$^{-4}$ & --	\\
 & 20 3 18 1 -- 19 2 17 1 & 218 091.4115 (0.0138) & 119 & 4.31$\times$10$^{-4}$	& -- \\	
 & 19 5 14 0 -- 18 6 13 1 & 234 491.3021 (0.0103) & 128 & 4.15$\times$10$^{-4}$	& -- \\
 & 19 6 14 0 -- 18 7 13 1 & 234 491.3021 (0.0103) & 128 & 4.15$\times$10$^{-4}$	& -- \\
 & 20 4 16 1 -- 19 5 15 1 & 235 490.7336 (0.0135) & 133 & 4.68$\times$10$^{-4}$	& trans-NHDCHO, CH$_{3}$NHCHO, NH$_{2}^{13}$CHO \\
 & 20 5 16 1 -- 19 6 15 1 & 235 490.7336 (0.0135) & 133 & 4.68$\times$10$^{-4}$	& trans-NHDCHO, CH$_{3}$NHCHO, NH$_{2}^{13}$CHO \\
 & 20 4 16 0 -- 19 5 15 1 & 235 548.3327 (0.0107) & 133 & 4.69$\times$10$^{-4}$ & CH$_{3}$NHCHO	\\
 & 20 5 16 0 -- 19 6 15 1 & 235 548.3327 (0.0107) & 133 & 4.69$\times$10$^{-4}$	& CH$_{3}$NHCHO \\
\hline															
CH$_{3}$C(O)NH$_{2}$, $\nu$=0	
 & 18 3 15 0 -- 17 4 14 0 	&	221 470.562	(0.016)\phantom{0}	&	113	&	6.01	$\times$10$^{-4}$ & --		\\
 & 18 4 15 0 -- 17 3 14 0 	&	221 470.562	(0.016)\phantom{0}	&	113	&	6.01	$\times$10$^{-4}$ & --		\\
 & 19 3 17 0 -- 18 2 16 0 	&	221 535.395	(0.0194)	&	116	&	6.14	$\times$10$^{-4}$ & --		\\
 & 19 2 17 0 -- 18 3 16 0 	&	221 535.395	(0.0194)	&	116	&	6.14	$\times$10$^{-4}$ & --		\\
 & 21 1 21 0 -- 20 1 20 0 	&	221 577.81\phantom{0}	(0.0291)	&	125	&	8.03	$\times$10$^{-4}$ & --	 	\\
 & 21 0 21 0 -- 20 0 20 0 	&	221 577.81\phantom{0}	(0.0291)	&	125	&	8.03	$\times$10$^{-4}$ & --		\\
 & 20 1 19 0 -- 19 2 18 0 	&	221 604.702	(0.0236)	&	118	&	2.22	$\times$10$^{-4}$ & --		\\
 & 20 2 19 0 -- 19 1 18 0 	&	221 604.702	(0.0236)	&	118	&	2.22	$\times$10$^{-4}$ & --		\\
 & 20 1 19 0 -- 19 1 18 0 	&	221 604.702	(0.0236)	&	118	&	5.39	$\times$10$^{-4}$ & --		\\
 & 20 2 19 0 -- 19 2 18 0 	&	221 604.702	(0.0236)	&	118	&	5.39	$\times$10$^{-4}$ & --		\\
 & 22 0 22 0 -- 21 0 21 0 	&	231 887.118	(0.0341)	&	137	&	4.32	$\times$10$^{-4}$ & --		\\
 & 22 1 22 0 -- 21 1 21 0 	&	231 887.118	(0.0341)	&	137	&	4.32	$\times$10$^{-4}$ & --		\\
 & 22 1 22 0 -- 21 0 21 0 	&	231 887.118	(0.0341)	&	137	&	5.01	$\times$10$^{-4}$ & --		\\
 & 22 0 22 0 -- 21 1 21 0 	&	231 887.118	(0.0341)	&	137	&	5.01	$\times$10$^{-4}$ & --		\\
 & 21 2 20 0 -- 20 1 19 0 	&	231 911.642	(0.0278)	&	130	&	8.52	$\times$10$^{-4}$ & --		\\
 & 21 1 20 0 -- 20 2 19 0 	&	231 911.642	(0.0278)	&	130	&	8.52	$\times$10$^{-4}$ & --		\\
 & 22 1 22 0 -- 21 1 21 0 	&	231 973.521	(0.0336)	&	131	&	1.24	$\times$10$^{-4}$ & --		\\
 & 22 0 22 0 -- 21 0 21 0 	&	231 973.521	(0.0336)	&	131	&	1.24	$\times$10$^{-4}$ & --		\\
 & 22 0 22 0 -- 21 1 21 0 	&	231 973.521	(0.0336)	&	131	&	8.11	$\times$10$^{-4}$ & --		\\
 & 22 1 22 0 -- 21 0 21 0 	&	231 973.521	(0.0336)	&	131	&	8.11	$\times$10$^{-4}$ & --		\\
\hline															
CH$_{3}$C(O)NH$_{2}$, $\nu$=1	&	16 7 9 1 -- 15 8 8 1 	&	217 631.256	(0.0129)	&	162	&	4.43	$\times$10$^{-4}$ & --		\\
 & 18 2 16 1 -- 17 3 15 1 	&	217 680.653	(0.0141)	&	142	&	3.95	$\times$10$^{-4}$ & HOCH$_{2}$CN		\\
 & 18 3 16 1 -- 17 3 15 1 	&	217 680.674	(0.0141)	&	142	&	2.15	$\times$10$^{-4}$ & HOCH$_{2}$CN		\\
 & 18 2 16 1 -- 17 2 15 1 	&	217 686.753	(0.0141)	&	142	&	2.15	$\times$10$^{-4}$ & HOCH$_{2}$CN		\\
 & 18 3 16 1 -- 17 2 15 1 	&	217 686.774	(0.0141)	&	142	&	3.95	$\times$10$^{-4}$ & HOCH$_{2}$CN		\\
 & 17 6 11 1 -- 16 7 10 1 	&	221 315.706	(0.0160)	&	168	&	5.39	$\times$10$^{-4}$ & CH$_{3}$C(O)CH$_{3}$		\\
 & 9 8 1 1 -- 8 7 1 1 	&	231 882.736	(0.0044)	&	93	&	5.40	$\times$10$^{-4}$ & --		\\
 & 22 1 21 1 -- 21 2 20 1	&	233 621.037	(0.0318)	&	197	&	9.04	$\times$10$^{-4}$ & --		\\
 & 22 2 21 1 -- 21 2 20 1	&	233 621.037	(0.0318)	&	197	&	9.04	$\times$10$^{-4}$ & --		\\
\hline															
CH$_{3}$NHCHO, $\nu$=0	&	20 3 17 0 -- 19 3 16 0 &	233 612.959	(0.0015)	&	125	&	5.65	$\times$10$^{-4}$ & --		\\
\hline															
CH$_{3}$NHCHO, $\nu$=1	&	20 8 12 0 -- 19 9 11 0 &	217 981.125	(0.0021)	&	215	&	4.15	$\times$10$^{-4}$ & --	\\
\hline															
NH$_{2}$C(O)NH$_{2}^{a}$	&	16 4 12 -- 15 5 11	&	221 615.758	(0.05)\phantom{00}	&	104	&	6.41	$\times$10$^{-4}$ & --		\\
 & 16 5 12 -- 15 4 11	&	221 615.739	(0.0113)	&	104	&	6.41	$\times$10$^{-4}$ & --		\\
\hline
HOCH$_{2}$CHO  & 9 5 5 - 8 4 4 & 217626.0610 (0.1200) & 40 & 1.59$\times$10$^{-4}$ & -- \\
 & 25 10 16 - 25 9 17 & 234765.736\phantom{0} (0.1200) & 242 & 2.04$\times$10$^{-4}$ & CH$_{2}$DOH (w) \\
 & 42 9 34 - 42 8 35 & 235311.3660 (0.1200) & 558 & 2.29$\times$10$^{-4}$ & -- \\
 & 24 10 15 - 24 9 16 & 235697.8890 (0.1200) & 228 & 2.03$\times$10$^{-4}$ & CH$_{3}$OCHO \\
\hline
CH$_{3}$COOH & 20 0 20 0 0 -- 19 0 19 0 0 & 218010.0224 (0.0018) & 	113	& 7.03$\times$10$^{-5}$ & -- \\
 & 20 1 20 0 0 -- 19 1 19 0 0 & 218 010.0224 (0.0018) & 113 & 7.03$\times$10$^{-5}$ & --  \\
 & 20 0 20 0 0 -- 19 1 19 0 0 & 218 010.0224 (0.0018) & 113 & 7.03$\times$10$^{-5}$ & --  \\
 & 20 1 20 0 0 -- 19 0 19 0 0 & 218 010.0224 (0.0018) & 113 & 7.03$\times$10$^{-5}$ & --  \\
 & 20 0 20 0 1 -- 19 0 19 0 2 & 218 044.2146 (0.0017) & 112 & 4.50$\times$10$^{-5}$ & CH$_{3}$NCO \\
 & 20 1 20 0 1 -- 19 1 19 0 2 & 218 044.2146 (0.0017) & 112 & 4.50$\times$10$^{-5}$ & CH$_{3}$NCO \\
 & 20 0 20 0 1 -- 19 1 19 0 2 & 218 044.2146 (0.0017) & 112 & 4.50$\times$10$^{-5}$ & CH$_{3}$NCO \\
 & 20 1 20 0 1 -- 19 0 19 0 2 & 218 044.2146 (0.0017) & 112 & 4.50$\times$10$^{-5}$ & CH$_{3}$NCO \\
\hline
NH$_{2}$C(O)CN$^{a}$	&	67 18 50 -- 67 17 51	&	217 621.387	(0.1500)	&	946	&	4.52	$\times$10$^{-4}$ & --		\\
 & 29 15 14 -- 29 14 15	&	217 621.351	(0.0076)	&	238	&	3.06	$\times$10$^{-4}$ & --		\\
 & 12 9 4 -- 11 8 3	&	218 496.949	(0.0094)	&	57	&	4.81	$\times$10$^{-4}$ & --		\\
 & 12 9 3 -- 11 8 4	&	218 496.953	(0.0094)	&	57	&	4.81	$\times$10$^{-4}$ & --		\\
 & 29 6 24 -- 28 5 23	&	218 930.010	(0.1500)	&	170	&	3.11	$\times$10$^{-4}$ & --		\\
\hline															
HOCH$_{2}$C(O)NH$_{2}^{a}$	&	38 1 37 -- 37 1 36	&	234 277.191	(0.0500)	&	227	&	6.12	$\times$10$^{-4}$ & --		\\
 & 38 1 37 -- 37 2 36	&	234 277.191	(0.0500)	&	227	&	7.19	$\times$10$^{-4}$ & --		\\
 & 38 2 37 -- 37 2 36	&	234 277.191	(0.0500)	&	227	&	6.12	$\times$10$^{-4}$ & --		\\
 & 38 2 37 -- 37 1 36	&	234 277.191	(0.0500)	&	227	&	7.19	$\times$10$^{-4}$ & --		\\
\end{longtable}
\textsuperscript{$^{a}$Of undetected species the listed lines are used to determine their upper limit column density.}
\end{landscape}

\section{Line fits}\label{ap:line_fits}

\begin{figure*}
\includegraphics[width=\hsize]{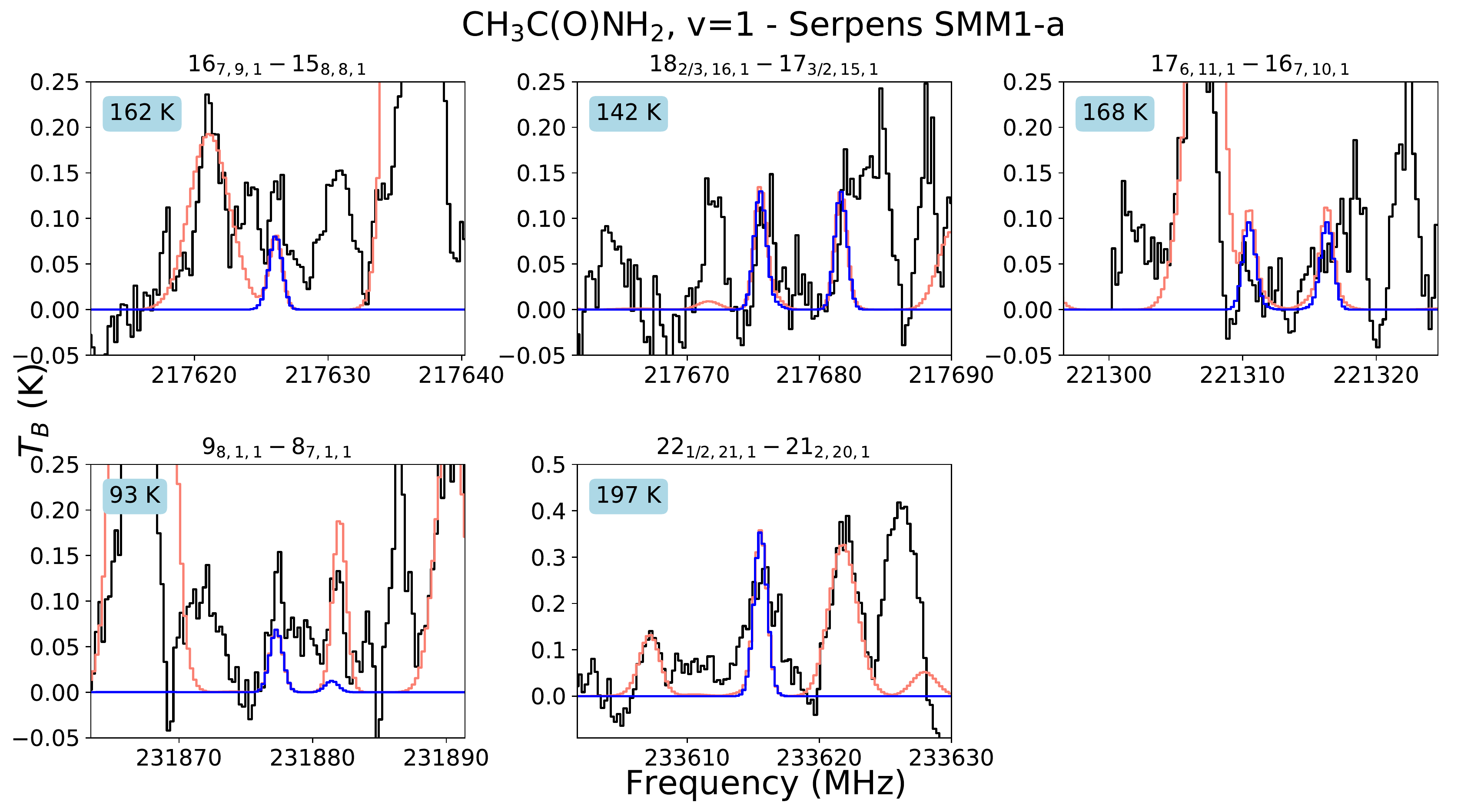}
\caption{Identified lines of CH$_{3}$C(O)NH$_{2}$, $\nu$=1 toward SMM1-a. The observed spectrum is plotted in black, with the synthetic spectrum of the species overplotted in blue, and the synthetic spectrum of all fitted species combined in red. The transition is indicated at the top of each panel and the upper state energy is given in the top left of each panel.}
\label{fig:lines_CH3CONH2v1}
\end{figure*}

\begin{figure*}
\includegraphics[width=\hsize]{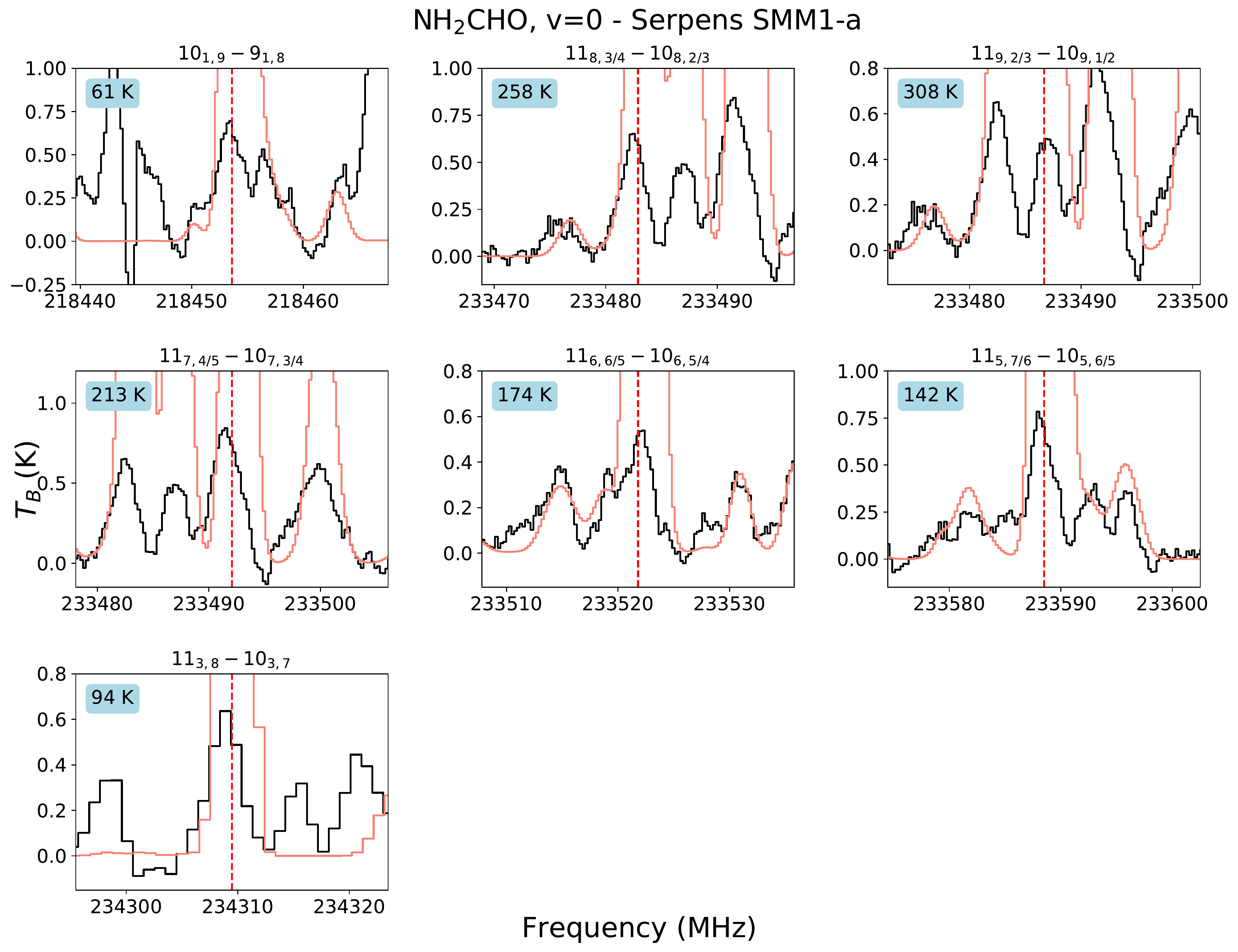}
\caption{Identified lines of NH$_{2}$CHO, $\nu$=0 toward SMM1-a. The observed spectrum is plotted in black and the spectral line is indicated with a red dashed line. Since these transitions are optically thick, no synthetic fit of the transition is made. However, the combined synthetic spectrum (red) does show the NH$_{2}$CHO spectral lines, which are determined from optically thin NH$_{2}^{13}$CHO measurements. The transition is indicated at the top of each panel and the upper state energy is given in the top left of each panel.}
\label{fig:lines_NH2CHO}
\end{figure*}

\begin{figure*}
\includegraphics[width=\hsize]{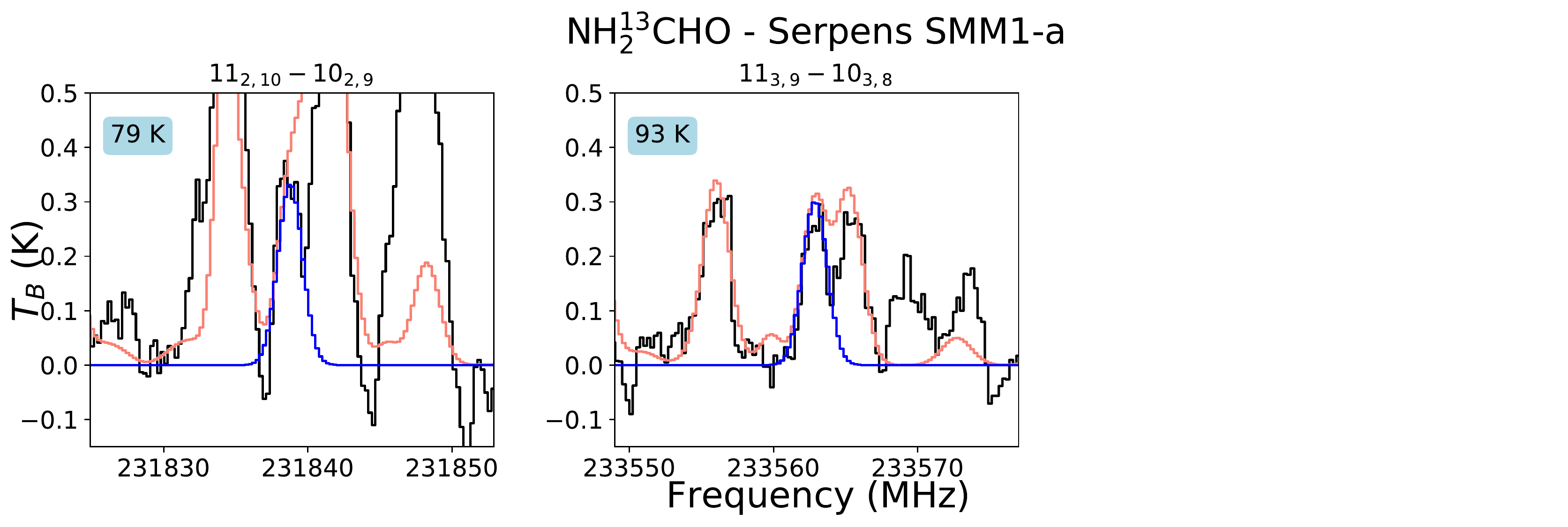}
\caption{Identified lines of NH$_{2}^{13}$CHO toward SMM1-a. The observed spectrum is plotted in black, with the synthetic spectrum of the species overplotted in blue, and the synthetic spectrum of all fitted species combined in red. The transition is indicated at the top of each panel and the upper state energy is given in the top left of each panel.}
\label{fig:lines_13-NH2CHO}
\end{figure*}

\begin{figure*}
\includegraphics[width=\hsize]{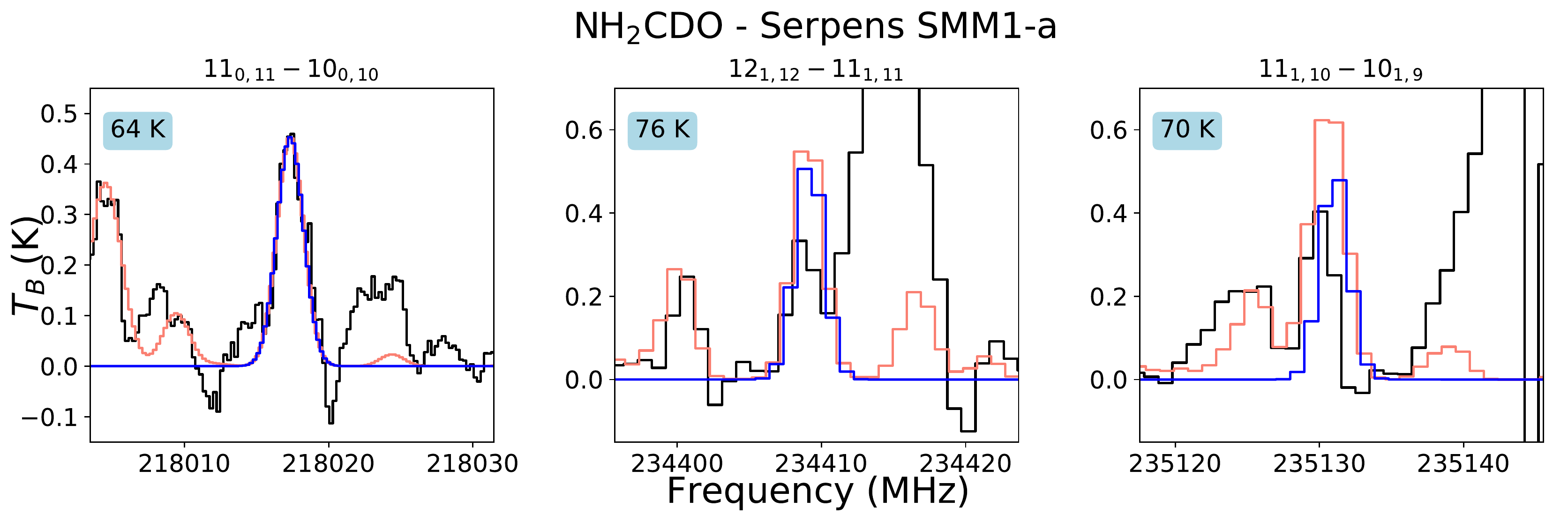}
\caption{Identified lines of NH$_{2}$CDO toward SMM1-a. The observed spectrum is plotted in black, with the synthetic spectrum of the species overplotted in blue, and the synthetic spectrum of all fitted species combined in red. The transition is indicated at the top of each panel and the upper state energy is given in the top left of each panel.}
\label{fig:lines_NH2CDO}
\end{figure*}

\begin{figure*}
\includegraphics[width=\hsize]{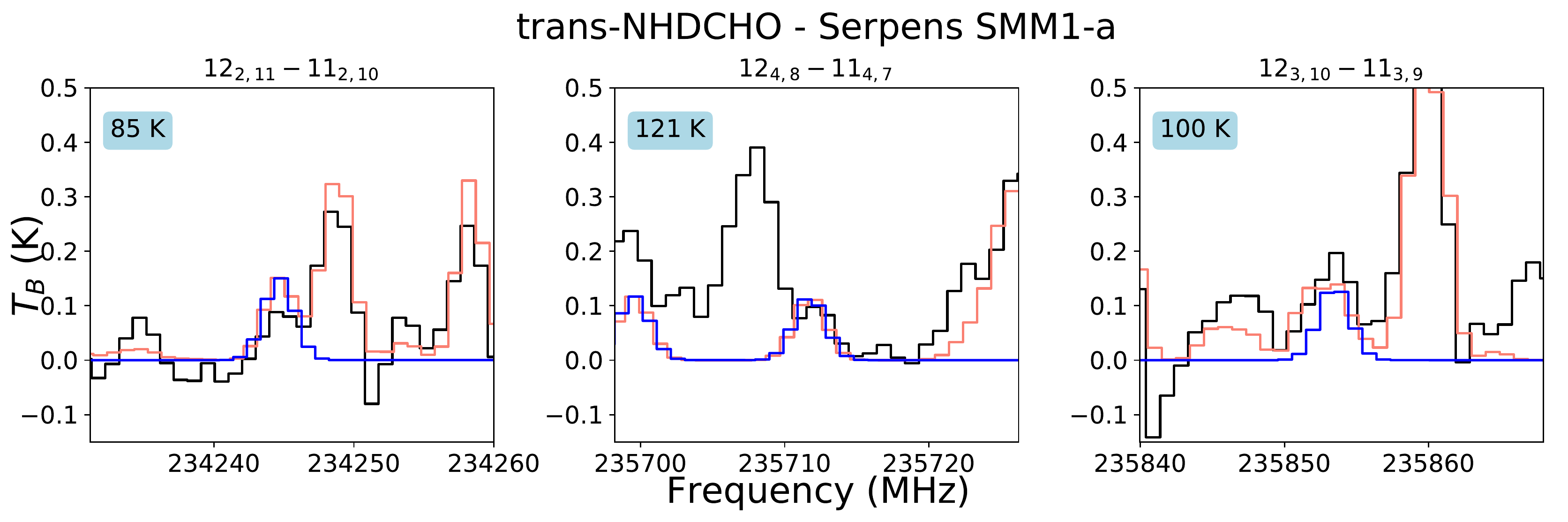}
\caption{Identified lines of trans-NHDCHO toward SMM1-a. The observed spectrum is plotted in black, with the synthetic spectrum of the species overplotted in blue, and the synthetic spectrum of all fitted species combined in red. The transition is indicated at the top of each panel and the upper state energy is given in the top left of each panel.}
\label{fig:lines_trans-NHDCHO}
\end{figure*}

\begin{figure*}
\includegraphics[width=\hsize]{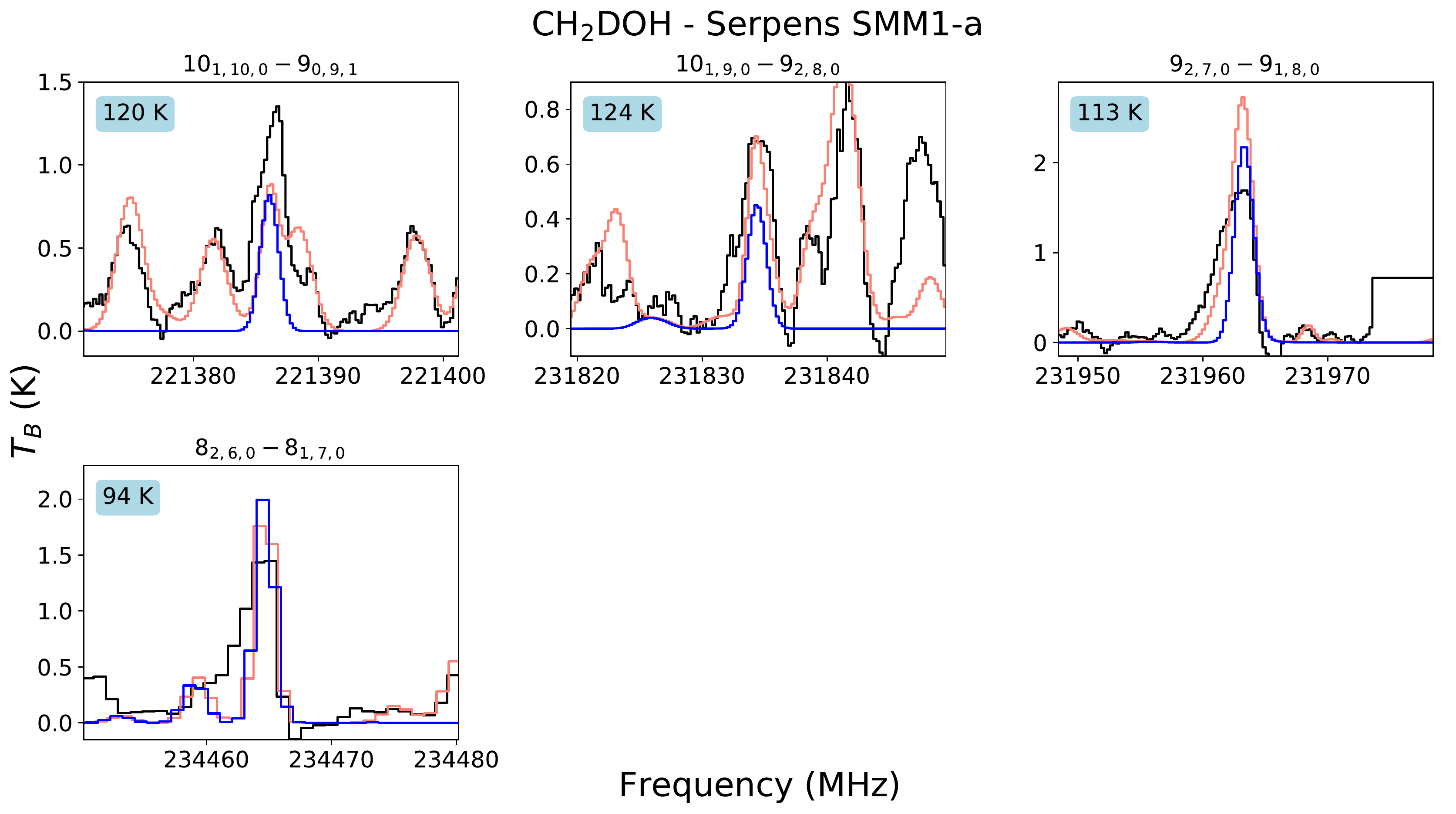}
\caption{Identified lines of CH$_{2}$DOH toward SMM1-a. The observed spectrum is plotted in black, with the synthetic spectrum of the species overplotted in blue, and the synthetic spectrum of all fitted species combined in red. The transition is indicated at the top of each panel and the upper state energy is given in the top left of each panel.}
\label{fig:lines_CH2DOH}
\end{figure*}

\begin{figure*}
\includegraphics[width=\hsize]{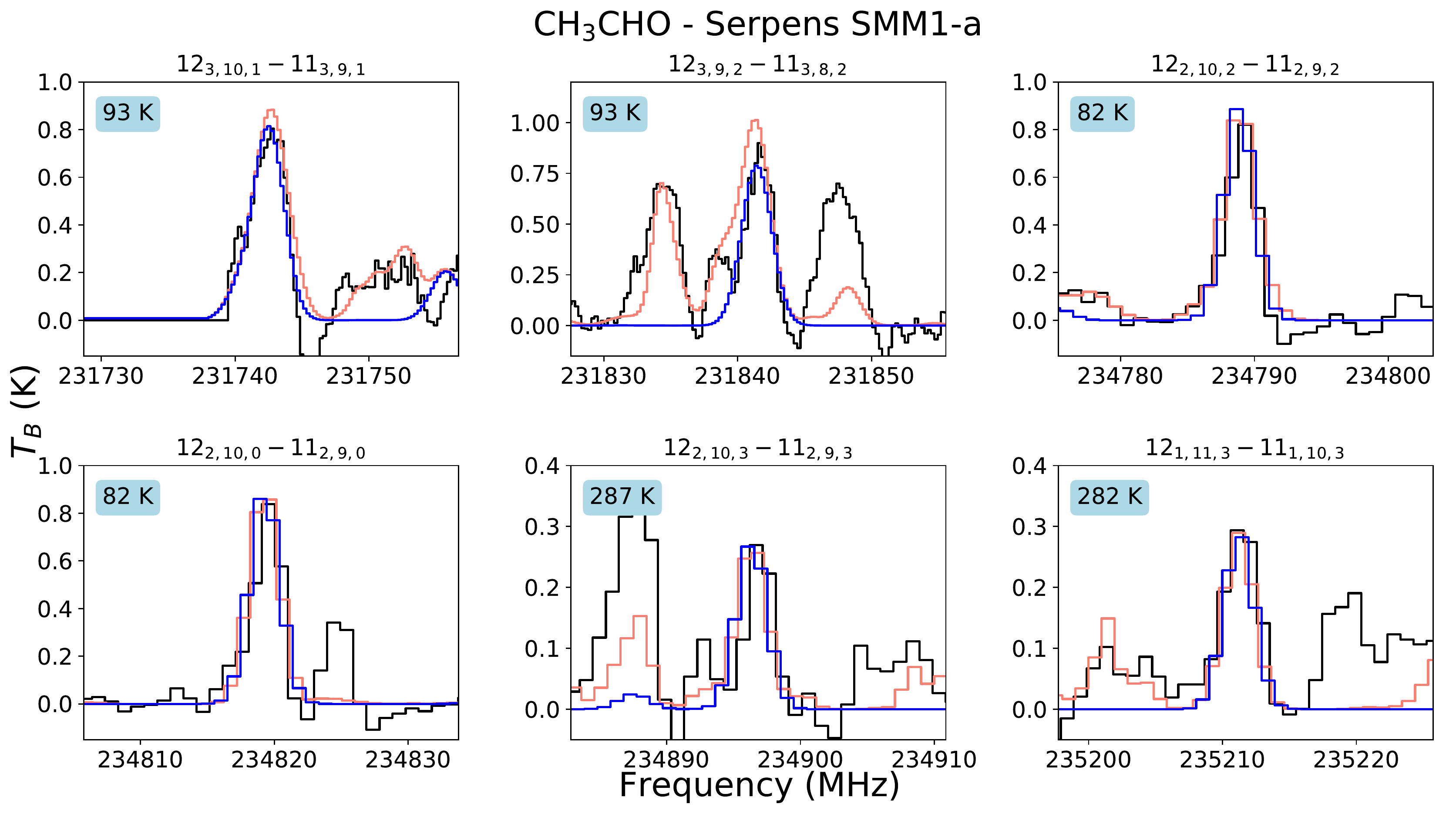}
\caption{Identified lines of CH$_{3}$CHO toward SMM1-a. The observed spectrum is plotted in black, with the synthetic spectrum of the species overplotted in blue, and the synthetic spectrum of all fitted species combined in red. The transition is indicated at the top of each panel and the upper state energy is given in the top left of each panel.}
\label{fig:lines_CH3CHO}
\end{figure*}

\begin{figure*}
\includegraphics[width=\hsize]{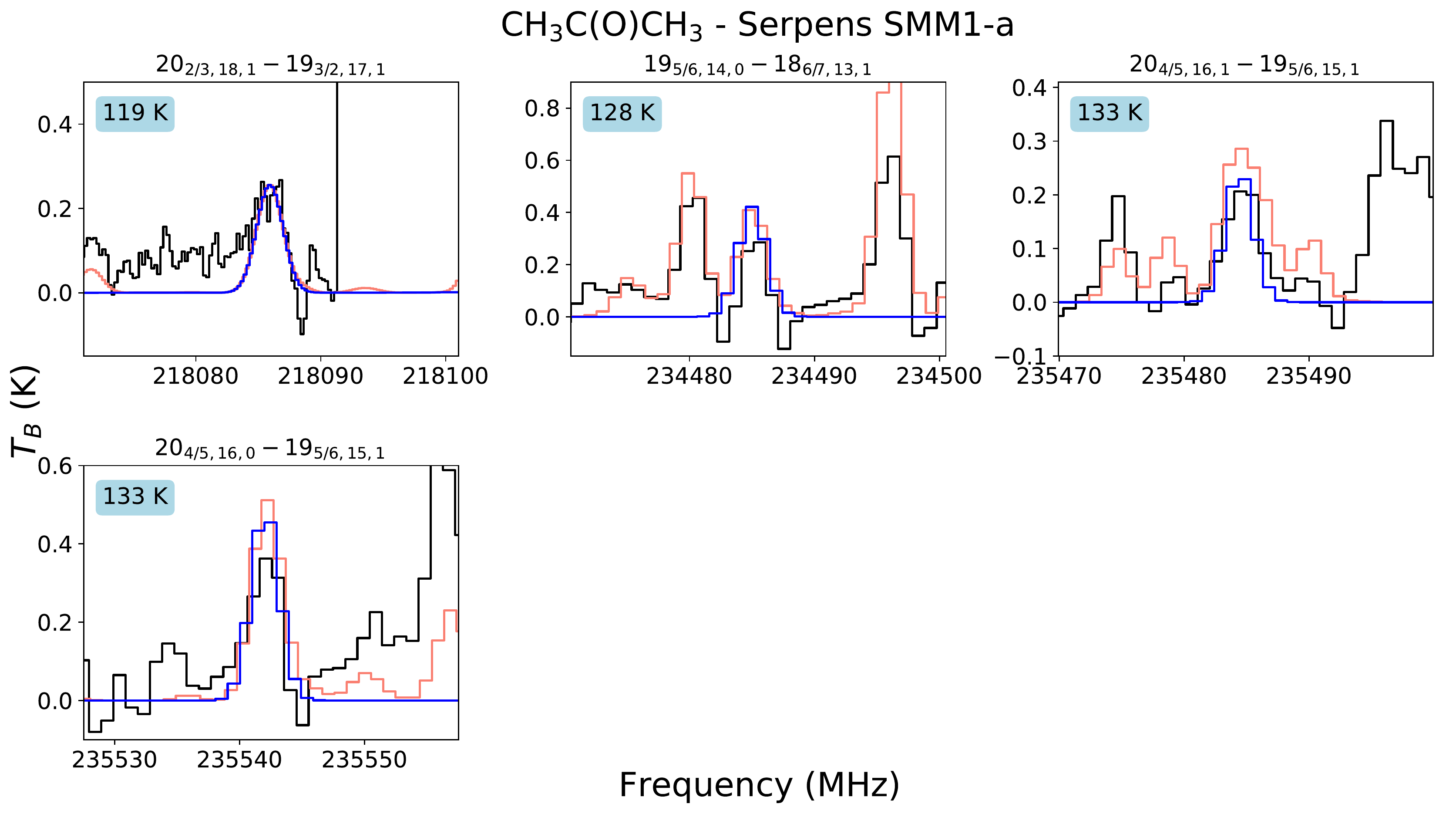}
\caption{Identified lines of CH$_{3}$C(O)CH$_{3}$ toward SMM1-a. The observed spectrum is plotted in black, with the synthetic spectrum overplotted in blue. The transition is indicated at the top of each panel and the upper state energy is given in the top left of each panel.}
\label{fig:lines_CH3COCH3}
\end{figure*}

\begin{figure*}
\includegraphics[width=\hsize]{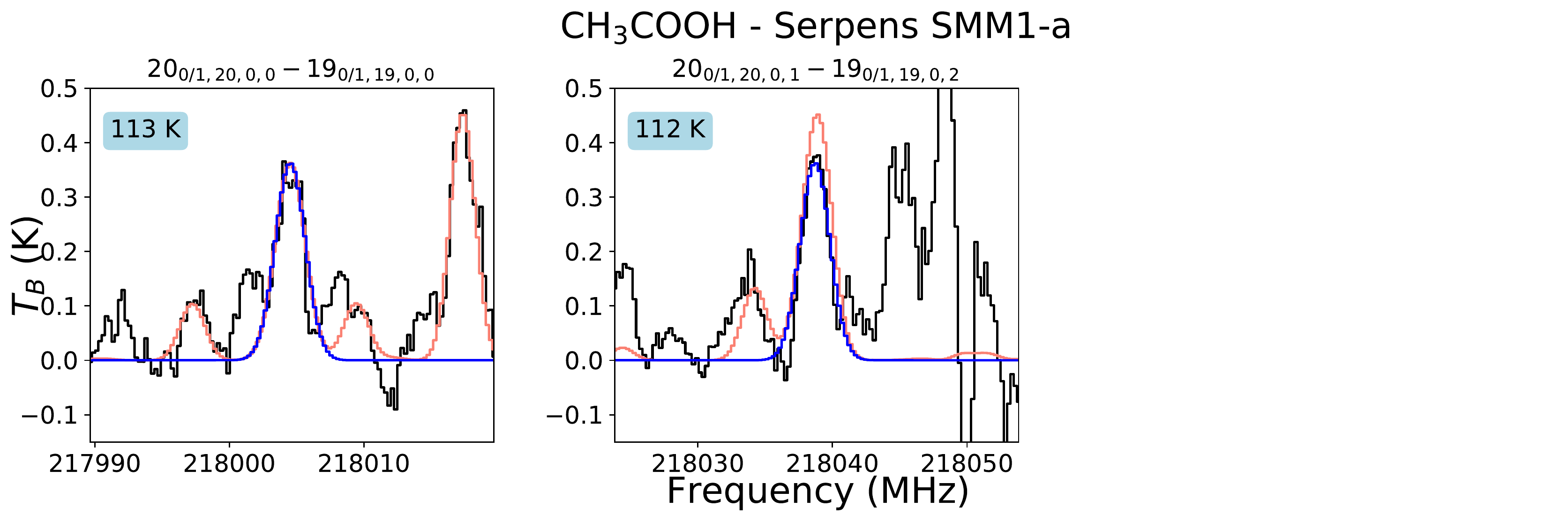}
\caption{Identified lines of CH$_{3}$COOH toward SMM1-a. The observed spectrum is plotted in black, with the synthetic spectrum of the species overplotted in blue, and the synthetic spectrum of all fitted species combined in red. The transition is indicated at the top of each panel and the upper state energy is given in the top left of each panel.}
\label{fig:lines_CH3COOH}
\end{figure*}

\begin{figure*}
\includegraphics[width=\hsize]{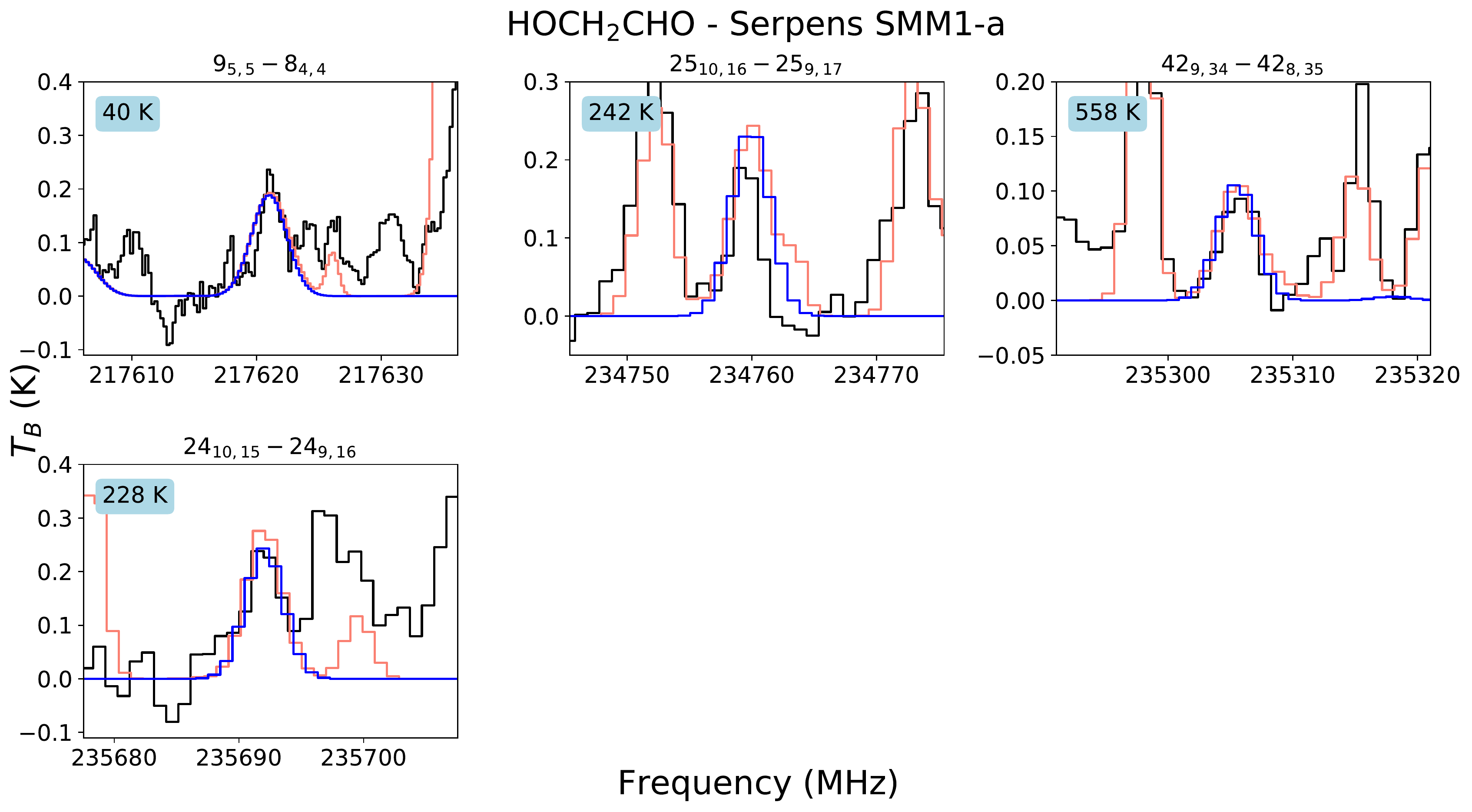}
\caption{Identified lines of HOCH$_{2}$CHO toward SMM1-a. The observed spectrum is plotted in black, with the synthetic spectrum of the species overplotted in blue, and the synthetic spectrum of all fitted species combined in red. The transition is indicated at the top of each panel and the upper state energy is given in the top left of each panel.}
\label{fig:lines_HOCH2CHO}
\end{figure*}

\section{Literature data}\label{ap:lit_data}

\begin{landscape}
\begin{longtable}{l c c c c c} 
\caption{Column densities taken from studies presented in the literature.}\\
\hline\hline  
Source & $N$(NH$_{2}$CHO) & $N$(CH$_{3}$CHO) & $N$(CH$_{3}$C(O)NH$_{2}$) & $N$(CH$_{3}$NHCHO) & $N$(NH$_{2}$C(O)NH$_{2}$) \\
\hline
& \% of H$_{2}$O & \% of H$_{2}$O & \% of H$_{2}$O & \% of H$_{2}$O & \% of H$_{2}$O \\
\hline
\endfirsthead
\multicolumn{6}{c}%
{\tablename\ \thetable\ -- \textit{Continued from previous page}} \\
\hline
Source & $N$(NH$_{2}$CHO) & $N$(CH$_{3}$CHO) & $N$(CH$_{3}$C(O)NH$_{2}$) & $N$(CH$_{3}$NHCHO) & $N$(NH$_{2}$C(O)NH$_{2}$) \\
\hline
& cm$^{-2}$ & cm$^{-2}$ & cm$^{-2}$ & cm$^{-2}$ & cm$^{-2}$ \\
\hline
\endhead
\hline \multicolumn{6}{r}{\textit{Continued on next page}} \\
\endfoot
\hline
\endlastfoot
\hline
67P/C-G$^{	a	}$	&		1					&		3.7					&		$\leq$2.2					&			--				&			--				\\
46P/Wirtanen$^{	b	}$	&	(0.06	$\pm$	0.01)		&	(0.015	$\pm$	0.002)			&			--				&			--				&			--				\\
\hline
& cm$^{-2}$ & cm$^{-2}$ & cm$^{-2}$ & cm$^{-2}$ & cm$^{-2}$ \\
\hline
G+0.693$^{	c	}$	&	(1.5	$\pm$	0.3)$\times$10$^{15	}$	&	(6.3	$\pm$	1.4)$\times$10$^{14	}$	&			--				&			--				&	(6.3	$\pm$	1.0)$\times$10$^{12	}$	\\
B1-c$^{	d	}$	&	(3.8	$\pm$	1.9)$\times$10$^{15	}$	&	(5.1	$\pm$	2.4)$\times$10$^{14	}$	&			--				&			--				&			--				\\
S68N$^{	d	}$	&	(9.8	$\pm$	4.2)$\times$10$^{14	}$	&	(2.7	$\pm$	0.3)$\times$10$^{14	}$	&			--				&			--				&			--				\\
Per-emb 44$^{	e	}$	&	(2.5	$\pm$	1.7)$\times$10$^{16	}$	&	(1.9	$\pm$	0.7)$\times$10$^{15	}$	&			--				&			--				&			--				\\
Per-emb 12 B$^{	e	}$	&	(1.7	$\pm$	0.9)$\times$10$^{16	}$	&	(1.1	$\pm$	0.4)$\times$10$^{15	}$	&			--				&			--				&			--				\\
Per-emb 13$^{	e	}$	&	(1.1	$\pm$	0.6)$\times$10$^{16	}$	&	(2.3	$\pm$	0.9)$\times$10$^{14	}$	&			--				&			--				&			--				\\
Per-emb 27$^{	e	}$	&	(2.9	$\pm$	1.8)$\times$10$^{16	}$	&	(2.1	$\pm$	0.8)$\times$10$^{15	}$	&			--				&			--				&			--				\\
Per-emb 29$^{	e	}$	&	(2.1	$\pm$	1.3)$\times$10$^{15	}$	&	(3.5	$\pm$	1.3)$\times$10$^{14	}$	&			--				&			--				&			--				\\
IRAS 16293B$^{	f	}$	&	(1.2	$\pm$	0.2)$\times$10$^{17	}$	&	(1.0	$\pm$	0.2)$\times$10$^{16	}$	&		$\leq$2.5$\times$10$^{15	}$					&			--				&			--				\\
G328 Shock-A$^{	g	}$	&	(3.2	$\pm$	0.6)$\times$10$^{16	}$	&	(9.3	$\pm$	1.9)$\times$10$^{16	}$	&			--				&			--				&			--				\\
G328 Shock-B$^{	g	}$	&	(4.0	$\pm$	0.8)$\times$10$^{16	}$	&	(1.6	$\pm$	0.3)$\times$10$^{17	}$	&			--				&			--				&			--				\\
G328 Envelope$^{	g	}$	&	(6.8	$\pm$	1.4)$\times$10$^{15	}$	&	(4.2	$\pm$	0.8)$\times$10$^{15	}$	&			--				&			--				&			--				\\
W 3(H2O)$^{	h	}$	&	(1.3	$\pm$	0.3)$\times$10$^{15	}$	&	(3.5	$\pm$	0.7)$\times$10$^{13	}$	&			--				&			--				&			--				\\
MM1-i$^{	i	}$	&			--				&	(5.3	$\pm$	0.9)$\times$10$^{17	}$	&	(4.7	$\pm$	0.7)$\times$10$^{16	}$	&	(7.5	$\pm$	2.8)$\times$10$^{16	}$	&			--				\\
MM1-ii$^{	i	}$	&			--				&	(1.4	$\pm$	0.4)$\times$10$^{18	}$	&	(7.4	$\pm$	1.1)$\times$10$^{16	}$	&			--				&			--				\\
MM1-iii$^{	i	}$	&			--				&	(2.0	$\pm$	0.3)$\times$10$^{18	}$	&	(1.7	$\pm$	0.4)$\times$10$^{17	}$	&			--				&			--				\\
MM1-iv$^{	i	}$	&			--				&	(8.1	$\pm$	1.2)$\times$10$^{17	}$	&	(4.0	$\pm$	1.2)$\times$10$^{17	}$	&	(1.0	$\pm$	0.4)$\times$10$^{17	}$	&			--				\\
MM1-v$^{	i	}$	&			--				&	(1.7	$\pm$	0.2)$\times$10$^{18	}$	&	(2.3	$\pm$	0.4)$\times$10$^{17	}$	&	(3.6	$\pm$	0.7)$\times$10$^{17	}$	&	(5.0	$\pm$	2.0)$\times$10$^{16	}$	\\
MM1-vi$^{	i	}$	&			--				&	(1.7	$\pm$	0.2)$\times$10$^{18	}$	&	(1.6	$\pm$	0.2)$\times$10$^{17	}$	&	(2.0	$\pm$	0.6)$\times$10$^{17	}$	&			--				\\
MM1-vii$^{	i	}$	&			--				&	(2.4	$\pm$	0.3)$\times$10$^{18	}$	&	(1.6	$\pm$	0.2)$\times$10$^{17	}$	&	(2.0	$\pm$	0.6)$\times$10$^{17	}$	&			--				\\
MM1-viii$^{	i	}$	&			--				&	(9.9	$\pm$	1.9)$\times$10$^{17	}$	&	(7.1	$\pm$	1.1)$\times$10$^{16	}$	&	(1.4	$\pm$	0.5)$\times$10$^{17	}$	&			--				\\
MM1-ix$^{	i	}$	&			--				&	(3.0	$\pm$	0.3)$\times$10$^{17	}$	&	(1.6	$\pm$	0.3)$\times$10$^{16	}$	&	(4.8	$\pm$	1.4)$\times$10$^{16	}$	&			--				\\
MM1-nmf$^{	i	}$	&			--				&	(9.9	$\pm$	1.9)$\times$10$^{17	}$	&	(6.8	$\pm$	1.3)$\times$10$^{16	}$	&	(1.0	$\pm$	0.5)$\times$10$^{17	}$	&			--				\\
MM2-i$^{	i	}$	&			--				&	(2.1	$\pm$	0.6)$\times$10$^{17	}$	&	(5.9	$\pm$	1.6)$\times$10$^{16	}$	&			--				&			--				\\
MM2-ii$^{	i	}$	&			--				&	(2.4	$\pm$	0.7)$\times$10$^{17	}$	&	(8.8	$\pm$	1.6)$\times$10$^{16	}$	&			--				&			--				\\
G31.41+0.31$^{	j	}$	&			(1.7	$\pm$	0.6)$\times$10$^{17	}$				&	--	&	(8.0	$\pm$	4.0)$\times$10$^{16	}$	&	(3.7	$\pm$	1.6)$\times$10$^{16	}$	&			--				\\
GAL 034.3+00.2$^{	k	}$	&	(5.3	$\pm$	0.7)$\times$10$^{13	}$	&	(1.7	$\pm$	0.3)$\times$10$^{13	}$	&			--				&			--				&			--				\\
Orion KL A $^{	l	}$	&	(1.6	$\pm$	0.5)$\times$10$^{15	}$	&	(4.0	$\pm$	1.0)$\times$10$^{15	}$	&	(1.2	$\pm$	0.4)$\times$10$^{15	}$	&			--				&			--				\\
Orion KL B $^{	l	}$	&	(1.4	$\pm$	0.3)$\times$10$^{15	}$	&	(7.0	$\pm$	3.0)$\times$10$^{14	}$	&		$\leq$4$\times$10$^{15	}$					&			--				&			--				\\
NGC 7538 IRS1$^{	h	}$	&	(5.7	$\pm$	1.1)$\times$10$^{14	}$	&	(2.8	$\pm$	0.6)$\times$10$^{13	}$	&			--				&			--				&			--				\\
NGC6334-29 $^{	k	}$	&	(7.6	$\pm$	1.1)$\times$10$^{13	}$	&	(1.3	$\pm$	0.3)$\times$10$^{13	}$	&			--				&			--				&			--				\\
AFGL 4176$^{	m	}$	&	(1.5	$\pm$	0.8)$\times$10$^{16	}$	&	(1.0	$\pm$	0.1)$\times$10$^{16	}$	&			--				&			--				&			--				\\
GAL 31.41+0.31$^{	k	}$	&	(6.4	$\pm$	0.6)$\times$10$^{13	}$	&	(7.1	$\pm$	1.0)$\times$10$^{12	}$	&			--				&			--				&			--				\\
GAL 10.47+00.03$^{	k	}$	&	(6.7	$\pm$	0.5)$\times$10$^{14	}$	&	(1.1	$\pm$	0.1)$\times$10$^{14	}$	&			--				&			--				&			--				\\
G10.6-0.4 hc1$^{	n	}$	&	(4.0	$\pm$	2.6)$\times$10$^{15	}$	&	(4.2	$\pm$	0.4)$\times$10$^{16	}$	&			--				&			--				&			--				\\
G10.6-0.4 hc2$^{	n	}$	&	(6.3	$\pm$	0.4)$\times$10$^{15	}$	&	(1.3	$\pm$	0.2)$\times$10$^{16	}$	&			--				&			--				&			--				\\
Sgr B2(M)$^{	o	}$	&	(5.8	$\pm$	1.2)$\times$10$^{14	}$	&	(1.4	$\pm$	0.3)$\times$10$^{16	}$	&			--				&			--				&			--				\\
Sgr B2(N1S)$^{	p	}$	&			(2.9	$\pm$	0.9)$\times$10$^{18	}$				&	--	&	(4.1	$\pm$	1.2)$\times$10$^{17	}$	&	(2.6	$\pm$	0.8)$\times$10$^{17	}$	&	(2.7	$\pm$	0.8)$\times$10$^{16	}$	\\
Sgr B2(N2)$^{	q	}$	&	(3.5	$\pm$	1.1)$\times$10$^{18	}$	&	(5.3	$\pm$	1.1)$\times$10$^{17	}$	&	(1.4	$\pm$	0.7)$\times$10$^{17	}$	&	(1.0	$\pm$	0.2)$\times$10$^{17	}$	&			--				\\
Sgr B2(N) (1)$^{	o	}$	&	(1.4	$\pm$	0.3)$\times$10$^{17	}$	&	(1.4	$\pm$	0.3)$\times$10$^{18	}$	&			--				&			--				&			--				\\
Sgr B2(N) (2)$^{	o	}$	&	(8.0	$\pm$	1.6)$\times$10$^{16	}$	&	(9.0	$\pm$	1.8)$\times$10$^{17	}$	&			--				&			--				&			--				\\
Sgr B2(N) cold $^{	r	}$	&				(1.6	$\pm$	0.7)$\times$10$^{14	}$				&   --	&	(5.2	$\pm$	3.5)$\times$10$^{13	}$	&			--				&			--				\\
Sgr B2(N) warm$^{	r	}$	&			(4.0	$\pm$	1.2)$\times$10$^{14	}$				&	--	&	(6.4	$\pm$	4.7)$\times$10$^{14	}$	&			--				&			--				\\
\label{tab:literature}
\end{longtable}
\textsuperscript{Data are taken from:
a\citep{goesmann2015,altwegg2017};
b\citep{biver2021};
c\citep{requena-torres2008,zeng2018,jimenez-serra2020};
d\citep{vangelder2020,nazari2021};
e\citep{yang2021};
f\citep{coutens2016,lykke2017,ligterink2018a};
g\citep{csengeri2019};
h\citep{bisschop2007};
i\citep{ligterink2020b};
j\citep{colzi2021};
k\citep{widicus-weaver2017};
l\citep{cernicharo2016};
m\citep{bogelund2019b};
n\citep{law2021};
o\citep{belloche2013};
p\citep{belloche2019};
q\citep{belloche2017,sanz-novo2020};
r\citep{halfen2011}}
\end{landscape}

\end{suppinfo}

\bibliography{bibliography}

\providecommand{\latin}[1]{#1}
\makeatletter
\providecommand{\doi}
  {\begingroup\let\do\@makeother\dospecials
  \catcode`\{=1 \catcode`\}=2 \doi@aux}
\providecommand{\doi@aux}[1]{\endgroup\texttt{#1}}
\makeatother
\providecommand*\mcitethebibliography{\thebibliography}
\csname @ifundefined\endcsname{endmcitethebibliography}
  {\let\endmcitethebibliography\endthebibliography}{}
\begin{mcitethebibliography}{116}
\providecommand*\natexlab[1]{#1}
\providecommand*\mciteSetBstSublistMode[1]{}
\providecommand*\mciteSetBstMaxWidthForm[2]{}
\providecommand*\mciteBstWouldAddEndPuncttrue
  {\def\EndOfBibitem{\unskip.}}
\providecommand*\mciteBstWouldAddEndPunctfalse
  {\let\EndOfBibitem\relax}
\providecommand*\mciteSetBstMidEndSepPunct[3]{}
\providecommand*\mciteSetBstSublistLabelBeginEnd[3]{}
\providecommand*\EndOfBibitem{}
\mciteSetBstSublistMode{f}
\mciteSetBstMaxWidthForm{subitem}{(\alph{mcitesubitemcount})}
\mciteSetBstSublistLabelBeginEnd
  {\mcitemaxwidthsubitemform\space}
  {\relax}
  {\relax}

\bibitem[Lucas-Lenard and Lipmann(1971)Lucas-Lenard, and
  Lipmann]{lucas-lenard1971}
Lucas-Lenard,~J.; Lipmann,~F. Protein biosynthesis. \emph{Annual review of
  biochemistry} \textbf{1971}, \emph{40}, 409--448\relax
\mciteBstWouldAddEndPuncttrue
\mciteSetBstMidEndSepPunct{\mcitedefaultmidpunct}
{\mcitedefaultendpunct}{\mcitedefaultseppunct}\relax
\EndOfBibitem
\bibitem[McKee \latin{et~al.}(2018)McKee, Solano, Saydjari, Bennett, Hud, and
  Orlando]{mckee2018}
McKee,~A.~D.; Solano,~M.; Saydjari,~A.; Bennett,~C.~J.; Hud,~N.~V.;
  Orlando,~T.~M. A Possible Path to Prebiotic Peptides Involving Silica and
  Hydroxy Acid-Mediated Amide Bond Formation. \emph{ChemBioChem} \textbf{2018},
  \emph{19}, 1913--1917\relax
\mciteBstWouldAddEndPuncttrue
\mciteSetBstMidEndSepPunct{\mcitedefaultmidpunct}
{\mcitedefaultendpunct}{\mcitedefaultseppunct}\relax
\EndOfBibitem
\bibitem[Imai \latin{et~al.}(1999)Imai, Honda, Hatori, Brack, and
  Matsuno]{imai1999}
Imai,~E.-i.; Honda,~H.; Hatori,~K.; Brack,~A.; Matsuno,~K. Elongation of
  oligopeptides in a simulated submarine hydrothermal system. \emph{Science}
  \textbf{1999}, \emph{283}, 831--833\relax
\mciteBstWouldAddEndPuncttrue
\mciteSetBstMidEndSepPunct{\mcitedefaultmidpunct}
{\mcitedefaultendpunct}{\mcitedefaultseppunct}\relax
\EndOfBibitem
\bibitem[Lemke \latin{et~al.}(2009)Lemke, Rosenbauer, and Bird]{lemke2009}
Lemke,~K.~H.; Rosenbauer,~R.~J.; Bird,~D.~K. Peptide synthesis in early Earth
  hydrothermal systems. \emph{Astrobiology} \textbf{2009}, \emph{9},
  141--146\relax
\mciteBstWouldAddEndPuncttrue
\mciteSetBstMidEndSepPunct{\mcitedefaultmidpunct}
{\mcitedefaultendpunct}{\mcitedefaultseppunct}\relax
\EndOfBibitem
\bibitem[Simakov \latin{et~al.}(1996)Simakov, Kuzicheva, Mal'Ko, and
  Dodonova]{simakov1996}
Simakov,~M.; Kuzicheva,~E.; Mal'Ko,~I.; Dodonova,~N.~Y. Abiogenic synthesis of
  oligopeptides in solid state under action of vacuum ultraviolet light
  (100--200 nm). \emph{Advances in Space Research} \textbf{1996}, \emph{18},
  61--64\relax
\mciteBstWouldAddEndPuncttrue
\mciteSetBstMidEndSepPunct{\mcitedefaultmidpunct}
{\mcitedefaultendpunct}{\mcitedefaultseppunct}\relax
\EndOfBibitem
\bibitem[Sugahara and Mimura(2014)Sugahara, and Mimura]{sugahara2014}
Sugahara,~H.; Mimura,~K. Glycine oligomerization up to triglycine by shock
  experiments simulating comet impacts. \emph{Geochemical Journal}
  \textbf{2014}, \emph{48}, 51--62\relax
\mciteBstWouldAddEndPuncttrue
\mciteSetBstMidEndSepPunct{\mcitedefaultmidpunct}
{\mcitedefaultendpunct}{\mcitedefaultseppunct}\relax
\EndOfBibitem
\bibitem[Kaiser \latin{et~al.}(2013)Kaiser, Stockton, Kim, Jensen, and
  Mathies]{kaiser2013}
Kaiser,~R.; Stockton,~A.; Kim,~Y.; Jensen,~E.; Mathies,~R. On the formation of
  dipeptides in interstellar model ices. \emph{The Astrophysical Journal}
  \textbf{2013}, \emph{765}, 111\relax
\mciteBstWouldAddEndPuncttrue
\mciteSetBstMidEndSepPunct{\mcitedefaultmidpunct}
{\mcitedefaultendpunct}{\mcitedefaultseppunct}\relax
\EndOfBibitem
\bibitem[{Rubin} \latin{et~al.}(1971){Rubin}, {Swenson}, {Benson}, {Tigelaar},
  and {Flygare}]{rubin1971}
{Rubin},~R.~H.; {Swenson},~J.,~G.~W.; {Benson},~R.~C.; {Tigelaar},~H.~L.;
  {Flygare},~W.~H. {Microwave Detection of Interstellar Formamide}.
  \emph{The Astrophysical Journal Letters} \textbf{1971}, \emph{169}, L39\relax
\mciteBstWouldAddEndPuncttrue
\mciteSetBstMidEndSepPunct{\mcitedefaultmidpunct}
{\mcitedefaultendpunct}{\mcitedefaultseppunct}\relax
\EndOfBibitem
\bibitem[Coutens \latin{et~al.}(2016)Coutens, J{\o}rgensen, Van~der Wiel,
  M{\"u}ller, Lykke, Bjerkeli, Bourke, Calcutt, Drozdovskaya, Favre,
  \latin{et~al.} others]{coutens2016}
Coutens,~A.; J{\o}rgensen,~J.~K.; Van~der Wiel,~M. H.~D.; M{\"u}ller,~H.;
  Lykke,~J.~M.; Bjerkeli,~P.; Bourke,~T.; Calcutt,~H.; Drozdovskaya,~M.;
  Favre,~C., \latin{et~al.}  The ALMA-PILS survey: First detections of
  deuterated formamide and deuterated isocyanic acid in the interstellar
  medium. \emph{Astronomy \& Astrophysics} \textbf{2016}, \emph{590}, L6\relax
\mciteBstWouldAddEndPuncttrue
\mciteSetBstMidEndSepPunct{\mcitedefaultmidpunct}
{\mcitedefaultendpunct}{\mcitedefaultseppunct}\relax
\EndOfBibitem
\bibitem[Biver \latin{et~al.}(2021)Biver, Bockel{\'e}e-Morvan, Boissier,
  Moreno, Crovisier, Lis, Colom, Cordiner, Milam, Roth, \latin{et~al.}
  others]{biver2021}
Biver,~N.; Bockel{\'e}e-Morvan,~D.; Boissier,~J.; Moreno,~R.; Crovisier,~J.;
  Lis,~D.; Colom,~P.; Cordiner,~M.; Milam,~S.; Roth,~N., \latin{et~al.}
  Molecular composition of comet 46P/Wirtanen from millimetre-wave
  spectroscopy. \emph{Astronomy \& Astrophysics} \textbf{2021}, \emph{648},
  A49\relax
\mciteBstWouldAddEndPuncttrue
\mciteSetBstMidEndSepPunct{\mcitedefaultmidpunct}
{\mcitedefaultendpunct}{\mcitedefaultseppunct}\relax
\EndOfBibitem
\bibitem[{Hollis} \latin{et~al.}(2006){Hollis}, {Lovas}, {Remijan}, {Jewell},
  {Ilyushin}, and {Kleiner}]{hollis2006}
{Hollis},~J.~M.; {Lovas},~F.~J.; {Remijan},~A.~J.; {Jewell},~P.~R.;
  {Ilyushin},~V.~V.; {Kleiner},~I. {Detection of Acetamide
  (CH$_{3}$CONH$_{2}$): The Largest Interstellar Molecule with a Peptide Bond}.
  \emph{The Astrophysical Journal Letters} \textbf{2006}, \emph{643}, L25--L28\relax
\mciteBstWouldAddEndPuncttrue
\mciteSetBstMidEndSepPunct{\mcitedefaultmidpunct}
{\mcitedefaultendpunct}{\mcitedefaultseppunct}\relax
\EndOfBibitem
\bibitem[{Belloche} \latin{et~al.}(2017){Belloche}, {Meshcheryakov}, {Garrod},
  {Ilyushin}, {Alekseev}, {Motiyenko}, {Margul{\`e}s}, {M{\"u}ller}, and
  {Menten}]{belloche2017}
{Belloche},~A.; {Meshcheryakov},~A.~A.; {Garrod},~R.~T.; {Ilyushin},~V.~V.;
  {Alekseev},~E.~A.; {Motiyenko},~R.~A.; {Margul{\`e}s},~L.;
  {M{\"u}ller},~H.~S.~P.; {Menten},~K.~M. {Rotational spectroscopy, tentative
  interstellar detection, and chemical modeling of N-methylformamide}.
  \emph{Astronomy \& Astrophysics} \textbf{2017}, \emph{601}, A49\relax
\mciteBstWouldAddEndPuncttrue
\mciteSetBstMidEndSepPunct{\mcitedefaultmidpunct}
{\mcitedefaultendpunct}{\mcitedefaultseppunct}\relax
\EndOfBibitem
\bibitem[{Belloche} \latin{et~al.}(2019){Belloche}, {Garrod}, {M{\"u}ller},
  {Menten}, {Medvedev}, {Thomas}, and {Kisiel}]{belloche2019}
{Belloche},~A.; {Garrod},~R.~T.; {M{\"u}ller},~H.~S.~P.; {Menten},~K.~M.;
  {Medvedev},~I.; {Thomas},~J.; {Kisiel},~Z. {Re-exploring Molecular Complexity
  with ALMA (ReMoCA): interstellar detection of urea}. \emph{Astronomy \& Astrophysics}
  \textbf{2019}, \emph{628}, A10\relax
\mciteBstWouldAddEndPuncttrue
\mciteSetBstMidEndSepPunct{\mcitedefaultmidpunct}
{\mcitedefaultendpunct}{\mcitedefaultseppunct}\relax
\EndOfBibitem
\bibitem[Ligterink \latin{et~al.}(2020)Ligterink, Grimaudo, Moreno-Garc{\'\i}a,
  Lukmanov, Tulej, Leya, Lindner, Wurz, Cockell, Ehrenfreund, \latin{et~al.}
  others]{ligterink2020a}
Ligterink,~N.~F.; Grimaudo,~V.; Moreno-Garc{\'\i}a,~P.; Lukmanov,~R.;
  Tulej,~M.; Leya,~I.; Lindner,~R.; Wurz,~P.; Cockell,~C.~S.; Ehrenfreund,~P.,
  \latin{et~al.}  ORIGIN: a novel and compact Laser Desorption--Mass
  Spectrometry system for sensitive in situ detection of amino acids on
  extraterrestrial surfaces. \emph{Scientific reports} \textbf{2020},
  \emph{10}, 1--10\relax
\mciteBstWouldAddEndPuncttrue
\mciteSetBstMidEndSepPunct{\mcitedefaultmidpunct}
{\mcitedefaultendpunct}{\mcitedefaultseppunct}\relax
\EndOfBibitem
\bibitem[Colzi \latin{et~al.}(2021)Colzi, Rivilla, Beltr{\'a}n,
  Jim{\'e}nez-Serra, Mininni, Melosso, Cesaroni, Fontani, Lorenzani,
  S{\'a}nchez-Monge, \latin{et~al.} others]{colzi2021}
Colzi,~L.; Rivilla,~V.; Beltr{\'a}n,~M.; Jim{\'e}nez-Serra,~I.; Mininni,~C.;
  Melosso,~M.; Cesaroni,~R.; Fontani,~F.; Lorenzani,~A.; S{\'a}nchez-Monge,~A.,
  \latin{et~al.}  The GUAPOS project II. A comprehensive study of peptide-like
  bond molecules. \emph{arXiv preprint arXiv:2107.11258} \textbf{2021}, \relax
\mciteBstWouldAddEndPunctfalse
\mciteSetBstMidEndSepPunct{\mcitedefaultmidpunct}
{}{\mcitedefaultseppunct}\relax
\EndOfBibitem
\bibitem[Jim{\'e}nez-Serra \latin{et~al.}(2020)Jim{\'e}nez-Serra,
  Mart{\'\i}n-Pintado, Rivilla, Rodr{\'\i}guez-Almeida, Alonso~Alonso, Zeng,
  Cocinero, Mart{\'\i}n, Requena-Torres, Mart{\'\i}n-Domenech, \latin{et~al.}
  others]{jimenez-serra2020}
Jim{\'e}nez-Serra,~I.; Mart{\'\i}n-Pintado,~J.; Rivilla,~V.~M.;
  Rodr{\'\i}guez-Almeida,~L.; Alonso~Alonso,~E.~R.; Zeng,~S.; Cocinero,~E.~J.;
  Mart{\'\i}n,~S.; Requena-Torres,~M.; Mart{\'\i}n-Domenech,~R., \latin{et~al.}
   Toward the RNA-world in the interstellar medium—Detection of urea and
  search of 2-amino-oxazole and simple sugars. \emph{Astrobiology}
  \textbf{2020}, \emph{20}, 1048--1066\relax
\mciteBstWouldAddEndPuncttrue
\mciteSetBstMidEndSepPunct{\mcitedefaultmidpunct}
{\mcitedefaultendpunct}{\mcitedefaultseppunct}\relax
\EndOfBibitem
\bibitem[Sanz-Novo \latin{et~al.}(2020)Sanz-Novo, Belloche, Alonso,
  Kolesnikov{\'a}, Garrod, Mata, M{\"u}ller, Menten, and Gong]{sanz-novo2020}
Sanz-Novo,~M.; Belloche,~A.; Alonso,~J.; Kolesnikov{\'a},~L.; Garrod,~R.;
  Mata,~S.; M{\"u}ller,~H.; Menten,~K.; Gong,~Y. Interstellar glycolamide: A
  comprehensive rotational study and an astronomical search in Sgr B2 (N).
  \emph{Astronomy \& Astrophysics} \textbf{2020}, \emph{639}, A135\relax
\mciteBstWouldAddEndPuncttrue
\mciteSetBstMidEndSepPunct{\mcitedefaultmidpunct}
{\mcitedefaultendpunct}{\mcitedefaultseppunct}\relax
\EndOfBibitem
\bibitem[Li \latin{et~al.}(2021)Li, Wang, Lu, Ilyushin, Motiyenko, Gou,
  Alekseev, Quan, Margules, Gao, \latin{et~al.} others]{li2021}
Li,~J.; Wang,~J.; Lu,~X.; Ilyushin,~V.; Motiyenko,~R.~A.; Gou,~Q.;
  Alekseev,~E.~A.; Quan,~D.; Margules,~L.; Gao,~F., \latin{et~al.}
  Propionamide (C2H5CONH2): The largest peptide-like molecule in space.
  \emph{arXiv preprint arXiv:2108.05001} \textbf{2021}, \relax
\mciteBstWouldAddEndPunctfalse
\mciteSetBstMidEndSepPunct{\mcitedefaultmidpunct}
{}{\mcitedefaultseppunct}\relax
\EndOfBibitem
\bibitem[Ligterink \latin{et~al.}(2020)Ligterink, El-Abd, Brogan, Hunter,
  Remijan, Garrod, and McGuire]{ligterink2020b}
Ligterink,~N.~F.; El-Abd,~S.~J.; Brogan,~C.~L.; Hunter,~T.~R.; Remijan,~A.~J.;
  Garrod,~R.~T.; McGuire,~B.~M. The family of amide molecules toward NGC 6334I.
  \emph{The Astrophysical Journal} \textbf{2020}, \emph{901}, 37\relax
\mciteBstWouldAddEndPuncttrue
\mciteSetBstMidEndSepPunct{\mcitedefaultmidpunct}
{\mcitedefaultendpunct}{\mcitedefaultseppunct}\relax
\EndOfBibitem
\bibitem[Ligterink \latin{et~al.}(2018)Ligterink, Terwisscha~van Scheltinga,
  Taquet, J{\o}rgensen, Cazaux, van Dishoeck, and Linnartz]{ligterink2018a}
Ligterink,~N.; Terwisscha~van Scheltinga,~J.; Taquet,~V.; J{\o}rgensen,~J.;
  Cazaux,~S.; van Dishoeck,~E.; Linnartz,~H. The formation of peptide-like
  molecules on interstellar dust grains. \emph{Monthly Notices of the Royal
  Astronomical Society} \textbf{2018}, \emph{480}, 3628--3643\relax
\mciteBstWouldAddEndPuncttrue
\mciteSetBstMidEndSepPunct{\mcitedefaultmidpunct}
{\mcitedefaultendpunct}{\mcitedefaultseppunct}\relax
\EndOfBibitem
\bibitem[Ortiz-Le{\'o}n \latin{et~al.}(2017)Ortiz-Le{\'o}n, Dzib, Kounkel,
  Loinard, Mioduszewski, Rodr{\'\i}guez, Torres, Pech, Rivera, Hartmann,
  \latin{et~al.} others]{ortiz-leon2017}
Ortiz-Le{\'o}n,~G.~N.; Dzib,~S.~A.; Kounkel,~M.~A.; Loinard,~L.;
  Mioduszewski,~A.~J.; Rodr{\'\i}guez,~L.~F.; Torres,~R.~M.; Pech,~G.;
  Rivera,~J.~L.; Hartmann,~L., \latin{et~al.}  The Gould’s Belt Distances
  Survey (GOBELINS). III. The Distance to the Serpens/Aquila Molecular Complex.
  \emph{The Astrophysical Journal} \textbf{2017}, \emph{834}, 143\relax
\mciteBstWouldAddEndPuncttrue
\mciteSetBstMidEndSepPunct{\mcitedefaultmidpunct}
{\mcitedefaultendpunct}{\mcitedefaultseppunct}\relax
\EndOfBibitem
\bibitem[Dionatos \latin{et~al.}(2013)Dionatos, J{\o}rgensen, Green, Herczeg,
  Evans, Kristensen, Lindberg, and Van~Dishoeck]{dionatos2013}
Dionatos,~O.; J{\o}rgensen,~J.~K.; Green,~J.~D.; Herczeg,~G.; Evans,~N.~J.;
  Kristensen,~L.; Lindberg,~J.; Van~Dishoeck,~E. Dust, ice and gas in time
  (DIGIT): Herschel and Spitzer spectro-imaging of SMM3 and SMM4 in Serpens.
  \emph{Astronomy \& Astrophysics} \textbf{2013}, \emph{558}, A88\relax
\mciteBstWouldAddEndPuncttrue
\mciteSetBstMidEndSepPunct{\mcitedefaultmidpunct}
{\mcitedefaultendpunct}{\mcitedefaultseppunct}\relax
\EndOfBibitem
\bibitem[Hull \latin{et~al.}(2016)Hull, Girart, Kristensen, Dunham,
  Rodr{\'\i}guez-Kamenetzky, Carrasco-Gonz{\'a}lez, Cort{\'e}s, Li, and
  Plambeck]{hull2016}
Hull,~C.~L.; Girart,~J.~M.; Kristensen,~L.~E.; Dunham,~M.~M.;
  Rodr{\'\i}guez-Kamenetzky,~A.; Carrasco-Gonz{\'a}lez,~C.; Cort{\'e}s,~P.~C.;
  Li,~Z.-Y.; Plambeck,~R.~L. An Extremely High Velocity Molecular Jet
  Surrounded by an Ionized Cavity in the Protostellar Source Serpens SMM1.
  \emph{The Astrophysical Journal Letters} \textbf{2016}, \emph{823}, L27\relax
\mciteBstWouldAddEndPuncttrue
\mciteSetBstMidEndSepPunct{\mcitedefaultmidpunct}
{\mcitedefaultendpunct}{\mcitedefaultseppunct}\relax
\EndOfBibitem
\bibitem[{\"O}berg \latin{et~al.}(2011){\"O}berg, Van~der Marel, Kristensen,
  and Van~Dishoeck]{oberg2011}
{\"O}berg,~K.~I.; Van~der Marel,~N.; Kristensen,~L.~E.; Van~Dishoeck,~E.~F.
  Complex molecules toward low-mass protostars: the Serpens core. \emph{The
  Astrophysical Journal} \textbf{2011}, \emph{740}, 14\relax
\mciteBstWouldAddEndPuncttrue
\mciteSetBstMidEndSepPunct{\mcitedefaultmidpunct}
{\mcitedefaultendpunct}{\mcitedefaultseppunct}\relax
\EndOfBibitem
\bibitem[Goicoechea \latin{et~al.}(2012)Goicoechea, Cernicharo, Karska,
  Herczeg, Polehampton, Wampfler, Kristensen, Van~Dishoeck, Etxaluze,
  Bern{\'e}, \latin{et~al.} others]{goicoechea2012}
Goicoechea,~J.~R.; Cernicharo,~J.; Karska,~A.; Herczeg,~G.; Polehampton,~E.;
  Wampfler,~S.~F.; Kristensen,~L.; Van~Dishoeck,~E.; Etxaluze,~M.;
  Bern{\'e},~O., \latin{et~al.}  The complete far-infrared and submillimeter
  spectrum of the Class 0 protostar Serpens SMM1 obtained with
  Herschel-Characterizing UV-irradiated shocks heating and chemistry.
  \emph{Astronomy \& Astrophysics} \textbf{2012}, \emph{548}, A77\relax
\mciteBstWouldAddEndPuncttrue
\mciteSetBstMidEndSepPunct{\mcitedefaultmidpunct}
{\mcitedefaultendpunct}{\mcitedefaultseppunct}\relax
\EndOfBibitem
\bibitem[Ligterink \latin{et~al.}(2021)Ligterink, Ahmadi, Coutens, Calcutt, van
  Dishoeck, Linnartz, J{\o}rgensen, Garrod, Bouwman, \latin{et~al.}
  others]{ligterink2020c}
Ligterink,~N.; Ahmadi,~A.; Coutens,~A.; Calcutt,~H.; van Dishoeck,~E.;
  Linnartz,~H.; J{\o}rgensen,~J.; Garrod,~R.; Bouwman,~J., \latin{et~al.}  The
  prebiotic molecular inventory of Serpens SMM1-I. An investigation of the
  isomers CH3NCO and HOCH2CN. \emph{Astronomy \& Astrophysics} \textbf{2021},
  \emph{647}, A87\relax
\mciteBstWouldAddEndPuncttrue
\mciteSetBstMidEndSepPunct{\mcitedefaultmidpunct}
{\mcitedefaultendpunct}{\mcitedefaultseppunct}\relax
\EndOfBibitem
\bibitem[Tychoniec \latin{et~al.}(2021)Tychoniec, van Dishoeck, van't Hoff, van
  Gelder, Tabone, Chen, Harsono, Hull, Hogerheijde, Murillo, \latin{et~al.}
  others]{tychoniec2021}
Tychoniec,~{\L}.; van Dishoeck,~E.~F.; van't Hoff,~M.~L.; van Gelder,~M.~L.;
  Tabone,~B.; Chen,~Y.; Harsono,~D.; Hull,~C.~L.; Hogerheijde,~M.~R.;
  Murillo,~N.~M., \latin{et~al.}  Which molecule traces what: chemical
  diagnostics of protostellar sources. \emph{arXiv preprint arXiv:2107.03696}
  \textbf{2021}, \relax
\mciteBstWouldAddEndPunctfalse
\mciteSetBstMidEndSepPunct{\mcitedefaultmidpunct}
{}{\mcitedefaultseppunct}\relax
\EndOfBibitem
\bibitem[Vastel \latin{et~al.}(2015)Vastel, Bottinelli, Caux, Glorian, and
  Boiziot]{vastel2015}
Vastel,~C.; Bottinelli,~S.; Caux,~E.; Glorian,~J.; Boiziot,~M. CASSIS: A tool
  to visualize and analyse instrumental and synthetic spectra. SF2A-2015:
  Proceedings of the Annual meeting of the French Society of Astronomy and
  Astrophysics. 2015; pp 313--316\relax
\mciteBstWouldAddEndPuncttrue
\mciteSetBstMidEndSepPunct{\mcitedefaultmidpunct}
{\mcitedefaultendpunct}{\mcitedefaultseppunct}\relax
\EndOfBibitem
\bibitem[M{\"u}ller \latin{et~al.}(2001)M{\"u}ller, Thorwirth, Roth, and
  Winnewisser]{muller2001}
M{\"u}ller,~H.~S.; Thorwirth,~S.; Roth,~D.; Winnewisser,~G. The Cologne
  database for molecular spectroscopy, CDMS. \emph{Astronomy \& Astrophysics}
  \textbf{2001}, \emph{370}, L49--L52\relax
\mciteBstWouldAddEndPuncttrue
\mciteSetBstMidEndSepPunct{\mcitedefaultmidpunct}
{\mcitedefaultendpunct}{\mcitedefaultseppunct}\relax
\EndOfBibitem
\bibitem[M{\"u}ller \latin{et~al.}(2005)M{\"u}ller, Schl{\"o}der, Stutzki, and
  Winnewisser]{muller2005}
M{\"u}ller,~H.~S.; Schl{\"o}der,~F.; Stutzki,~J.; Winnewisser,~G. The Cologne
  Database for Molecular Spectroscopy, CDMS: a useful tool for astronomers and
  spectroscopists. \emph{Journal of Molecular Structure} \textbf{2005},
  \emph{742}, 215--227\relax
\mciteBstWouldAddEndPuncttrue
\mciteSetBstMidEndSepPunct{\mcitedefaultmidpunct}
{\mcitedefaultendpunct}{\mcitedefaultseppunct}\relax
\EndOfBibitem
\bibitem[Endres \latin{et~al.}(2016)Endres, Schlemmer, Schilke, Stutzki, and
  M{\"u}ller]{endres2016}
Endres,~C.~P.; Schlemmer,~S.; Schilke,~P.; Stutzki,~J.; M{\"u}ller,~H.~S. The
  cologne database for molecular spectroscopy, CDMS, in the virtual atomic and
  molecular data centre, VAMDC. \emph{Journal of Molecular Spectroscopy}
  \textbf{2016}, \emph{327}, 95--104\relax
\mciteBstWouldAddEndPuncttrue
\mciteSetBstMidEndSepPunct{\mcitedefaultmidpunct}
{\mcitedefaultendpunct}{\mcitedefaultseppunct}\relax
\EndOfBibitem
\bibitem[Pickett \latin{et~al.}(1998)Pickett, Poynter, Cohen, Delitsky,
  Pearson, and M{\"u}ller]{pickett1998}
Pickett,~H.; Poynter,~R.; Cohen,~E.; Delitsky,~M.; Pearson,~J.; M{\"u}ller,~H.
  Submillimeter, millimeter, and microwave spectral line catalog. \emph{Journal
  of Quantitative Spectroscopy and Radiative Transfer} \textbf{1998},
  \emph{60}, 883--890\relax
\mciteBstWouldAddEndPuncttrue
\mciteSetBstMidEndSepPunct{\mcitedefaultmidpunct}
{\mcitedefaultendpunct}{\mcitedefaultseppunct}\relax
\EndOfBibitem
\bibitem[Schuller \latin{et~al.}(2009)Schuller, Menten, Contreras, Wyrowski,
  Schilke, Bronfman, Henning, Walmsley, Beuther, Bontemps, \latin{et~al.}
  others]{schuller2009}
Schuller,~F.; Menten,~K.; Contreras,~Y.; Wyrowski,~F.; Schilke,~P.;
  Bronfman,~L.; Henning,~T.; Walmsley,~C.; Beuther,~H.; Bontemps,~S.,
  \latin{et~al.}  ATLASGAL--The APEX telescope large area survey of the galaxy
  at 870 m. \emph{Astronomy \& Astrophysics} \textbf{2009}, \emph{504},
  415--427\relax
\mciteBstWouldAddEndPuncttrue
\mciteSetBstMidEndSepPunct{\mcitedefaultmidpunct}
{\mcitedefaultendpunct}{\mcitedefaultseppunct}\relax
\EndOfBibitem
\bibitem[Ahmadi \latin{et~al.}(2018)Ahmadi, Beuther, Mottram, Bosco, Linz,
  Henning, Winters, Kuiper, Pudritz, S{\'a}nchez-Monge, \latin{et~al.}
  others]{ahmadi2018}
Ahmadi,~A.; Beuther,~H.; Mottram,~J.; Bosco,~F.; Linz,~H.; Henning,~T.;
  Winters,~J.; Kuiper,~R.; Pudritz,~R.; S{\'a}nchez-Monge,~{\'A}.,
  \latin{et~al.}  Core fragmentation and Toomre stability analysis of W3
  (H2O)-A case study of the IRAM NOEMA large program CORE. \emph{Astronomy \&
  Astrophysics} \textbf{2018}, \emph{618}, A46\relax
\mciteBstWouldAddEndPuncttrue
\mciteSetBstMidEndSepPunct{\mcitedefaultmidpunct}
{\mcitedefaultendpunct}{\mcitedefaultseppunct}\relax
\EndOfBibitem
\bibitem[Ossenkopf and Henning(1994)Ossenkopf, and Henning]{ossenkopf1994}
Ossenkopf,~V.; Henning,~T. Dust opacities for protostellar cores.
  \emph{Astronomy and Astrophysics} \textbf{1994}, \emph{291}, 943--959\relax
\mciteBstWouldAddEndPuncttrue
\mciteSetBstMidEndSepPunct{\mcitedefaultmidpunct}
{\mcitedefaultendpunct}{\mcitedefaultseppunct}\relax
\EndOfBibitem
\bibitem[Calcutt \latin{et~al.}(2014)Calcutt, Viti, Codella, Beltr{\'a}n,
  Fontani, and Woods]{calcutt2014}
Calcutt,~H.; Viti,~S.; Codella,~C.; Beltr{\'a}n,~M.~T.; Fontani,~F.;
  Woods,~P.~M. A high-resolution study of complex organic molecules in hot
  cores. \emph{Monthly Notices of the Royal Astronomical Society}
  \textbf{2014}, \emph{443}, 3157--3173\relax
\mciteBstWouldAddEndPuncttrue
\mciteSetBstMidEndSepPunct{\mcitedefaultmidpunct}
{\mcitedefaultendpunct}{\mcitedefaultseppunct}\relax
\EndOfBibitem
\bibitem[Wilson(1999)]{wilson1999}
Wilson,~T. Isotopes in the interstellar medium and circumstellar envelopes.
  \emph{Reports on Progress in Physics} \textbf{1999}, \emph{62}, 143\relax
\mciteBstWouldAddEndPuncttrue
\mciteSetBstMidEndSepPunct{\mcitedefaultmidpunct}
{\mcitedefaultendpunct}{\mcitedefaultseppunct}\relax
\EndOfBibitem
\bibitem[Yan \latin{et~al.}(2019)Yan, Zhang, Henkel, Mufakharov, Jia, Tang, Wu,
  Li, Zeng, Wang, \latin{et~al.} others]{yan2019}
Yan,~Y.; Zhang,~J.; Henkel,~C.; Mufakharov,~T.; Jia,~L.; Tang,~X.; Wu,~Y.;
  Li,~J.; Zeng,~Z.; Wang,~Y., \latin{et~al.}  A Systematic TMRT Observational
  Study of Galactic 12C/13C Ratios from Formaldehyde. \emph{The Astrophysical
  Journal} \textbf{2019}, \emph{877}, 154\relax
\mciteBstWouldAddEndPuncttrue
\mciteSetBstMidEndSepPunct{\mcitedefaultmidpunct}
{\mcitedefaultendpunct}{\mcitedefaultseppunct}\relax
\EndOfBibitem
\bibitem[Zeng \latin{et~al.}(2020)Zeng, Zhang, Jim{\'e}nez-Serra, Tercero, Lu,
  Mart{\'\i}n-Pintado, De~Vicente, Rivilla, and Li]{zeng2020}
Zeng,~S.; Zhang,~Q.; Jim{\'e}nez-Serra,~I.; Tercero,~B.; Lu,~X.;
  Mart{\'\i}n-Pintado,~J.; De~Vicente,~P.; Rivilla,~V.; Li,~S. Cloud--cloud
  collision as drivers of the chemical complexity in Galactic Centre molecular
  clouds. \emph{Monthly Notices of the Royal Astronomical Society}
  \textbf{2020}, \emph{497}, 4896--4909\relax
\mciteBstWouldAddEndPuncttrue
\mciteSetBstMidEndSepPunct{\mcitedefaultmidpunct}
{\mcitedefaultendpunct}{\mcitedefaultseppunct}\relax
\EndOfBibitem
\bibitem[Goesmann \latin{et~al.}(2015)Goesmann, Rosenbauer, Bredeh{\"o}ft,
  Cabane, Ehrenfreund, Gautier, Giri, Kr{\"u}ger, Le~Roy, MacDermott,
  \latin{et~al.} others]{goesmann2015}
Goesmann,~F.; Rosenbauer,~H.; Bredeh{\"o}ft,~J.~H.; Cabane,~M.;
  Ehrenfreund,~P.; Gautier,~T.; Giri,~C.; Kr{\"u}ger,~H.; Le~Roy,~L.;
  MacDermott,~A.~J., \latin{et~al.}  Organic compounds on comet
  67P/Churyumov-Gerasimenko revealed by COSAC mass spectrometry. \emph{Science}
  \textbf{2015}, \emph{349}, aab0689\relax
\mciteBstWouldAddEndPuncttrue
\mciteSetBstMidEndSepPunct{\mcitedefaultmidpunct}
{\mcitedefaultendpunct}{\mcitedefaultseppunct}\relax
\EndOfBibitem
\bibitem[Altwegg \latin{et~al.}(2017)Altwegg, Balsiger, Berthelier, Bieler,
  Calmonte, Fuselier, Goesmann, Gasc, Gombosi, Le~Roy, \latin{et~al.}
  others]{altwegg2017}
Altwegg,~K.; Balsiger,~H.; Berthelier,~J.-J.; Bieler,~A.; Calmonte,~U.;
  Fuselier,~S.~A.; Goesmann,~F.; Gasc,~S.; Gombosi,~T.~I.; Le~Roy,~L.,
  \latin{et~al.}  Organics in comet 67P--a first comparative analysis of mass
  spectra from ROSINA--DFMS, COSAC and Ptolemy. \emph{Monthly Notices of the
  Royal Astronomical Society} \textbf{2017}, \emph{469}, S130--S141\relax
\mciteBstWouldAddEndPuncttrue
\mciteSetBstMidEndSepPunct{\mcitedefaultmidpunct}
{\mcitedefaultendpunct}{\mcitedefaultseppunct}\relax
\EndOfBibitem
\bibitem[Requena-Torres \latin{et~al.}(2008)Requena-Torres,
  Mart{\'\i}n-Pintado, Mart{\'\i}n, and Morris]{requena-torres2008}
Requena-Torres,~M.; Mart{\'\i}n-Pintado,~J.; Mart{\'\i}n,~S.; Morris,~M. The
  galactic center: The largest oxygen-bearing organic molecule repository.
  \emph{The Astrophysical Journal} \textbf{2008}, \emph{672}, 352\relax
\mciteBstWouldAddEndPuncttrue
\mciteSetBstMidEndSepPunct{\mcitedefaultmidpunct}
{\mcitedefaultendpunct}{\mcitedefaultseppunct}\relax
\EndOfBibitem
\bibitem[Zeng \latin{et~al.}(2018)Zeng, Jim{\'e}nez-Serra, Rivilla,
  Mart{\'\i}n, Mart{\'\i}n-Pintado, Requena-Torres, Armijos-Abenda{\~n}o,
  Riquelme, and Aladro]{zeng2018}
Zeng,~S.; Jim{\'e}nez-Serra,~I.; Rivilla,~V.; Mart{\'\i}n,~S.;
  Mart{\'\i}n-Pintado,~J.; Requena-Torres,~M.; Armijos-Abenda{\~n}o,~J.;
  Riquelme,~D.; Aladro,~R. Complex organic molecules in the Galactic Centre:
  the N-bearing family. \emph{Monthly Notices of the Royal Astronomical
  Society} \textbf{2018}, \emph{478}, 2962--2975\relax
\mciteBstWouldAddEndPuncttrue
\mciteSetBstMidEndSepPunct{\mcitedefaultmidpunct}
{\mcitedefaultendpunct}{\mcitedefaultseppunct}\relax
\EndOfBibitem
\bibitem[van Gelder \latin{et~al.}(2020)van Gelder, Tabone, van Dishoeck,
  Beuther, Boogert, o~Garatti, Klaassen, Linnartz, M{\"u}ller, Taquet,
  \latin{et~al.} others]{vangelder2020}
van Gelder,~M.; Tabone,~B.; van Dishoeck,~E.; Beuther,~H.; Boogert,~A.;
  o~Garatti,~A.~C.; Klaassen,~P.; Linnartz,~H.; M{\"u}ller,~H.; Taquet,~V.,
  \latin{et~al.}  Complex organic molecules in low-mass protostars on Solar
  System scales-I. Oxygen-bearing species. \emph{Astronomy \& Astrophysics}
  \textbf{2020}, \emph{639}, A87\relax
\mciteBstWouldAddEndPuncttrue
\mciteSetBstMidEndSepPunct{\mcitedefaultmidpunct}
{\mcitedefaultendpunct}{\mcitedefaultseppunct}\relax
\EndOfBibitem
\bibitem[Nazari \latin{et~al.}(2021)Nazari, van Gelder, van Dishoeck, Tabone,
  van’t Hoff, Ligterink, Beuther, Boogert, Caratti~o Garatti, Klaassen,
  \latin{et~al.} others]{nazari2021}
Nazari,~P.; van Gelder,~M.; van Dishoeck,~E.; Tabone,~B.; van’t Hoff,~M.;
  Ligterink,~N.; Beuther,~H.; Boogert,~A.; Caratti~o Garatti,~A.; Klaassen,~P.,
  \latin{et~al.}  Complex organic molecules in low-mass protostars on Solar
  System scales: II. Nitrogen-bearing species. \emph{Astronomy \& Astrophysics}
  \textbf{2021}, \emph{650}, A150\relax
\mciteBstWouldAddEndPuncttrue
\mciteSetBstMidEndSepPunct{\mcitedefaultmidpunct}
{\mcitedefaultendpunct}{\mcitedefaultseppunct}\relax
\EndOfBibitem
\bibitem[Yang \latin{et~al.}(2021)Yang, Sakai, Zhang, Murillo, Zhang, Higuchi,
  Zeng, L{\'o}pez-Sepulcre, Yamamoto, Lefloch, \latin{et~al.} others]{yang2021}
Yang,~Y.-L.; Sakai,~N.; Zhang,~Y.; Murillo,~N.~M.; Zhang,~Z.~E.;
  Higuchi,~A.~E.; Zeng,~S.; L{\'o}pez-Sepulcre,~A.; Yamamoto,~S.; Lefloch,~B.,
  \latin{et~al.}  The Perseus ALMA Chemistry Survey (PEACHES). I. The Complex
  Organic Molecules in Perseus Embedded Protostars. \emph{The Astrophysical
  Journal} \textbf{2021}, \emph{910}, 20\relax
\mciteBstWouldAddEndPuncttrue
\mciteSetBstMidEndSepPunct{\mcitedefaultmidpunct}
{\mcitedefaultendpunct}{\mcitedefaultseppunct}\relax
\EndOfBibitem
\bibitem[Lykke \latin{et~al.}(2017)Lykke, Coutens, J{\o}rgensen, Van~der Wiel,
  Garrod, M{\"u}ller, Bjerkeli, Bourke, Calcutt, Drozdovskaya, \latin{et~al.}
  others]{lykke2017}
Lykke,~J.; Coutens,~A.; J{\o}rgensen,~J.; Van~der Wiel,~M.; Garrod,~R.;
  M{\"u}ller,~H.; Bjerkeli,~P.; Bourke,~T.; Calcutt,~H.; Drozdovskaya,~M.,
  \latin{et~al.}  The ALMA-PILS survey: First detections of ethylene oxide,
  acetone and propanal toward the low-mass protostar IRAS 16293-2422.
  \emph{Astronomy \& Astrophysics} \textbf{2017}, \emph{597}, A53\relax
\mciteBstWouldAddEndPuncttrue
\mciteSetBstMidEndSepPunct{\mcitedefaultmidpunct}
{\mcitedefaultendpunct}{\mcitedefaultseppunct}\relax
\EndOfBibitem
\bibitem[Csengeri \latin{et~al.}(2019)Csengeri, Belloche, Bontemps, Wyrowski,
  Menten, and Bouscasse]{csengeri2019}
Csengeri,~T.; Belloche,~A.; Bontemps,~S.; Wyrowski,~F.; Menten,~K.;
  Bouscasse,~L. Search for high-mass protostars with ALMA revealed up to
  kilo-parsec scales (SPARKS)-II. Complex organic molecules and heavy water in
  shocks around a young high-mass protostar. \emph{Astronomy \& Astrophysics}
  \textbf{2019}, \emph{632}, A57\relax
\mciteBstWouldAddEndPuncttrue
\mciteSetBstMidEndSepPunct{\mcitedefaultmidpunct}
{\mcitedefaultendpunct}{\mcitedefaultseppunct}\relax
\EndOfBibitem
\bibitem[Bisschop \latin{et~al.}(2007)Bisschop, J{\o}rgensen, Van~Dishoeck, and
  De~Wachter]{bisschop2007}
Bisschop,~S.; J{\o}rgensen,~J.; Van~Dishoeck,~E.; De~Wachter,~E. Testing
  grain-surface chemistry in massive hot-core regions. \emph{Astronomy \&
  Astrophysics} \textbf{2007}, \emph{465}, 913--929\relax
\mciteBstWouldAddEndPuncttrue
\mciteSetBstMidEndSepPunct{\mcitedefaultmidpunct}
{\mcitedefaultendpunct}{\mcitedefaultseppunct}\relax
\EndOfBibitem
\bibitem[Weaver \latin{et~al.}(2017)Weaver, Laas, Zou, Kroll, Rad, Hays,
  Sanders, Lis, Cross, Wehres, \latin{et~al.} others]{widicus-weaver2017}
Weaver,~S. L.~W.; Laas,~J.~C.; Zou,~L.; Kroll,~J.~A.; Rad,~M.~L.; Hays,~B.~M.;
  Sanders,~J.~L.; Lis,~D.~C.; Cross,~T.~N.; Wehres,~N., \latin{et~al.}  Deep,
  broadband spectral line surveys of molecule-rich interstellar clouds.
  \emph{The Astrophysical Journal Supplement Series} \textbf{2017}, \emph{232},
  3\relax
\mciteBstWouldAddEndPuncttrue
\mciteSetBstMidEndSepPunct{\mcitedefaultmidpunct}
{\mcitedefaultendpunct}{\mcitedefaultseppunct}\relax
\EndOfBibitem
\bibitem[Cernicharo \latin{et~al.}(2016)Cernicharo, Kisiel, Tercero,
  Kolesnikov{\'a}, Medvedev, L{\'o}pez, Fortman, Winnewisser, De~Lucia, Alonso,
  \latin{et~al.} others]{cernicharo2016}
Cernicharo,~J.; Kisiel,~Z.; Tercero,~B.; Kolesnikov{\'a},~L.; Medvedev,~I.;
  L{\'o}pez,~A.; Fortman,~S.; Winnewisser,~M.; De~Lucia,~F.; Alonso,~J.,
  \latin{et~al.}  A rigorous detection of interstellar CH$_{3}$NCO: an
  important missing species in astrochemical networks. \emph{Astronomy \&
  Astrophysics} \textbf{2016}, \emph{587}, L4\relax
\mciteBstWouldAddEndPuncttrue
\mciteSetBstMidEndSepPunct{\mcitedefaultmidpunct}
{\mcitedefaultendpunct}{\mcitedefaultseppunct}\relax
\EndOfBibitem
\bibitem[{B{\o}gelund} \latin{et~al.}(2019){B{\o}gelund}, {Barr}, {Taquet},
  {Ligterink}, {Persson}, {Hogerheijde}, and {van Dishoeck}]{bogelund2019b}
{B{\o}gelund},~E.~G.; {Barr},~A.~G.; {Taquet},~V.; {Ligterink},~N. F.~W.;
  {Persson},~M.~V.; {Hogerheijde},~M.~R.; {van Dishoeck},~E.~F. {Molecular
  complexity on disc scales uncovered by ALMA. Chemical composition of the
  high-mass protostar AFGL 4176}. \emph{Astronomy \& Astrophysics} \textbf{2019}, \emph{628},
  A2\relax
\mciteBstWouldAddEndPuncttrue
\mciteSetBstMidEndSepPunct{\mcitedefaultmidpunct}
{\mcitedefaultendpunct}{\mcitedefaultseppunct}\relax
\EndOfBibitem
\bibitem[Law \latin{et~al.}(2021)Law, Zhang, {\"O}berg, Galv{\'a}n-Madrid,
  Keto, Liu, and Ho]{law2021}
Law,~C.~J.; Zhang,~Q.; {\"O}berg,~K.~I.; Galv{\'a}n-Madrid,~R.; Keto,~E.;
  Liu,~H.~B.; Ho,~P.~T. Subarcsecond Imaging of the Complex Organic Chemistry
  in Massive Star-forming Region G10. 6-0.4. \emph{The Astrophysical Journal}
  \textbf{2021}, \emph{909}, 214\relax
\mciteBstWouldAddEndPuncttrue
\mciteSetBstMidEndSepPunct{\mcitedefaultmidpunct}
{\mcitedefaultendpunct}{\mcitedefaultseppunct}\relax
\EndOfBibitem
\bibitem[Belloche \latin{et~al.}(2013)Belloche, M{\"u}ller, Menten, Schilke,
  and Comito]{belloche2013}
Belloche,~A.; M{\"u}ller,~H.~S.; Menten,~K.~M.; Schilke,~P.; Comito,~C. Complex
  organic molecules in the interstellar medium: IRAM 30 m line survey of
  Sagittarius B2 (N) and (M). \emph{Astronomy \& Astrophysics} \textbf{2013},
  \emph{559}, A47\relax
\mciteBstWouldAddEndPuncttrue
\mciteSetBstMidEndSepPunct{\mcitedefaultmidpunct}
{\mcitedefaultendpunct}{\mcitedefaultseppunct}\relax
\EndOfBibitem
\bibitem[{Halfen} \latin{et~al.}(2011){Halfen}, {Ilyushin}, and
  {Ziurys}]{halfen2011}
{Halfen},~D.~T.; {Ilyushin},~V.; {Ziurys},~L.~M. {Formation of Peptide Bonds in
  Space: A Comprehensive Study of Formamide and Acetamide in Sgr B2(N)}.
  \emph{The Astrophysical Journal} \textbf{2011}, \emph{743}, 60\relax
\mciteBstWouldAddEndPuncttrue
\mciteSetBstMidEndSepPunct{\mcitedefaultmidpunct}
{\mcitedefaultendpunct}{\mcitedefaultseppunct}\relax
\EndOfBibitem
\bibitem[Jacobsen \latin{et~al.}(2018)Jacobsen, J{\o}rgensen, Van~der Wiel,
  Calcutt, Bourke, Brinch, Coutens, Drozdovskaya, Kristensen, M{\"u}ller,
  \latin{et~al.} others]{jacobsen2018}
Jacobsen,~S.; J{\o}rgensen,~J.; Van~der Wiel,~M.; Calcutt,~H.; Bourke,~T.;
  Brinch,~C.; Coutens,~A.; Drozdovskaya,~M.~N.; Kristensen,~L.; M{\"u}ller,~H.,
  \latin{et~al.}  The ALMA-PILS survey: 3D modeling of the envelope, disks and
  dust filament of IRAS 16293--2422. \emph{Astronomy \& Astrophysics}
  \textbf{2018}, \emph{612}, A72\relax
\mciteBstWouldAddEndPuncttrue
\mciteSetBstMidEndSepPunct{\mcitedefaultmidpunct}
{\mcitedefaultendpunct}{\mcitedefaultseppunct}\relax
\EndOfBibitem
\bibitem[Qu{\'e}nard \latin{et~al.}(2018)Qu{\'e}nard, Jim{\'e}nez-Serra, Viti,
  Holdship, and Coutens]{quenard2018}
Qu{\'e}nard,~D.; Jim{\'e}nez-Serra,~I.; Viti,~S.; Holdship,~J.; Coutens,~A.
  Chemical modelling of complex organic molecules with peptide-like bonds in
  star-forming regions. \emph{Monthly Notices of the Royal Astronomical
  Society} \textbf{2018}, \emph{474}, 2796--2812\relax
\mciteBstWouldAddEndPuncttrue
\mciteSetBstMidEndSepPunct{\mcitedefaultmidpunct}
{\mcitedefaultendpunct}{\mcitedefaultseppunct}\relax
\EndOfBibitem
\bibitem[Charnley(2004)]{charnley2004}
Charnley,~S. Acetaldehyde in star-forming regions. \emph{Advances in Space
  Research} \textbf{2004}, \emph{33}, 23--30\relax
\mciteBstWouldAddEndPuncttrue
\mciteSetBstMidEndSepPunct{\mcitedefaultmidpunct}
{\mcitedefaultendpunct}{\mcitedefaultseppunct}\relax
\EndOfBibitem
\bibitem[Garrod \latin{et~al.}(2021)Garrod, Jin, Matis, Jones, Willis, and
  Herbst]{garrod2021}
Garrod,~R.~T.; Jin,~M.; Matis,~K.~A.; Jones,~D.; Willis,~E.~R.; Herbst,~E.
  Formation of complex organic molecules in hot molecular cores through
  nondiffusive grain-surface and ice-mantle chemistry. \emph{arXiv preprint
  arXiv:2110.09743} \textbf{2021}, \relax
\mciteBstWouldAddEndPunctfalse
\mciteSetBstMidEndSepPunct{\mcitedefaultmidpunct}
{}{\mcitedefaultseppunct}\relax
\EndOfBibitem
\bibitem[Codella \latin{et~al.}(2017)Codella, Ceccarelli, Caselli, Balucani,
  Barone, Fontani, Lefloch, Podio, Viti, Feng, \latin{et~al.}
  others]{codella2017}
Codella,~C.; Ceccarelli,~C.; Caselli,~P.; Balucani,~N.; Barone,~V.;
  Fontani,~F.; Lefloch,~B.; Podio,~L.; Viti,~S.; Feng,~S., \latin{et~al.}
  Seeds of Life in Space (SOLIS)-II. Formamide in protostellar shocks: Evidence
  for gas-phase formation. \emph{Astronomy \& Astrophysics} \textbf{2017},
  \emph{605}, L3\relax
\mciteBstWouldAddEndPuncttrue
\mciteSetBstMidEndSepPunct{\mcitedefaultmidpunct}
{\mcitedefaultendpunct}{\mcitedefaultseppunct}\relax
\EndOfBibitem
\bibitem[Quan and Herbst(2007)Quan, and Herbst]{quan2007}
Quan,~D.; Herbst,~E. Possible gas-phase syntheses for seven neutral molecules
  studied recently with the Green Bank Telescope. \emph{Astronomy \&
  Astrophysics} \textbf{2007}, \emph{474}, 521--527\relax
\mciteBstWouldAddEndPuncttrue
\mciteSetBstMidEndSepPunct{\mcitedefaultmidpunct}
{\mcitedefaultendpunct}{\mcitedefaultseppunct}\relax
\EndOfBibitem
\bibitem[Redondo \latin{et~al.}(2014)Redondo, Barrientos, and
  Largo]{redondo2014}
Redondo,~P.; Barrientos,~C.; Largo,~A. Peptide bond formation through gas-phase
  reactions in the interstellar medium: formamide and acetamide as prototypes.
  \emph{The Astrophysical Journal} \textbf{2014}, \emph{793}, 32\relax
\mciteBstWouldAddEndPuncttrue
\mciteSetBstMidEndSepPunct{\mcitedefaultmidpunct}
{\mcitedefaultendpunct}{\mcitedefaultseppunct}\relax
\EndOfBibitem
\bibitem[Yang and Pan(2015)Yang, and Pan]{yang2015}
Yang,~Z.; Pan,~N. Computational studies of ion--neutral reactions of
  astrochemical relevance: Formation of hydrogen peroxide, acetamide, and amino
  acetonitrile. \emph{International Journal of Mass Spectrometry}
  \textbf{2015}, \emph{378}, 364--368\relax
\mciteBstWouldAddEndPuncttrue
\mciteSetBstMidEndSepPunct{\mcitedefaultmidpunct}
{\mcitedefaultendpunct}{\mcitedefaultseppunct}\relax
\EndOfBibitem
\bibitem[Foo \latin{et~al.}(2018)Foo, Sur{\'a}nyi, Guljas, Sz{\H{o}}ri, Villar,
  Viskolcz, Csizmadia, R{\'a}gyanszki, and Fiser]{foo2018}
Foo,~L.; Sur{\'a}nyi,~A.; Guljas,~A.; Sz{\H{o}}ri,~M.; Villar,~J.~J.;
  Viskolcz,~B.; Csizmadia,~I.~G.; R{\'a}gyanszki,~A.; Fiser,~B. Formation of
  acetamide in interstellar medium. \emph{Molecular Astrophysics}
  \textbf{2018}, \emph{13}, 1--5\relax
\mciteBstWouldAddEndPuncttrue
\mciteSetBstMidEndSepPunct{\mcitedefaultmidpunct}
{\mcitedefaultendpunct}{\mcitedefaultseppunct}\relax
\EndOfBibitem
\bibitem[Kothari \latin{et~al.}(2020)Kothari, Zhu, Babi, Galant,
  R{\'a}gyanszki, and Csizmadia]{kothari2020}
Kothari,~A.; Zhu,~L.; Babi,~J.; Galant,~N.; R{\'a}gyanszki,~A.; Csizmadia,~I.
  Ketene and Ammonia Forming Acetamide in the Interstellar Medium.
  \emph{Journal of Undergraduate Life Sciences} \textbf{2020}, \emph{14}\relax
\mciteBstWouldAddEndPuncttrue
\mciteSetBstMidEndSepPunct{\mcitedefaultmidpunct}
{\mcitedefaultendpunct}{\mcitedefaultseppunct}\relax
\EndOfBibitem
\bibitem[Berger(1961)]{berger1961}
Berger,~R. The proton irradiation of methane, ammonia, and water at 77 K.
  \emph{Proceedings of the National Academy of Sciences of the United States of
  America} \textbf{1961}, \emph{47}, 1434\relax
\mciteBstWouldAddEndPuncttrue
\mciteSetBstMidEndSepPunct{\mcitedefaultmidpunct}
{\mcitedefaultendpunct}{\mcitedefaultseppunct}\relax
\EndOfBibitem
\bibitem[Bernstein \latin{et~al.}(1995)Bernstein, Sandford, Allamandola, Chang,
  and Scharberg]{bernstein1995}
Bernstein,~M.~P.; Sandford,~S.~A.; Allamandola,~L.~J.; Chang,~S.;
  Scharberg,~M.~A. Organic compounds produced by photolysis of realistic
  interstellar and cometary ice analogs containing methanol. \emph{The
  Astrophysical Journal} \textbf{1995}, \emph{454}, 327\relax
\mciteBstWouldAddEndPuncttrue
\mciteSetBstMidEndSepPunct{\mcitedefaultmidpunct}
{\mcitedefaultendpunct}{\mcitedefaultseppunct}\relax
\EndOfBibitem
\bibitem[Henderson and Gudipati(2015)Henderson, and Gudipati]{henderson2015}
Henderson,~B.~L.; Gudipati,~M.~S. Direct detection of complex organic products
  in ultraviolet (Ly$\alpha$) and electron-irradiated astrophysical and
  cometary ice analogs using two-step laser ablation and ionization mass
  spectrometry. \emph{The Astrophysical Journal} \textbf{2015}, \emph{800},
  66\relax
\mciteBstWouldAddEndPuncttrue
\mciteSetBstMidEndSepPunct{\mcitedefaultmidpunct}
{\mcitedefaultendpunct}{\mcitedefaultseppunct}\relax
\EndOfBibitem
\bibitem[Agarwal \latin{et~al.}(1985)Agarwal, Schutte, Greenberg, Ferris,
  Briggs, Connor, Van~de Bult, and Baas]{agarwal1985}
Agarwal,~V.; Schutte,~W.; Greenberg,~J.; Ferris,~J.; Briggs,~R.; Connor,~S.;
  Van~de Bult,~C.; Baas,~F. Photochemical reactions in interstellar grains
  photolysis of CO, NH 3, and H 2 O. \emph{Origins of Life and Evolution of the
  Biosphere} \textbf{1985}, \emph{16}, 21--40\relax
\mciteBstWouldAddEndPuncttrue
\mciteSetBstMidEndSepPunct{\mcitedefaultmidpunct}
{\mcitedefaultendpunct}{\mcitedefaultseppunct}\relax
\EndOfBibitem
\bibitem[Rimola \latin{et~al.}(2018)Rimola, Skouteris, Balucani, Ceccarelli,
  Enrique-Romero, Taquet, and Ugliengo]{rimola2018}
Rimola,~A.; Skouteris,~D.; Balucani,~N.; Ceccarelli,~C.; Enrique-Romero,~J.;
  Taquet,~V.; Ugliengo,~P. Can formamide be formed on interstellar ice? An
  atomistic perspective. \emph{ACS Earth and Space Chemistry} \textbf{2018},
  \emph{2}, 720--734\relax
\mciteBstWouldAddEndPuncttrue
\mciteSetBstMidEndSepPunct{\mcitedefaultmidpunct}
{\mcitedefaultendpunct}{\mcitedefaultseppunct}\relax
\EndOfBibitem
\bibitem[Haupa \latin{et~al.}(2019)Haupa, Tarczay, and Lee]{haupa2019}
Haupa,~K.~A.; Tarczay,~G.; Lee,~Y.-P. Hydrogen abstraction/addition tunneling
  reactions elucidate the interstellar H2NCHO/HNCO ratio and H2 formation.
  \emph{Journal of the American Chemical Society} \textbf{2019}, \emph{141},
  11614--11620\relax
\mciteBstWouldAddEndPuncttrue
\mciteSetBstMidEndSepPunct{\mcitedefaultmidpunct}
{\mcitedefaultendpunct}{\mcitedefaultseppunct}\relax
\EndOfBibitem
\bibitem[Bulak \latin{et~al.}(2021)Bulak, Paardekooper, Fedoseev, and
  Linnartz]{bulak2021}
Bulak,~M.; Paardekooper,~D.; Fedoseev,~G.; Linnartz,~H. Photolysis of
  acetonitrile in a water-rich ice as a source of complex organic molecules:
  CH3CN and H2O: CH3CN ices. \emph{Astronomy \& Astrophysics} \textbf{2021},
  \emph{647}, A82\relax
\mciteBstWouldAddEndPuncttrue
\mciteSetBstMidEndSepPunct{\mcitedefaultmidpunct}
{\mcitedefaultendpunct}{\mcitedefaultseppunct}\relax
\EndOfBibitem
\bibitem[Duvernay \latin{et~al.}(2010)Duvernay, Dufauret, Danger, Theul{\'e},
  Borget, and Chiavassa]{duvernay2010}
Duvernay,~F.; Dufauret,~V.; Danger,~G.; Theul{\'e},~P.; Borget,~F.;
  Chiavassa,~T. Chiral molecule formation in interstellar ice analogs:
  alpha-aminoethanol NH2CH (CH3) OH. \emph{Astronomy \& Astrophysics}
  \textbf{2010}, \emph{523}, A79\relax
\mciteBstWouldAddEndPuncttrue
\mciteSetBstMidEndSepPunct{\mcitedefaultmidpunct}
{\mcitedefaultendpunct}{\mcitedefaultseppunct}\relax
\EndOfBibitem
\bibitem[Haupa \latin{et~al.}(2020)Haupa, Ong, and Lee]{haupa2020}
Haupa,~K.~A.; Ong,~W.-S.; Lee,~Y.-P. Hydrogen abstraction in astrochemistry:
  formation of˙ CH 2 CONH 2 in the reaction of H atom with acetamide (CH 3
  CONH 2) and photolysis of˙ CH 2 CONH 2 to form ketene (CH 2 CO) in solid
  para-hydrogen. \emph{Physical Chemistry Chemical Physics} \textbf{2020},
  \emph{22}, 6192--6201\relax
\mciteBstWouldAddEndPuncttrue
\mciteSetBstMidEndSepPunct{\mcitedefaultmidpunct}
{\mcitedefaultendpunct}{\mcitedefaultseppunct}\relax
\EndOfBibitem
\bibitem[Lopez-Sepulcre \latin{et~al.}(2019)Lopez-Sepulcre, Balucani,
  Ceccarelli, Codella, Dulieu, and Theule]{lopez-sepulcre2019}
Lopez-Sepulcre,~A.; Balucani,~N.; Ceccarelli,~C.; Codella,~C.; Dulieu,~F.;
  Theule,~P. Interstellar formamide (NH2CHO), a key prebiotic precursor.
  \emph{ACS Earth and Space Chemistry} \textbf{2019}, \emph{3},
  2122--2137\relax
\mciteBstWouldAddEndPuncttrue
\mciteSetBstMidEndSepPunct{\mcitedefaultmidpunct}
{\mcitedefaultendpunct}{\mcitedefaultseppunct}\relax
\EndOfBibitem
\bibitem[Tielens(1983)]{tielens1983}
Tielens,~A. Surface chemistry of deuterated molecules. \emph{Astronomy and
  Astrophysics} \textbf{1983}, \emph{119}, 177--184\relax
\mciteBstWouldAddEndPuncttrue
\mciteSetBstMidEndSepPunct{\mcitedefaultmidpunct}
{\mcitedefaultendpunct}{\mcitedefaultseppunct}\relax
\EndOfBibitem
\bibitem[Skouteris \latin{et~al.}(2017)Skouteris, Vazart, Ceccarelli, Balucani,
  Puzzarini, and Barone]{skouteris2017}
Skouteris,~D.; Vazart,~F.; Ceccarelli,~C.; Balucani,~N.; Puzzarini,~C.;
  Barone,~V. New quantum chemical computations of formamide deuteration support
  gas-phase formation of this prebiotic molecule. \emph{Monthly Notices of the
  Royal Astronomical Society: Letters} \textbf{2017}, \emph{468}, L1--L5\relax
\mciteBstWouldAddEndPuncttrue
\mciteSetBstMidEndSepPunct{\mcitedefaultmidpunct}
{\mcitedefaultendpunct}{\mcitedefaultseppunct}\relax
\EndOfBibitem
\bibitem[Fedoseev \latin{et~al.}(2016)Fedoseev, Chuang, van Dishoeck, Ioppolo,
  and Linnartz]{fedoseev2016}
Fedoseev,~G.; Chuang,~K.-J.; van Dishoeck,~E.~F.; Ioppolo,~S.; Linnartz,~H.
  Simultaneous hydrogenation and UV-photolysis experiments of NO in CO-rich
  interstellar ice analogues; linking HNCO, OCN-, NH2CHO, and NH2OH.
  \emph{Monthly Notices of the Royal Astronomical Society} \textbf{2016},
  \emph{460}, 4297--4309\relax
\mciteBstWouldAddEndPuncttrue
\mciteSetBstMidEndSepPunct{\mcitedefaultmidpunct}
{\mcitedefaultendpunct}{\mcitedefaultseppunct}\relax
\EndOfBibitem
\bibitem[{Milam} \latin{et~al.}(2005){Milam}, {Savage}, {Brewster}, {Ziurys},
  and {Wyckoff}]{milam2005}
{Milam},~S.~N.; {Savage},~C.; {Brewster},~M.~A.; {Ziurys},~L.~M.; {Wyckoff},~S.
  {The $^{12}$C/$^{13}$C Isotope Gradient Derived from Millimeter Transitions
  of CN: The Case for Galactic Chemical Evolution}. \emph{The Astrophysical Journal} \textbf{2005},
  \emph{634}, 1126--1132\relax
\mciteBstWouldAddEndPuncttrue
\mciteSetBstMidEndSepPunct{\mcitedefaultmidpunct}
{\mcitedefaultendpunct}{\mcitedefaultseppunct}\relax
\EndOfBibitem
\bibitem[Frigge \latin{et~al.}(2018)Frigge, Zhu, Turner, Abplanalp, Bergantini,
  Sun, Chen, Chang, and Kaiser]{frigge2018}
Frigge,~R.; Zhu,~C.; Turner,~A.~M.; Abplanalp,~M.~J.; Bergantini,~A.;
  Sun,~B.-J.; Chen,~Y.-L.; Chang,~A.~H.; Kaiser,~R.~I. A Vacuum Ultraviolet
  Photoionization Study on the Formation of N-methyl Formamide (HCONHCH3) in
  Deep Space: A Potential Interstellar Molecule with a Peptide Bond. \emph{The
  Astrophysical Journal} \textbf{2018}, \emph{862}, 84\relax
\mciteBstWouldAddEndPuncttrue
\mciteSetBstMidEndSepPunct{\mcitedefaultmidpunct}
{\mcitedefaultendpunct}{\mcitedefaultseppunct}\relax
\EndOfBibitem
\bibitem[Booth \latin{et~al.}(2021)Booth, Walsh, van Scheltinga, van Dishoeck,
  Ilee, Hogerheijde, Kama, and Nomura]{booth2021}
Booth,~A.~S.; Walsh,~C.; van Scheltinga,~J.~T.; van Dishoeck,~E.~F.;
  Ilee,~J.~D.; Hogerheijde,~M.~R.; Kama,~M.; Nomura,~H. An inherited complex
  organic molecule reservoir in a warm planet-hosting disk. \emph{Nature
  Astronomy} \textbf{2021}, 1--7\relax
\mciteBstWouldAddEndPuncttrue
\mciteSetBstMidEndSepPunct{\mcitedefaultmidpunct}
{\mcitedefaultendpunct}{\mcitedefaultseppunct}\relax
\EndOfBibitem
\bibitem[Drozdovskaya \latin{et~al.}(2019)Drozdovskaya, van Dishoeck, Rubin,
  J{\o}rgensen, and Altwegg]{drozdovskaya2019}
Drozdovskaya,~M.~N.; van Dishoeck,~E.~F.; Rubin,~M.; J{\o}rgensen,~J.~K.;
  Altwegg,~K. Ingredients for solar-like systems: protostar IRAS 16293-2422 B
  versus comet 67P/Churyumov--Gerasimenko. \emph{Monthly Notices of the Royal
  Astronomical Society} \textbf{2019}, \emph{490}, 50--79\relax
\mciteBstWouldAddEndPuncttrue
\mciteSetBstMidEndSepPunct{\mcitedefaultmidpunct}
{\mcitedefaultendpunct}{\mcitedefaultseppunct}\relax
\EndOfBibitem
\bibitem[Belloche \latin{et~al.}(2020)Belloche, Maury, Maret, Anderl, Bacmann,
  Andr{\'e}, Bontemps, Cabrit, Codella, Gaudel, \latin{et~al.}
  others]{belloche2020}
Belloche,~A.; Maury,~A.; Maret,~S.; Anderl,~S.; Bacmann,~A.; Andr{\'e},~P.;
  Bontemps,~S.; Cabrit,~S.; Codella,~C.; Gaudel,~M., \latin{et~al.}
  Questioning the spatial origin of complex organic molecules in young
  protostars with the CALYPSO survey. \emph{Astronomy \& Astrophysics}
  \textbf{2020}, \relax
\mciteBstWouldAddEndPunctfalse
\mciteSetBstMidEndSepPunct{\mcitedefaultmidpunct}
{}{\mcitedefaultseppunct}\relax
\EndOfBibitem
\bibitem[Chyba and Sagan(1992)Chyba, and Sagan]{chyba1992}
Chyba,~C.; Sagan,~C. Endogenous production, exogenous delivery and impact-shock
  synthesis of organic molecules: an inventory for the origins of life.
  \emph{Nature} \textbf{1992}, \emph{355}, 125--132\relax
\mciteBstWouldAddEndPuncttrue
\mciteSetBstMidEndSepPunct{\mcitedefaultmidpunct}
{\mcitedefaultendpunct}{\mcitedefaultseppunct}\relax
\EndOfBibitem
\bibitem[Pearson \latin{et~al.}(2012)Pearson, Yu, and Drouin]{pearson2012}
Pearson,~J.~C.; Yu,~S.; Drouin,~B.~J. The ground state torsion rotation
  spectrum of CH2DOH. \emph{Journal of Molecular Spectroscopy} \textbf{2012},
  \emph{280}, 119--133\relax
\mciteBstWouldAddEndPuncttrue
\mciteSetBstMidEndSepPunct{\mcitedefaultmidpunct}
{\mcitedefaultendpunct}{\mcitedefaultseppunct}\relax
\EndOfBibitem
\bibitem[Jacq \latin{et~al.}(1993)Jacq, Walmsley, Mauersberger, Anderson,
  Herbst, and De~Lucia]{jacq1993}
Jacq,~T.; Walmsley,~C.; Mauersberger,~R.; Anderson,~T.; Herbst,~E.;
  De~Lucia,~F. Detection of interstellar CH2DOH. \emph{Astronomy and
  Astrophysics} \textbf{1993}, \emph{271}, 276\relax
\mciteBstWouldAddEndPuncttrue
\mciteSetBstMidEndSepPunct{\mcitedefaultmidpunct}
{\mcitedefaultendpunct}{\mcitedefaultseppunct}\relax
\EndOfBibitem
\bibitem[Quade and Suenram(1980)Quade, and Suenram]{quade1980}
Quade,~C.~R.; Suenram,~R. The microwave spectrum of CH2DOH. \emph{The Journal
  of Chemical Physics} \textbf{1980}, \emph{73}, 1127--1131\relax
\mciteBstWouldAddEndPuncttrue
\mciteSetBstMidEndSepPunct{\mcitedefaultmidpunct}
{\mcitedefaultendpunct}{\mcitedefaultseppunct}\relax
\EndOfBibitem
\bibitem[Mukhopadhyay and Sastry(1997)Mukhopadhyay, and
  Sastry]{mukhopadhyay1997}
Mukhopadhyay,~I.; Sastry,~K. Q-branch microwave transitions in the torsional
  ground state of methanol-D1. \emph{Spectrochimica Acta Part A: Molecular and
  Biomolecular Spectroscopy} \textbf{1997}, \emph{53}, 2061--2065\relax
\mciteBstWouldAddEndPuncttrue
\mciteSetBstMidEndSepPunct{\mcitedefaultmidpunct}
{\mcitedefaultendpunct}{\mcitedefaultseppunct}\relax
\EndOfBibitem
\bibitem[Su and Quade(1989)Su, and Quade]{su1989}
Su,~C.~F.; Quade,~C.~R. Microwave detection of direct trans to gauche
  transitions in CH2DOH. \emph{Journal of Molecular Spectroscopy}
  \textbf{1989}, \emph{134}, 290--296\relax
\mciteBstWouldAddEndPuncttrue
\mciteSetBstMidEndSepPunct{\mcitedefaultmidpunct}
{\mcitedefaultendpunct}{\mcitedefaultseppunct}\relax
\EndOfBibitem
\bibitem[El~Hilali \latin{et~al.}(2011)El~Hilali, Coudert, Konov, and
  Klee]{elhilali2011}
El~Hilali,~A.; Coudert,~L.; Konov,~I.; Klee,~S. Analysis of the torsional
  spectrum of monodeuterated methanol CH2DOH. \emph{The Journal of chemical
  physics} \textbf{2011}, \emph{135}, 194309\relax
\mciteBstWouldAddEndPuncttrue
\mciteSetBstMidEndSepPunct{\mcitedefaultmidpunct}
{\mcitedefaultendpunct}{\mcitedefaultseppunct}\relax
\EndOfBibitem
\bibitem[Johnson and Strandberg(1952)Johnson, and Strandberg]{johnson1952}
Johnson,~H.; Strandberg,~M. The Microwave Spectrum of Ketene. \emph{The Journal
  of Chemical Physics} \textbf{1952}, \emph{20}, 687--695\relax
\mciteBstWouldAddEndPuncttrue
\mciteSetBstMidEndSepPunct{\mcitedefaultmidpunct}
{\mcitedefaultendpunct}{\mcitedefaultseppunct}\relax
\EndOfBibitem
\bibitem[Fabricant \latin{et~al.}(1977)Fabricant, Krieger, and
  Muenter]{fabricant1977}
Fabricant,~B.; Krieger,~D.; Muenter,~J. Molecular beam electric resonance study
  of formaldehyde, thioformaldehyde, and ketene. \emph{The Journal of Chemical
  Physics} \textbf{1977}, \emph{67}, 1576--1586\relax
\mciteBstWouldAddEndPuncttrue
\mciteSetBstMidEndSepPunct{\mcitedefaultmidpunct}
{\mcitedefaultendpunct}{\mcitedefaultseppunct}\relax
\EndOfBibitem
\bibitem[Brown \latin{et~al.}(1990)Brown, Godfrey, McNaughton, Pierlot, and
  Taylor]{brown1990}
Brown,~R.~D.; Godfrey,~P.~D.; McNaughton,~D.; Pierlot,~A.~P.; Taylor,~W.~H.
  Microwave spectrum of ketene. \emph{Journal of Molecular Spectroscopy}
  \textbf{1990}, \emph{140}, 340--352\relax
\mciteBstWouldAddEndPuncttrue
\mciteSetBstMidEndSepPunct{\mcitedefaultmidpunct}
{\mcitedefaultendpunct}{\mcitedefaultseppunct}\relax
\EndOfBibitem
\bibitem[Kleiner \latin{et~al.}(1996)Kleiner, Lovas, and
  Godefroid]{kleiner1996}
Kleiner,~I.; Lovas,~F.~J.; Godefroid,~M. Microwave spectra of molecules of
  astrophysical interest. XXIII. Acetaldehyde. \emph{Journal of Physical and
  Chemical Reference Data} \textbf{1996}, \emph{25}, 1113--1210\relax
\mciteBstWouldAddEndPuncttrue
\mciteSetBstMidEndSepPunct{\mcitedefaultmidpunct}
{\mcitedefaultendpunct}{\mcitedefaultseppunct}\relax
\EndOfBibitem
\bibitem[Motiyenko \latin{et~al.}(2012)Motiyenko, Tercero, Cernicharo, and
  Margul{\`e}s]{motiyenko2012}
Motiyenko,~R.; Tercero,~B.; Cernicharo,~J.; Margul{\`e}s,~L. Rotational
  spectrum of formamide up to 1 THz and first ISM detection of its $\nu$12
  vibrational state. \emph{Astronomy \& Astrophysics} \textbf{2012},
  \emph{548}, A71\relax
\mciteBstWouldAddEndPuncttrue
\mciteSetBstMidEndSepPunct{\mcitedefaultmidpunct}
{\mcitedefaultendpunct}{\mcitedefaultseppunct}\relax
\EndOfBibitem
\bibitem[Kryvda \latin{et~al.}(2009)Kryvda, Gerasimov, Dyubko, Alekseev, and
  Motiyenko]{kryvda2009}
Kryvda,~A.; Gerasimov,~V.; Dyubko,~S.; Alekseev,~E.; Motiyenko,~R. New
  measurements of the microwave spectrum of formamide. \emph{Journal of
  Molecular Spectroscopy} \textbf{2009}, \emph{254}, 28--32\relax
\mciteBstWouldAddEndPuncttrue
\mciteSetBstMidEndSepPunct{\mcitedefaultmidpunct}
{\mcitedefaultendpunct}{\mcitedefaultseppunct}\relax
\EndOfBibitem
\bibitem[Blanco \latin{et~al.}(2006)Blanco, L{\'o}pez, Lesarri, and
  Alonso]{blanco2006}
Blanco,~S.; L{\'o}pez,~J.~C.; Lesarri,~A.; Alonso,~J.~L. Microsolvation of
  formamide: a rotational study. \emph{Journal of the American Chemical
  Society} \textbf{2006}, \emph{128}, 12111--12121\relax
\mciteBstWouldAddEndPuncttrue
\mciteSetBstMidEndSepPunct{\mcitedefaultmidpunct}
{\mcitedefaultendpunct}{\mcitedefaultseppunct}\relax
\EndOfBibitem
\bibitem[Vorob'eva and Dyubko(1994)Vorob'eva, and Dyubko]{vorobeva1994}
Vorob'eva,~E.; Dyubko,~S. Submillimeter rotational spectrum of the formamide
  molecule: Ground and first excited vibrational states. \emph{Radiophysics and
  quantum electronics} \textbf{1994}, \emph{37}, 155--158\relax
\mciteBstWouldAddEndPuncttrue
\mciteSetBstMidEndSepPunct{\mcitedefaultmidpunct}
{\mcitedefaultendpunct}{\mcitedefaultseppunct}\relax
\EndOfBibitem
\bibitem[Moskienko and Dyubko(1991)Moskienko, and Dyubko]{moskienko1991}
Moskienko,~E.; Dyubko,~S. Submillimeter rotational spectrum of formamide in the
  ground vibrational state. \emph{Radiophysics and quantum electronics}
  \textbf{1991}, \emph{34}, 181--183\relax
\mciteBstWouldAddEndPuncttrue
\mciteSetBstMidEndSepPunct{\mcitedefaultmidpunct}
{\mcitedefaultendpunct}{\mcitedefaultseppunct}\relax
\EndOfBibitem
\bibitem[Gardner \latin{et~al.}(1980)Gardner, Godfrey, and
  Williams]{gardner1980}
Gardner,~F.; Godfrey,~P.; Williams,~D. Observations of the 12C and 13C isotopes
  of formamide at 19 cm. \emph{Monthly Notices of the Royal Astronomical
  Society} \textbf{1980}, \emph{193}, 713--721\relax
\mciteBstWouldAddEndPuncttrue
\mciteSetBstMidEndSepPunct{\mcitedefaultmidpunct}
{\mcitedefaultendpunct}{\mcitedefaultseppunct}\relax
\EndOfBibitem
\bibitem[Hirota \latin{et~al.}(1974)Hirota, Sugisaki, Nielsen, and
  S{\o}rensen]{hirota1974}
Hirota,~E.; Sugisaki,~R.; Nielsen,~C.~J.; S{\o}rensen,~G.~O. Molecular
  structure and internal motion of formamide from microwave spectrum.
  \emph{Journal of Molecular Spectroscopy} \textbf{1974}, \emph{49},
  251--267\relax
\mciteBstWouldAddEndPuncttrue
\mciteSetBstMidEndSepPunct{\mcitedefaultmidpunct}
{\mcitedefaultendpunct}{\mcitedefaultseppunct}\relax
\EndOfBibitem
\bibitem[Kukolich and Nelson(1971)Kukolich, and Nelson]{kukolich1971b}
Kukolich,~S.~G.; Nelson,~A. High resolution rotational spectra of formamide.
  \emph{Chemical Physics Letters} \textbf{1971}, \emph{11}, 383--384\relax
\mciteBstWouldAddEndPuncttrue
\mciteSetBstMidEndSepPunct{\mcitedefaultmidpunct}
{\mcitedefaultendpunct}{\mcitedefaultseppunct}\relax
\EndOfBibitem
\bibitem[Kutsenko \latin{et~al.}(2013)Kutsenko, Motiyenko, Margul{\`e}s, and
  Guillemin]{kutsenko2013}
Kutsenko,~A.; Motiyenko,~R.; Margul{\`e}s,~L.; Guillemin,~J.-C. The extended
  spectroscopic database for deuterated species of formamide up to 1 THz.
  \emph{Astronomy \& Astrophysics} \textbf{2013}, \emph{549}, A128\relax
\mciteBstWouldAddEndPuncttrue
\mciteSetBstMidEndSepPunct{\mcitedefaultmidpunct}
{\mcitedefaultendpunct}{\mcitedefaultseppunct}\relax
\EndOfBibitem
\bibitem[Peter and Dreizler(1965)Peter, and Dreizler]{peter1965}
Peter,~R.; Dreizler,~H. Das Mikrowellenspektrum von Aceton im
  Torsionsgrundzustand. \emph{Zeitschrift f{\"u}r Naturforschung A}
  \textbf{1965}, \emph{20}, 301--312\relax
\mciteBstWouldAddEndPuncttrue
\mciteSetBstMidEndSepPunct{\mcitedefaultmidpunct}
{\mcitedefaultendpunct}{\mcitedefaultseppunct}\relax
\EndOfBibitem
\bibitem[Vacherand \latin{et~al.}(1986)Vacherand, Van~Eijck, Burie, and
  Demaison]{vacherand1986}
Vacherand,~J.; Van~Eijck,~B.; Burie,~J.; Demaison,~J. The rotational spectrum
  of acetone: internal rotation and centrifugal distortion analysis.
  \emph{Journal of Molecular Spectroscopy} \textbf{1986}, \emph{118},
  355--362\relax
\mciteBstWouldAddEndPuncttrue
\mciteSetBstMidEndSepPunct{\mcitedefaultmidpunct}
{\mcitedefaultendpunct}{\mcitedefaultseppunct}\relax
\EndOfBibitem
\bibitem[Oldag and Sutter(1992)Oldag, and Sutter]{oldag1992}
Oldag,~F.; Sutter,~D.~H. The Rotational Zeeman Effect in Acetone, its g-Tensor,
  its Magnetic Susceptibility Anisotropics and its Molecular Electric
  Quadrupole Moment Tensor; A High Resolution Microwave Fourier Transform
  Study. \emph{Zeitschrift f{\"u}r Naturforschung A} \textbf{1992}, \emph{47},
  527--532\relax
\mciteBstWouldAddEndPuncttrue
\mciteSetBstMidEndSepPunct{\mcitedefaultmidpunct}
{\mcitedefaultendpunct}{\mcitedefaultseppunct}\relax
\EndOfBibitem
\bibitem[Groner \latin{et~al.}(2002)Groner, Albert, Herbst, De~Lucia, Lovas,
  Drouin, and Pearson]{groner2002}
Groner,~P.; Albert,~S.; Herbst,~E.; De~Lucia,~F.~C.; Lovas,~F.~J.;
  Drouin,~B.~J.; Pearson,~J.~C. Acetone: laboratory assignments and predictions
  through 620 GHz for the vibrational-torsional ground state. \emph{The
  Astrophysical Journal Supplement Series} \textbf{2002}, \emph{142}, 145\relax
\mciteBstWouldAddEndPuncttrue
\mciteSetBstMidEndSepPunct{\mcitedefaultmidpunct}
{\mcitedefaultendpunct}{\mcitedefaultseppunct}\relax
\EndOfBibitem
\bibitem[{Ilyushin} \latin{et~al.}(2004){Ilyushin}, {Alekseev}, {Dyubko},
  {Kleiner}, and {Hougen}]{ilyushin2004}
{Ilyushin},~V.~V.; {Alekseev},~E.~A.; {Dyubko},~S.~F.; {Kleiner},~I.;
  {Hougen},~J.~T. {Ground and first excited torsional states of acetamide}.
  \emph{Journal of Molecular Spectroscopy} \textbf{2004}, \emph{227},
  115--139\relax
\mciteBstWouldAddEndPuncttrue
\mciteSetBstMidEndSepPunct{\mcitedefaultmidpunct}
{\mcitedefaultendpunct}{\mcitedefaultseppunct}\relax
\EndOfBibitem
\bibitem[Ilyushin \latin{et~al.}(2013)Ilyushin, Endres, Lewen, Schlemmer, and
  Drouin]{ilyushin2013}
Ilyushin,~V.~V.; Endres,~C.~P.; Lewen,~F.; Schlemmer,~S.; Drouin,~B.~J.
  Submillimeter wave spectrum of acetic acid. \emph{Journal of Molecular
  Spectroscopy} \textbf{2013}, \emph{290}, 31--41\relax
\mciteBstWouldAddEndPuncttrue
\mciteSetBstMidEndSepPunct{\mcitedefaultmidpunct}
{\mcitedefaultendpunct}{\mcitedefaultseppunct}\relax
\EndOfBibitem
\bibitem[{Remijan} \latin{et~al.}(2014){Remijan}, {Snyder}, {McGuire}, {Kuo},
  {Looney}, {Friedel}, {Golubiatnikov}, {Lovas}, {Ilyushin}, {Alekseev},
  {Dyubko}, {McCall}, and {Hollis}]{remijan2014}
{Remijan},~A.~J.; {Snyder},~L.~E.; {McGuire},~B.~A.; {Kuo},~H.-L.;
  {Looney},~L.~W.; {Friedel},~D.~N.; {Golubiatnikov},~G.~Y.; {Lovas},~F.~J.;
  {Ilyushin},~V.~V.; {Alekseev},~E.~A.; {Dyubko},~S.~F.; {McCall},~B.~J.;
  {Hollis},~J.~M. {Observational Results of a Multi-telescope Campaign in
  Search of Interstellar Urea [(NH$_{2}$)$_{2}$CO]}. \emph{The Astrophysical Journal} \textbf{2014},
  \emph{783}, 77\relax
\mciteBstWouldAddEndPuncttrue
\mciteSetBstMidEndSepPunct{\mcitedefaultmidpunct}
{\mcitedefaultendpunct}{\mcitedefaultseppunct}\relax
\EndOfBibitem
\bibitem[Brown \latin{et~al.}(1975)Brown, Godfrey, and Storey]{brown1975}
Brown,~R.; Godfrey,~P.; Storey,~J. The microwave spectrum of urea.
  \emph{Journal of Molecular Spectroscopy} \textbf{1975}, \emph{58},
  445--450\relax
\mciteBstWouldAddEndPuncttrue
\mciteSetBstMidEndSepPunct{\mcitedefaultmidpunct}
{\mcitedefaultendpunct}{\mcitedefaultseppunct}\relax
\EndOfBibitem
\bibitem[Kasten and Dreizler(1986)Kasten, and Dreizler]{kasten1986}
Kasten,~W.; Dreizler,~H. A highly resolved rotational transition of urea
  measured for radioastronomical searches. Analysis of the nitrogen quadrupole
  coupling. \emph{Zeitschrift f{\"u}r Naturforschung A} \textbf{1986},
  \emph{41}, 1173--1174\relax
\mciteBstWouldAddEndPuncttrue
\mciteSetBstMidEndSepPunct{\mcitedefaultmidpunct}
{\mcitedefaultendpunct}{\mcitedefaultseppunct}\relax
\EndOfBibitem
\bibitem[Kretschmer \latin{et~al.}(1996)Kretschmer, Consalvo, Knaack, Schade,
  Stahl, and Dreizler]{kretschmer1996}
Kretschmer,~U.; Consalvo,~D.; Knaack,~A.; Schade,~W.; Stahl,~W.; Dreizler,~H.
  The 14N quadrupole hyperfine structure in the rotational spectrum of laser
  vaporized urea observed by molecular beam Fourier transform microwave
  spectroscopy. \emph{Molecular Physics} \textbf{1996}, \emph{87},
  1159--1168\relax
\mciteBstWouldAddEndPuncttrue
\mciteSetBstMidEndSepPunct{\mcitedefaultmidpunct}
{\mcitedefaultendpunct}{\mcitedefaultseppunct}\relax
\EndOfBibitem
\bibitem[Christiansen(2005)]{christiansen2005}
Christiansen,~J.~J. The microwave spectrum of cyanoformamide. \emph{Journal of
  Molecular Spectroscopy} \textbf{2005}, \emph{231}, 131--136\relax
\mciteBstWouldAddEndPuncttrue
\mciteSetBstMidEndSepPunct{\mcitedefaultmidpunct}
{\mcitedefaultendpunct}{\mcitedefaultseppunct}\relax
\EndOfBibitem
\bibitem[Winnewisser \latin{et~al.}(2005)Winnewisser, Medvedev, De~Lucia,
  Herbst, Koput, Sastry, and Butler]{winnewisser2005}
Winnewisser,~M.; Medvedev,~I.~R.; De~Lucia,~F.~C.; Herbst,~E.; Koput,~J.;
  Sastry,~K.; Butler,~R.~A. The millimeter-and submillimeter-wave spectrum of
  cyanoformamide. \emph{The Astrophysical Journal Supplement Series}
  \textbf{2005}, \emph{159}, 189\relax
\mciteBstWouldAddEndPuncttrue
\mciteSetBstMidEndSepPunct{\mcitedefaultmidpunct}
{\mcitedefaultendpunct}{\mcitedefaultseppunct}\relax
\EndOfBibitem
\end{mcitethebibliography}

\end{document}